\documentclass[12pt,preprint]{aastex}

\usepackage{amssymb}
\usepackage{graphicx}

\begin{document}

\title{Infrared Luminosities and Aromatic Features in the
24\,$\mu$m Flux Limited Sample of 5MUSES} 

\author{Yanling Wu\altaffilmark{1}, George Helou\altaffilmark{1}, Lee
  Armus\altaffilmark{2}, Diane Cormier\altaffilmark{3}, Yong
  Shi\altaffilmark{1}, Daniel Dale\altaffilmark{4}, Kalliopi
  Dasyra\altaffilmark{5}, J.D. Smith\altaffilmark{6}, Casey
  Papovich\altaffilmark{7}, Bruce Draine\altaffilmark{8}, Nurur
  Rahman\altaffilmark{9}, Sabrina Stierwalt\altaffilmark{2}, Dario
  Fadda\altaffilmark{10}, G. Lagache\altaffilmark{11}, Edward L.
  Wright\altaffilmark{12}}

\altaffiltext{1}{Infrared Processing and Analysis Center, California
  Institute of Technology, 1200 E. California Blvd, Pasadena, CA
  91125} 

\altaffiltext{2}{Spitzer Science Center, California Institute of
  Technology, 1200 E California Blvd, Pasadena, CA, 91125}

\altaffiltext{3}{Laboratoire AIM, CEA/DSM-CNRS-Universite Paris
  Diderot, Irfu/Service d'Astrophysique, CEA Saclay, 91191
  Gif-sur-Yvette}

\altaffiltext{4}{Department of Physics \& Astronomy, University of Wyoming, USA}

\altaffiltext{5}{Irfu/Service d' Astrophysique, CEA Saclay, France}

\altaffiltext{6}{Ritter Astrophysical Observatory, University of
  Toledo, Toledo, OH, 43606}

\altaffiltext{7}{George P. and Cynthia Woods Mitchell Institute for
  Fundamental Physics and Astronomy, Department of Physics and
  Astronomy, Texas A\&M University, College Station, TX 77843-4242 }

\altaffiltext{8}{Department of Astrophysical Sciences, 108 Peyton
  Hall, Princeton University, Princeton, NJ, 08544}

\altaffiltext{9}{Department of Astronomy, University of Maryland,
  College Park, MD, 20742}

\altaffiltext{10}{NASA Herschel Science Center, California Institute
  of Technology, 1200 E California Blvd, Pasadena, CA, 91125}

\altaffiltext{11}{Institut d'Astrophysique Spatiale (IAS), Batiment
  121, F-91405 Orsay, France}

\altaffiltext{12}{Department of Physics \& Astronomy, University of
  California, P O Box 951547, Los Angeles, 90095}

\email{yanling@ipac.caltech.edu, gxh@ipac.caltech.edu,
  lee@ipac.caltech.edu, diane.cormier@cea.fr, yong@ipac.caltech.edu,
  ddale@uwyo.edu, kalliopi.dasyra@cea.fr, jd.smith@utoledo.edu,
  papovich@physics.tamu.edu, draine@astro.princeton.edu,
  nurur@astro.umd.edu, sabrina@ipac.caltech.edu,
  fadda@ipac.caltech.edu, guilaine.lagache@ias.u-psud.fr, wright@astro.ucla.edu}

\begin{abstract}
  We study a 24\,$\mu$m selected sample of 330 galaxies observed with
  the Infrared Spectrograph for the 5\,mJy Unbiased Spitzer
  Extragalactic Survey.  We estimate accurate total infrared
  luminosities by combining mid-IR spectroscopy and mid-to-far
  infrared photometry, and by utilizing new empirical spectral
  templates from {\em Spitzer} data.  The infrared luminosities of
  this sample range mostly from 10$^9$L$_\odot$ to
  $10^{13.5}$L$_\odot$, with 83\% in the range
  10$^{10}$L$_\odot$$<$L$_{\rm IR}$$<$10$^{12}$L$_\odot$. The
  redshifts range from 0.008 to 4.27, with a median of 0.144. The
  equivalent widths of the 6.2\,$\mu$m aromatic feature have a bimodal
  distribution. We use the 6.2\,$\mu$m PAH EW to classify our objects
  as SB-dominated (44\%), SB-AGN composite (22\%), and AGN-dominated
  (34\%). The high EW objects (SB-dominated) tend to have steeper
  mid-IR to far-IR spectral slopes and lower L$_{\rm IR}$ and
  redshifts. The low EW objects (AGN-dominated) tend to have less
  steep spectral slopes and higher L$_{\rm IR}$ and redshifts.  This
  dichotomy leads to a gross correlation between EW and slope, which
  does not hold within either group.  AGN dominated sources tend to
  have lower log(L$_{\rm PAH 7.7\mu m}$/L$_{\rm PAH 11.3\mu m}$)
  ratios than star-forming galaxies, possibly due to preferential
  destruction of the smaller aromatics by the AGN. The log(L$_{\rm PAH
    7.7\mu m}$/L$_{\rm PAH 11.3\mu m}$) ratios for star-forming
  galaxies are lower in our sample than the ratios measured from the
  nuclear spectra of nearby normal galaxies, most probably indicating
  a difference in the ionization state or grain size distribution
  between the nuclear regions and the entire galaxy.  Finally, we
  provide a calibration relating the monochromatic 5.8, 8, 14 and
  24\,$\mu$m continuum or Aromatic Feature luminosity to L$_{\rm IR}$
  for different types of objects.

\end{abstract}
\keywords{galaxies: active, galaxies: starburst, galaxies: evolution, infrared radiation, surveys}

\section{Introduction}

Infrared bright galaxies play critical roles in galaxy formation and
evolution. The {\em InfraRed Astronomical Satellite (IRAS)} facilitated
the study of an important group of objects, the Ultra Luminous
InfraRed Galaxies (ULIRGs) \citep{Soifer89, Sanders96}, which were
first hinted at by ground based observations of \citet{Rieke72}. Studies
from the {\em Infrared Space Observatory (ISO)} \citep{Elbaz99} and
the {\em Spitzer} Space Telescope \citep{Houck05, Yan07} later
revealed that LIRGs and ULIRGs are much more common at high redshift
than in the local Universe. The number density of IR luminous galaxies
evolves strongly with redshift to at least z$\sim$1
\citep{Lefloch05}. The fraction of galaxies powered by star formation
versus AGN is still controversial, but is crucial for determining
unbiased luminosity functions for various categories of objects and
understanding the evolution process.

The superb sensitivity of the {\em Spitzer} Space Telescope
\citep{Werner04} has led to the discovery of new populations of faint,
high-redshift galaxies with extreme IR/optical colors
\citep{Dickinson04, Houck05, Weedman06, Yan07, Caputi07, Dey08,
  Dasyra09}. However, these studies often have at least one other
constraint than the mid-IR flux limit, usually a minimum R band
magnitude or an IRAC-based color selection, designed to favor sources
in specific redshift ranges, or with high luminosity. The 5
Millijanksy Unbiased {\em Spitzer} Extragalactic Survey (5MUSES) is an
infrared selected sample. A major advantage of 5MUSES is its simple
selection: f$_\nu$(24$\mu$m)$>$5mJy. This relatively bright flux limit
allows for a more detailed study of the infrared properties, filling in
the gap between local galaxies and high redshift samples, and helping
to improve the modeling of galaxy populations and their evolution.

In order to advance our understanding of the properties and evolution
of galaxies, it is crucial to obtain accurate estimates of their
bolometric luminosities. Several studies have shown that monochromatic
luminosities in the mid-IR can be used to estimate L$_{\rm IR}$
\citep{Sajina07, Bavouzet08, Rieke09, Calzetti10}, and the
uncertainties on these estimates decrease significantly when
far-infrared (FIR) fluxes are available \citep{Kartaltepe10}. However,
the spectral energy distribution (SED) of star-forming galaxies, AGN
and ULIRGs display a wide range of shapes \citep{Weedman05, Brandl06,
  Smith07, Armus07, Hao07, Wu09, Veilleux09}. Applying these methods
without knowing a source's spectral type could cause significant
biases in luminosity estimates between types of objects and seriously
mislead the interpretations. The 5 to 36\,$\mu$m spectra obtained by
the Infrared Spectrograph (IRS) \citep{Houck04} for the 5MUSES sample allows for aromatic
feature identification, excitation line analysis, and decomposition
into star formation and AGN components, thus providing essential
information for classifying the origin of the luminosity.

The mid-IR is home to a set of broad emission line features, which are
thought to originate from Polycyclic Aromatic Hydrocarbons
\citep{Puget85, Allamandola89}. PAHs are organic molecules that are
ubiquitous in our own Galaxy \citep{Peeters02} and nearby star-forming
galaxies \citep{Helou01, Smith07}. In total, they can contribute a
significant fraction (10\% or more) of the total infrared luminosity
in star-forming galaxies. PAHs are weak in low metallicity galaxies
\citep{Madden06, Wu06, Engelbracht08}, or in galaxies with powerful
\citep{Roche91, Weedman05, Armus07, Desai07, Wu09} or even weak AGN
\citep{Smith07, Dale09}.  The PAH features, including their profiles,
central wavelengths and band-to-band intensity ratios have been studied
in detail by \citet{Peeters02}, \citet{Smith07} and most recently
reviewed by \citet{Tielens08}. The 6.2\,$\mu$m feature and the
7.7\,$\mu$m complex are attributed to vibrational modes of the carbon
skeleton.  The 8.6\,$\mu$m feature is attributed to in-plane C-H
bending, while the features at 11.3\,$\mu$m and 12.7\,$\mu$m are
identified as out-of-plane C-H bending modes.  It is generally thought
that charged PAHs radiate more strongly in the C-C vibrational modes,
while neutral PAHs radiate strongly in the out-of-plane C-H bending
modes at 11.3\,$\mu$m and 12.7\,$\mu$m.  The fraction of the power
radiated by PAH in the different bands following single-photon
heating depends on both the PAH ionization and on the size of the PAH
\citep{Draine07}. Thus the observed variations in the PAH band-to-band
ratios can reflect variations in physical conditions \citep{Smith07,
  Galliano08, Gordon08, Odowd09}.

Because PAH emission can be very prominent in star-forming systems, it
has often been used as a relatively extinction-free diagnostic tool to
constrain star formation. Detailed studies on the properties of PAH
features locally \citep{Spoon07, Desai07} and at higher redshift
\citep{Yan05, Houck05, Huang09} reveal differences in the PAH Equivalength Widths (EWs) and
L$_{\rm PAH}$/L$_{\rm IR}$ ratios. This might indicate that some
evolution in the PAH properties occurs with redshift, or that sample
selection effects make for large variations in the aromatic feature
properties. However, one cannot simply apply our knowledge from the
local universe to high redshift galaxies, or make fair comparisons
between the two unless truly equivalent samples have been
studied. Current analysis on the PAH properties are based on {\em ISO}
or {\em Spitzer} observations of relatively bright objects, which have
been selected because of previously known optical or {\em IRAS}
criteria. Thus it is crucial to have a complete or at least unbiased
census of galaxies in order to understand the galaxy evolution process
and its relation to the aromatic feature emission.

In this paper, we study the properties of PAH emission and IR
luminosities. This is the first of a series of papers to study the IR
selected representative sample of 5MUSES. Helou et al. (in
preparation) will address the general properties of the sample and how
it bridges the gap between local and high-z galaxies. Yong et al. (in
preparation) will present the correlations between old stars and
current star formation. Detailed population modelling will also be
performed to address the bimodal distribution of the PAH EWs
discovered in this study.  In Section 2, we briefly describe the
sample selection, data reduction and measurements of spectral
features. We introduce our library of empirical IR SED templates built
upon {\em Spitzer} observations in Section 3, and derive the total
infrared luminosities for 5MUSES galaxies. We also discuss how well
one can constrain the IR SED if only mid-IR data are available. In
Section 4, we study the properties of PAH emission from our flux
limited sample.  Finally, we present our conclusions in Section
5. Using the IR luminosities we derived in Section 3 and the PAH
luminosities from Section 4, we discuss estimation of L$_{\rm IR}$
from PAH luminosity or monochromatic continuum luminosity in the
Appendix. Throughout this work, we assume a $\Lambda$CDM cosmology
with H$_{0}$=70\,km\,s$^{-1}$\,Mpc$^{-1}$, $\Omega_m$=0.27 and
$\Omega_\lambda$=0.73.

\section{Observations and Data Analysis}
\subsection{The Sample}

5MUSES is a mid-IR spectroscopic survey of a 24\,$\mu$m flux-limited
(f$_{24\mu m}>$5\,mJy) representative sample of 330 galaxies. The
galaxies are selected from the SWIRE fields \citep{Lonsdale03},
including Elais-N1 (9.5\,deg$^2$), Elais-N2 (5.3\,deg$^2$), Lockman
Hole (11.6\,deg$^2$) and XMM (9.2\,deg$^2$), in addition to the {\em
  Spitzer} Extragalactic First Look Survey (XFLS, 5.0\,deg$^2$) field
\citep{Fadda06}. It provides a representative sample at intermediate
redshift ($<z>\sim$0.144) which bridges the gap between the bright,
nearby star-forming galaxies \citep{Kennicutt03, Smith07, Dale09},
local ULIRGs \citep{Armus07, Desai07,Veilleux09} and the much fainter
and more distant sources pursued in most z$\sim$2 IRS follow-up work
to date \citep{Houck05, Yan07}. The full details of the sample,
including selection criteria and observation strategy are covered in
Helou et al. (2010, in prep).

\subsection{Observation and Data Reduction}

Because of its selection in the SWIRE and XFLS fields, IRAC
3.6-8.0\,$\mu$m photometry is available for the entire 5MUSES
sample. In addition to the MIPS 24\,$\mu$m photometry used to select
this sample, 90\% of our sources have also been detected at MIPS
70\,$\mu$m and 54\% have been detected at MIPS
160\,$\mu$m. Low-resolution spectra (R=64$\sim$128) of all 330
galaxies in 5MUSES have been obtained with the Short-Low (SL:
5.5-14.5\,$\mu$m and Long-Low (LL: 14-35\,$\mu$m) modules of the IRS
using the staring mode observations. The integration time on each
object was estimated based on its 24\,$\mu$m flux densities and
typically ranges from 300-960 seconds (see Table
\ref{integration_time}).  A sub-set of the 5MUSES sample has also been
observed with the high-resolution modules of the IRS, which will be
covered in a future paper.

The low-resolution IRS data were processed by the {\em Spitzer}
Science Center data reduction pipeline version S17. The
two-dimensional image data were converted to slopes after
linearization correction, subtraction of darks, cosmic-ray removal,
stray light and flat field correction. The post-pipeline reduction of
the spectral data started from the pipeline products basic
calibrated data (bcd) files. We took the median of all images from the
off-source part of the slit (off-order and off-nod) and then
subtracted it from the image on the source. Then we combined all the
background-subtracted images at one nod position and took the
mean. The resulting images were then cleaned with the IRSCLEAN
package\footnote{For more details, check
  http://ssc.spitzer.caltech.edu/dataanalysistools/tools/irsclean/} to
remove bad pixels and apply rogue pixel correction.

We used the {\em Spitzer} IRS Custom Extractor (SPICE)\footnote{For
  more details, check
  http://ssc.spitzer.caltech.edu/dataanalysistools/tools/spice/}
software to extract the spectra.  With a flux limit of 5\,mJy at
24\,$\mu$m, we chose to use the optimal extraction with point-source
calibration because it significantly improved the S/N ratios for our
sources. When using the optimal method, each pixel was weighted by its
position, based on the spatial profile of a bright calibration star.
The outputs from SPICE produced one spectrum per order at each nod
position, which were then combined. We also trimmed the ends of each
order where the noise rose quickly. Finally, the flux-calibrated
spectra of each order (including the 1st, 2nd and 3rd orders) and
module were merged without applying any scaling factor between SL and
LL, and yielded a single spectrum per source. This spectrum was used
to estimate aromatic feature fluxes, continuum flux densities at
various wavelengths and line fluxes.

\subsection{Data Analysis}
\subsubsection{The PAH Fluxes and Equivalent  Widths}
\begin{figure}
  \epsscale{1.}
  \plotone{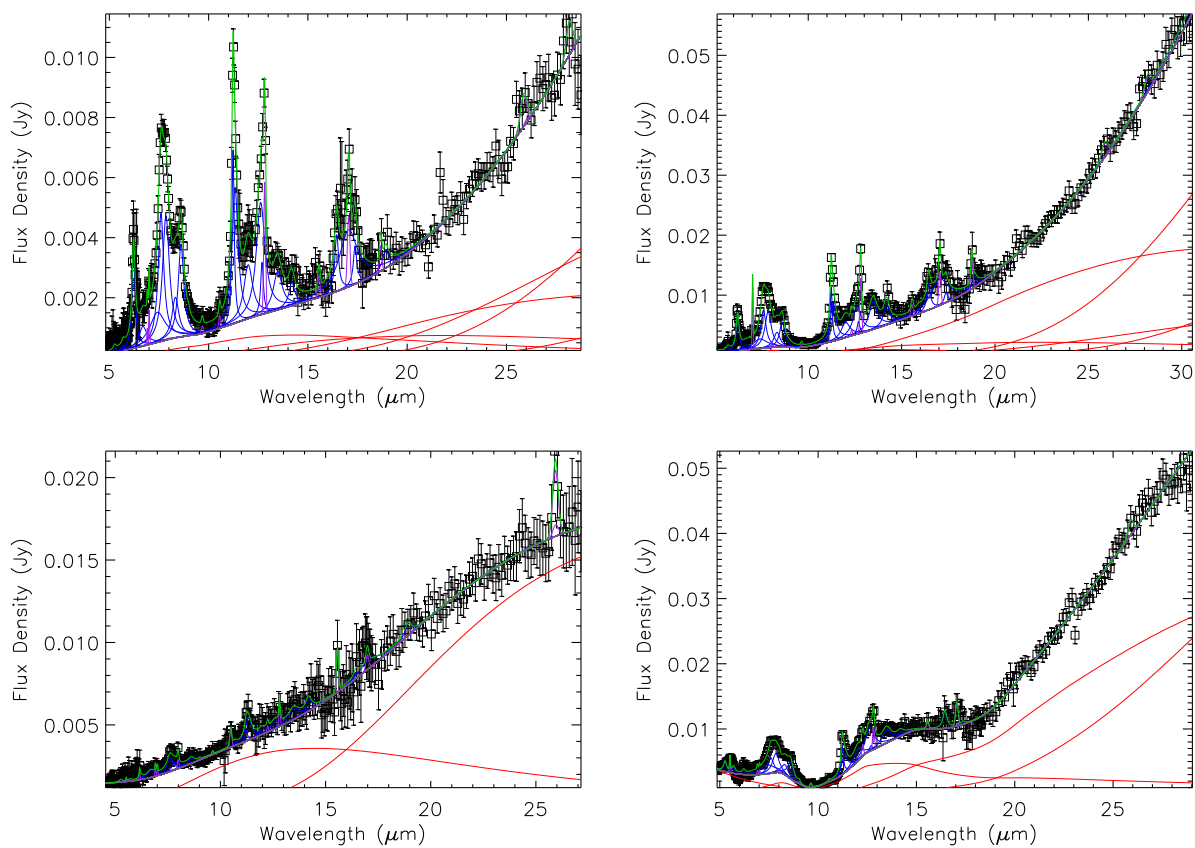}
  \caption{Examples of PAHFIT decomposition of 5MUSES spectra (black
    squares) with strong, moderate, weak PAH emission and with
    silicate absorption. The best fit SED (green) is composed of
    thermal dust continua (red), PAH features (blue), stellar light
    (magenta) and emission lines (purple).}
  \label{fig:pahfitsample}
\end{figure}

To study the properties of PAH emission in our sample, we have used
two methods to estimate the feature strength. The first method defines
a local continuum or ``plateau'' under the emission features at 6.2
and 11.3\,$\mu$m by fitting a spline function to selected points, and
measures the features above the continuum. The wavelength limits for
the integration of the features are approximately 5.95-6.55\,$\mu$m
for the 6.2\,$\mu$m PAH and 10.80-11.80 for the 11.3\,$\mu$m
PAH. We have not taken into account the possibility of water ice or
HAC absorption in our measurement of the 6.2\,$\mu$m PAH EW because
these features are known to be important mainly in strongly obscured
local ULIRGs \citep{Spoon04}; thus neglecting this component does not
significantly change the 6.2\,$\mu$m PAH EW. Although the 9.7\,$\mu$m
silicate feature could affect the measurement on the 11.3\,$\mu$m PAH,
our sample has very few deeply obscured sources. The PAH EWs are
derived by dividing the integrated flux over the average continuum
flux in each feature range. This PAH EW measured from the spline
fitting method is defined as the ``apparent PAH EW'' and is directly
comparable to the studies in the literature such as \citet{Peeters02,
  Spoon07, Armus07, Desai07, Pope08} and \citet{Dale09}. In the second
method, we use the PAHFIT software \citep{Smith07} to measure the PAHs
in our sample (see Figure \ref{fig:pahfitsample} for examples). In
PAHFIT, the PAH features are fit with Drude profiles, which have
extended wings that account for a significant fraction of the
underlying plateau \citep{Smith07}. As has been shown in
\citet{Smith07} and \citet{Galliano08}, although the PAHFIT method
gives higher values of PAH integrated fluxes or EWs due to the lower
continuum adopted than the ``apparent PAH EW'' method, the two methods
yield consistent results on trends, such as the variations of
band-to-band PAH luminosity ratios. Throughout this paper, when we
refer to PAH EWs, we mean the apparent PAH EWs measured from the
spline fitting method and they are used to classify object types. When
we refer to PAH flux or luminosity, we mean the values derived from
PAHFIT.

\subsubsection{The Fine-Structure Line Fluxes}
The mid-IR has a rich suite of fine-structure lines. [SIV]10.51$\mu$m,
[NeII]12.81$\mu$m, [NeIII]15.55$\mu$m, [SIII]18.71/33.48$\mu$m and
[SiII]34.82$\mu$m are the most frequently detected fine-structure
lines in the spectral range covered by the IRS. The high-excitation
line of [OIV]25.89$\mu$m has often been detected in low metallicity
galaxies, starburst galaxies or AGN, excited by the photoionization
and/or shocks associated with intense star formation or nuclear
activity, while the [NeV]14.32/24.32$\mu$m lines are frequently
detected in AGN-dominated sources and serve as unambiguous indicators
of an AGN.

We use the ISAP package in SMART \citep{Higdon04} to measure the
strength of the fine-structure lines. A Gaussian profile is adopted to
fit the lines above a local continuum. The continuum is derived by
linear fitting except for the [NeII]12.81$\mu$m line, which is blended
with the 12.7\,$\mu$m PAH feature. The continuum underlying the [NeII]
line is fit with a 2nd-order polynomial. The integrated fitted flux
above the continuum is taken as the total flux of the line.  Upper
limits are derived by measuring the flux with a height of three times
the local {\em rms} and a width equal to the instrument resolution. In
this paper, we only use the flux ratio of [NeIII]/[NeII] to compare
with the PAH strength, while the tabulated line fluxes will be
presented and discussed in a future paper.

\section{The Infrared Luminosities of the 5MUSES Sample}

Several SED libraries have been built to capture the variation in the
shape of IR SEDs and to estimate L$_{\rm IR}$ \citep{Dale02, Chary01,
  Draine07, Rieke09}. In the absence of multi-wavelength data,
monochromatic luminosities have also been widely used to estimate
L$_{\rm IR}$ \citep{Sajina07, Bavouzet08, Rieke09, Kartaltepe10}.  The
5MUSES sample has mid-IR spectra, in addition to the IRAC and MIPS
photometry, which allows us to account properly for variations in the
SED shape and obtain more accurate estimates of L$_{\rm IR}$.

\subsection{Constructing an SED Template Library}

In order to cover a wide range of SED shapes to fit the 5MUSES
sources, we have built an IR template library based on the recent
observations obtained from {\em Spitzer}.  The library encompasses 83
ULIRGs observed by the IRS GTO sample \citep{Armus07}; 75 normal
star-forming galaxies from Spitzer Infrared Nearby Galaxies Survey
(SINGS, \citet{Kennicutt03}); and 136 PG and 2MASS quasars
\citep{Shi07}. The templates in the library consist of SEDs derived
from IRS spectra and/or IRAC and MIPS photometry. For both the ULIRG
and PG/2MASS sources, full 1-1000\,$\mu$m SED have been obtained by
Marshall et al. (2010, in preparation) and Shi et al.(2010, in
preparation) from IRS, MIPS and IRAS observations. For the SINGS
galaxies, \citet{Dale07} have provided SED fits to the MIPS 24, 70 and
160\,$\mu$m photometry using the \citet{Dale02} templates. However,
these templates do not sample the full variation of the strength of
PAH features in the 5-15\,$\mu$m regime, due to the limited mid-IR
spectra available when the templates were created. As a result, when
we use SINGS galaxies as templates, we use their FIR SED from the fits
of \citet{Dale07}, while in the mid-IR, we use the observed IRAC
photometry integrated from the whole galaxy. This extensive template
library provides a good coverage on the variations of IR SEDs. 33\% of
our sources are best-fit with SINGS-type templates and 38\% are
best-fit with quasar-type templates. The remaining sources are
best-fit by ULIRG-type templates. The type of the best-fit template
also correlates well with the 6.2\,$\mu$m PAH EWs. SB-dominated
sources are normally best fit by SINGS-type templates and
AGN-dominated sources are best-fit with quasar-type templates. For
SB-AGN composite sources, the best-fit templates are divided among
ULIRG, SINGS and quasar-type templates (48\%, 37\% and 15\%
respectively).

\subsection{Estimating L$_{\rm IR}$ using {\em Spitzer} data}

Out of the 330 sources in 5MUSES, 280 galaxies have redshifts from
optical or mid-IR spectroscopy. We are in the process of obtaining
spec-z for the remaining 50 sources. We have estimated redshifts for
11 out of these 50 objects from silicate features or very weak PAH
features, but do not include them in the discussion of this paper
because of the large associated uncertainties.  For the 280 objects,
we use a combination of synthetic IRAC photometry obtained from the
rest-frame IRS spectra, as well as the observed MIPS photometry to
compare with the corresponding synthetic photometry from the SED
templates and estimate total L$_{\rm IR}$. We select the best-fit
template by minimizing $\chi^2$ and we use progressively more detailed
and accurate L$_{\rm IR}$ estimation methods for 5MUSES source with
more photometry available. The final SED is composed of the IRS
spectrum in the mid-IR and the best-fit template SED in the FIR.  In
the remainder of this section, we describe our method for estimating
L$_{\rm IR}$ and the associated uncertainties.

\subsubsection{Sources with MIPS FIR photometry}

\begin{figure}
  \epsscale{0.9}
  \plotone{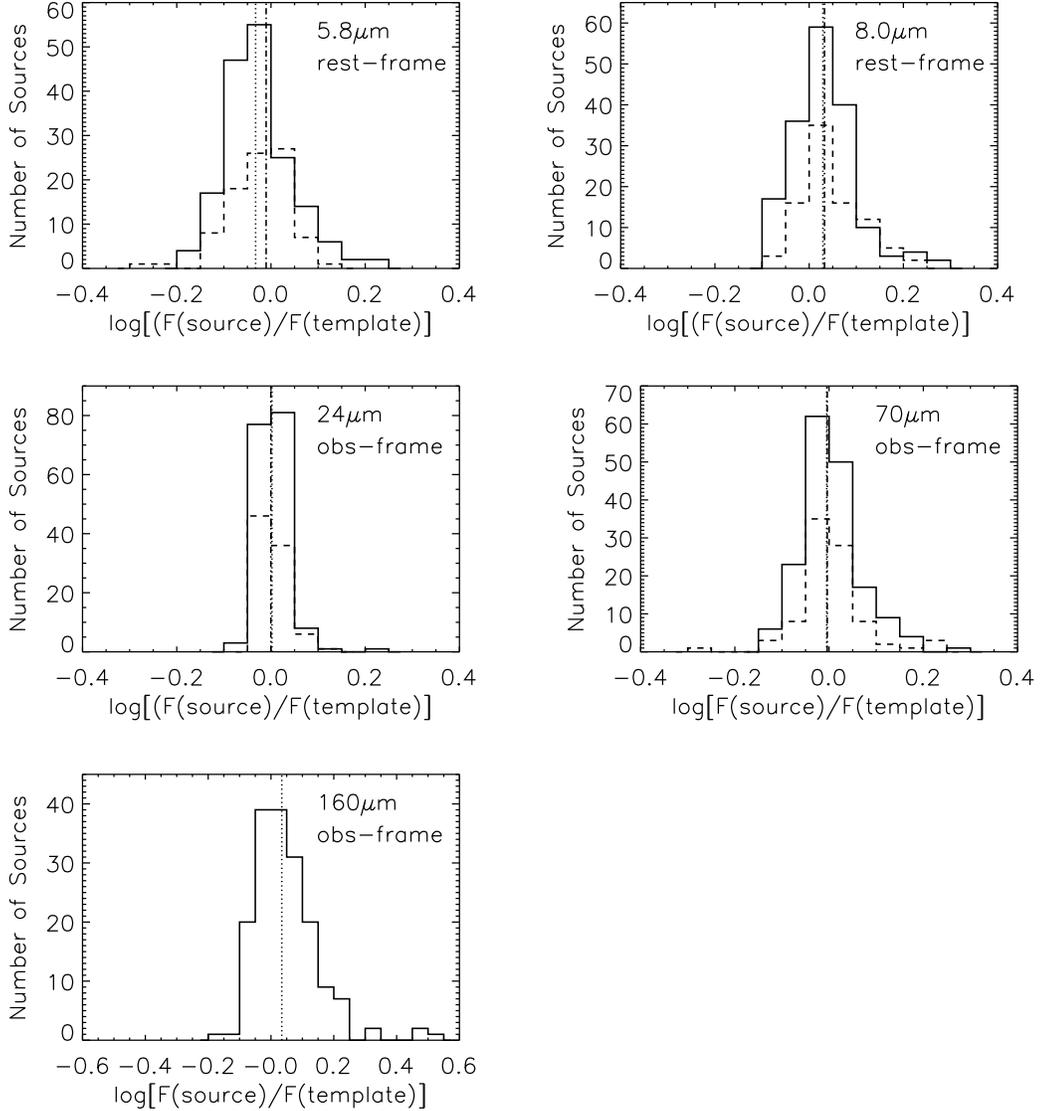}
  \caption{The histogram of the flux ratios between the source and the
    best-fit template, for galaxies with MIPS 70 and 160\,$\mu$m
    detections (solid line) and for galaxies only with MIPS 70\,$\mu$m
    detections (dashed line). The dotted and dash-dotted lines
    indicate the medians of the flux ratios for sources with 70 and
    160\,$\mu$m detections and sources with only 70\,$\mu$m
    detections. All five bands appear to peak around a
    log[(F(source)/F(template)] ratio of 0 with rather narrow
    distributions. The 1$\sigma$ deviations in F(source)/F(template)
    for the 5.8, 8.0, 24 and 70 160\,$\mu$m bands are 0.07, 0.07,
    0.03, 0.06 and 0.10 dex respectively for sources with 70 and
    160\,$\mu$m detections and 0.07, 0.07, 0.03 and 0.07 dex for
    sources with only 70\,$\mu$m detections.}
  \label{fig:compare_phot_70_160}
\end{figure}

For sources with FIR detection at MIPS 70 and 160\,$\mu$m, we use five
data points to fit their SEDs. The first two data points are
rest-frame IRAC 5.8 and 8.0\,$\mu$m\footnote{5MUSES-312 has a redshift
  of 4.27 and for this source, we only use its MIPS 70 and 160\,$\mu$m
  fluxes during the SED fitting.} derived by convolving the rest-frame
5MUSES spectrum with the filter response curves of IRAC 5.8 and
8.0\,$\mu$m. The other three data points are the observed MIPS 24, 70
and 160\,$\mu$m photometry for each 5MUSES source. The corresponding
data points from the templates are derived in the following way: For
ULIRG and PG/2MASS templates, the 5.8 and 8.0\,$\mu$m fluxes are
derived in the same manner as 5MUSES sources. The 24, 70 and
160\,$\mu$m data points are derived by convolving the template SED at
matching redshift with the MIPS 24, 70 and 160\,$\mu$m filter response
curves. For SINGS templates, we use directly the observed IRAC 5.8 and
8.0\,$\mu$m photometry as the first two data points, which are
essentially at rest-frame for all SINGS objects. Then we move the
SINGS SEDs given by Dale et al. (2007) to the redshift of the 5MUSES
source and derive the corresponding observed-frame MIPS 24, 70 and
160\,$\mu$m photometry.  During the SED fitting, we weight the data
points by their wavelength since the majority of the energy is emitted
at FIR for IR selected sources, and look for the template that fits
each 5MUSES source best by minimizing the $\chi^2$. A comparison of
the ratio of flux densities at rest-frame 5.8, 8.0\,$\mu$m and the
observed MIPS 24, 70 and 160\,$\mu$m photometry from the source and
the best-fit template can be found in Figure
\ref{fig:compare_phot_70_160} (solid line). The dispersion in the
ratio of the observed photometry over the photometry from the best-fit
template (F$_{\rm source}$/F$_{\rm template}$) in each band is 0.07,
0.07, 0.03, 0.06 and 0.10 dex respectively.

For sources with FIR detection only at MIPS 70\,$\mu$m, we apply the
same technique to fit the SED. We use the rest-frame IRAC 5.8 and
8.0\,$\mu$m photometry and the observed MIPS 24 and 70\,$\mu$m data in
our fitting. The upper limit at 160\,$\mu$m is used to exclude
templates for which the synthetic photometry exceeds the 2$\sigma$
upper limit of the source. A comparison of the ratio of flux densities
at rest-frame 5.8, 8.0\,$\mu$m and the observed MIPS 24 and 70\,$\mu$m
photometry from the source and the best-fit template can be found in
Figure \ref{fig:compare_phot_70_160} (dashed line). The dispersion in
the ratio of F$_{\rm source}$/F$_{\rm template}$ in each band is 0.07,
0.07, 0.03 and 0.07 dex respectively.

Once the best-fit template is identified, we derive L$_{\rm IR}$ as
explained in the last paragraph of the next section.

\subsubsection{Sources without MIPS FIR photometry}

19 objects do not have FIR detection even at 70\,$\mu$m. For these
sources, we select the IR SED based on the mid-IR spectra. Our method
is to fit the IRS spectrum of the 5MUSES source with the mid-IR
spectra of the templates in the corresponding wavelength regime and
adopt the SED of the best-fit template. The templates of which the
synthetic photometry exceed the 2$\sigma$ upper limits at MIPS 70 and
160\,$\mu$m bands are excluded.  As will be shown in Section 3.2.3,
this IRS-only method might underestimate the L$_{\rm IR}$ for cold
sources by $\sim$20\%, while it shows no significant offset for warm
sources\footnote{Warm sources are defined to have $f_{\rm 24\mu
    m}/f_{\rm 70\mu m}>$0.2, derived from the definition of $f_{\rm
    25\mu m}/f_{\rm 60\mu m}>$0.2 by \citet{Sanders88}}. All of the 19
objects in this category show SEDs with high ($f_{\rm 24\mu m}/f_{\rm
  70\mu m}$)$_{\rm obs}$ ratios\footnote{The 70\,$\mu$m flux densities
  are upper limits.}; thus they are more likely to be warm
sources. This suggests that our approach of using the IRS spectrum to
find the best-fit SED is unlikely to result in significant biases on
L$_{\rm IR}$.

\begin{figure}
  \epsscale{1.0}
  \plotone{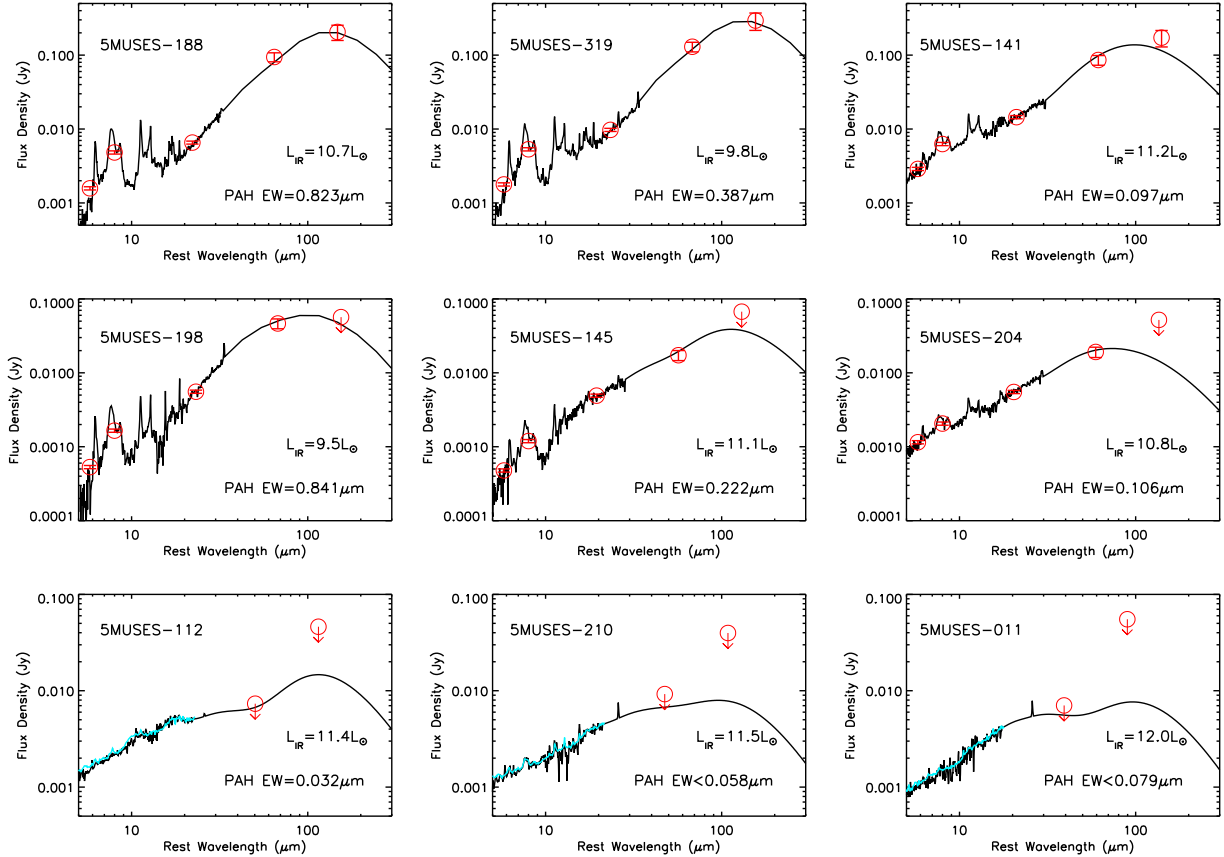}
  \caption{SEDs of a sample of nine 5MUSES sources. The final SED
    (black) is composed of the IRS spectrum in the mid-IR and the
    best-fit template in the FIR. The observed data are shown as red
    circles. The three sources in the top panel are fit with 5 data
    points (IRAC 5.8, 8.0\,$\mu$m and MIPS 24, 70 and
    160\,$\mu$m). The three sources in the middle panel are fit with 4
    data points (IRAC 5.8, 8.0\,$\mu$m and MIPS 24 and 70\,$\mu$m).
    The three sources in the bottom panel are fit with only the IRS
    spectra. The blue line is the mid-IR SED of the best-fit template
    for the sources fit with the IRS spectra.}
  \label{fig:sedfit}
\end{figure}

\begin{figure}
  \epsscale{1.6}
  \plottwo{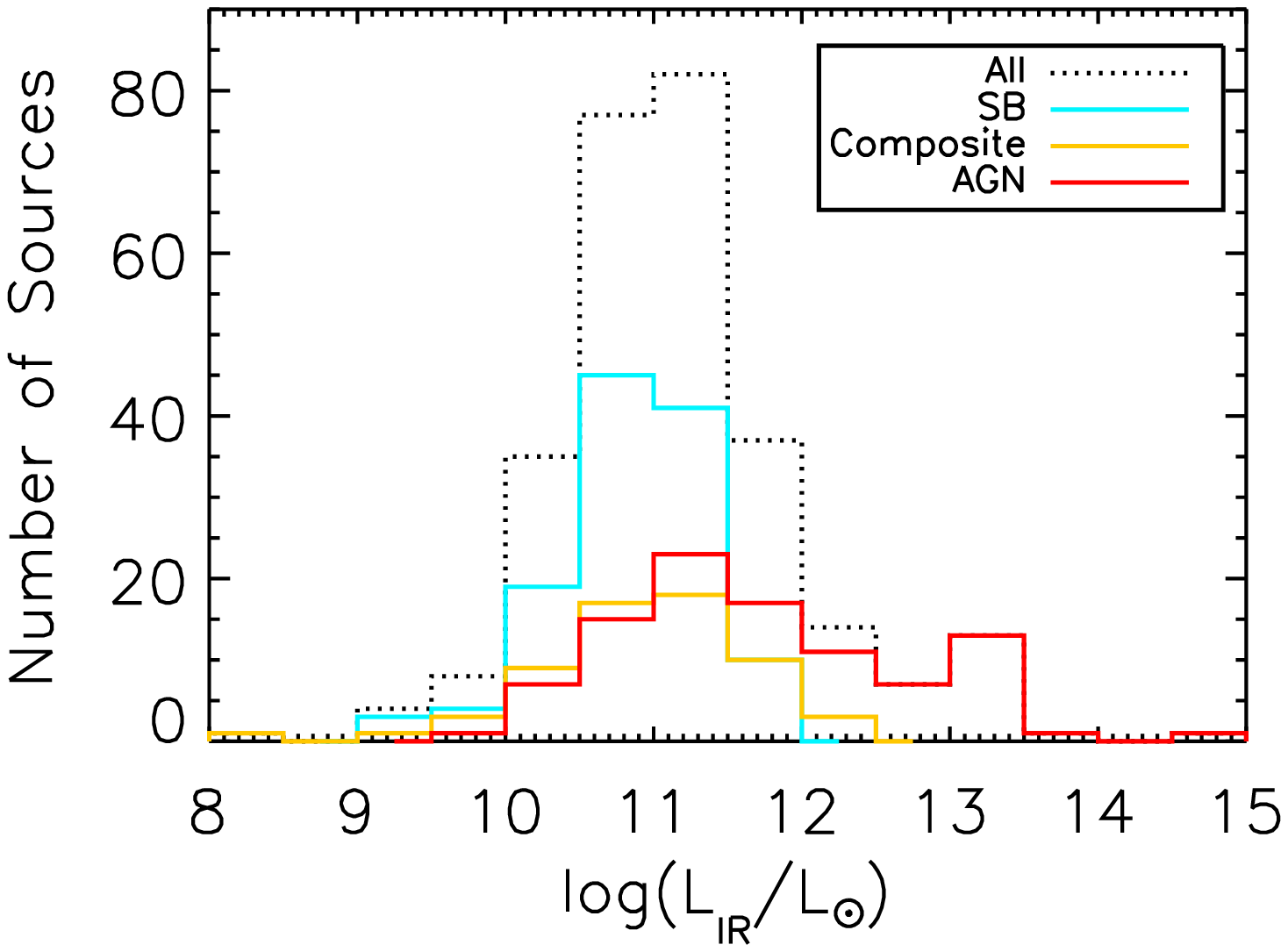}{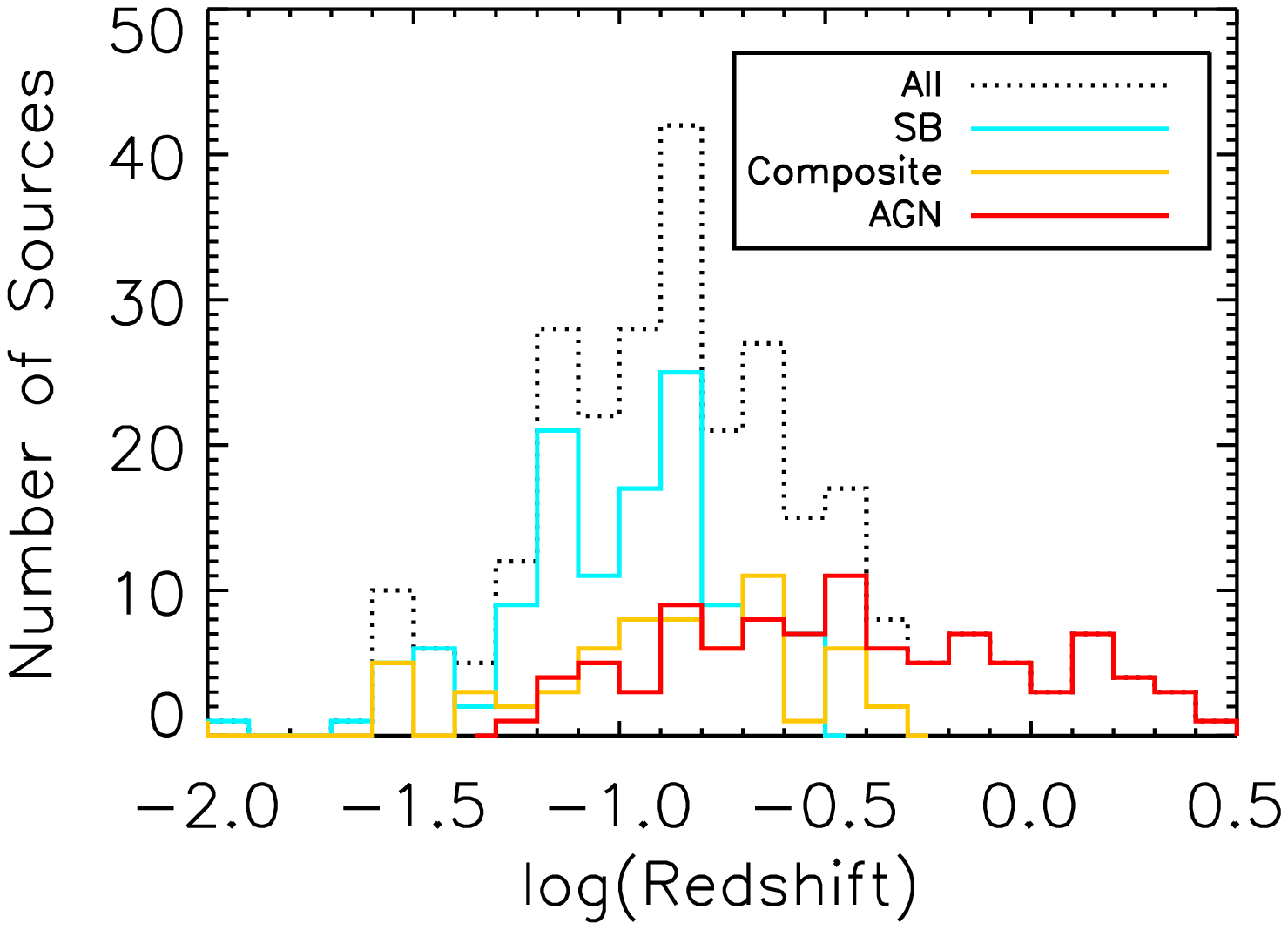}
  \caption{a) Top panel: The distribution of the total infrared
    luminosity of the 5MUSES sample. The dotted line represents the
    whole sample with known redshift. The blue, yellow and red solid
    lines represent the SB, composite and AGN sources in the
    sample. The SBs and AGN dominate the lower and higher end of the
    luminosity distribution respectively. b) Bottom panel: The
    distribution of the redshifts of the 5MUSES sample. The symbols
    are the same as in a).}
  \label{fig:LIR_z}
\end{figure}

Finally, for each source, we visually inspect the fitting results. We
find that a range of templates could fit the SED well. We construct
the final 5-1000\,$\mu$m SED of a galaxy by combining its IRS spectrum
in the mid-IR with the best-fit template SED at FIR. The total IR
luminosity is derived by integrating under this SED curve. The
uncertainty is derived from the standard deviation among the six best
fits. We show examples of our SED fitting results in Figure
\ref{fig:sedfit} and the distribution of L$_{\rm IR}$ is shown Figure
\ref{fig:LIR_z} a . We also show the distribution of L$_{\rm IR}$ for
each type of objects, e.g. starburst, AGN and composite (defined in
detail in Section 4.1) in this figure. The derived L$_{\rm IR}$ for
each source and its uncertainty is tabulated in Table
\ref{tab:data}. The distribution for the redshifts of 5MUSES objects
are shown in Figure \ref{fig:LIR_z} b.

\subsubsection{How well can we constrain IR SED from mid-IR?}

As has been shown above as well as in \citet{Kartaltepe10}, the
availability of longer wavelength data greatly reduces the uncertainty
in the estimate of L$_{\rm IR}$. We need to quantify how well one can
constrain the SED of a galaxy if only the continuum shape up to
$\sim$30\,$\mu$m is available. In Figure \ref{fig:comp_irs_phot}, we
show the comparison of the IRS predicted L$_{\rm IR}^{\rm IRS}$ and
the L$_{\rm IR}^{\rm phot}$ estimated from photometric data points
(IRAC 5.8, 8.0\,$\mu$m and MIPS 24, 70 and 160\,$\mu$m). For the
IRS-only method, we use only the IRS spectrum and do not employ any
longer wavelength information (70 and 160\,$\mu$m fluxes or upper
limits) in our SED fitting, with the goal of testing solely the power
of using mid-IR SED to predict FIR SED.  We find that L$_{\rm IR}$
estimated from mid-to-FIR photometry are on average 10\% higher than
L$_{\rm IR}$ estimated from the IRS-only method, with a considerable
scatter of 0.14\,dex. It is worth noting that L$_{\rm IR}^{\rm IRS}$
deviates from L$_{\rm IR}^{\rm phot}$ by more than 0.2\,dex for 20\%
of the sources while 5\% of the sources deviate by more than 0.3\,dex.
We further divide the sources into two groups: cold sources and warm
sources, based on the ratio of $f_{\rm 24\mu m}/f_{\rm 70\mu m}$. Cold
sources ($f_{\rm 24\mu m}/f_{\rm 70\mu m}<$0.2) show an average
underestimate of 17\% when using the IRS-only method and the 1$\sigma$
scatter is 0.16\,dex, while warm sources ($f_{\rm 24\mu m}/f_{\rm
  70\mu m}>$0.2) do not show systematic offset in the estimated
L$_{\rm IR}$ from the two methods, with a scatter of 0.18\,dex. This
comparison suggest that in IR-selected samples, the mid-IR spectrum
could be used as an important indicator for L$_{\rm IR}$ when no
longer wavelength data are available. However, for cold sources, the
IRS-only method might underestimate L$_{\rm IR}$ by $\sim$17\%, due to
the lack of information on the peak of the SED. For warm sources,
although L$_{\rm IR}$ estimated from the IRS-only method generally
agrees with the value estimated from mid-to-far IR SED, the associated
uncertainty is rather large. Thus for an individual galaxy, the
L$_{\rm IR}$ predicted by its mid-IR SED could be a factor of 1.5 off
from its intrinsic value for a significant subset of the population.

\subsection{Estimating the total IR Luminosity from a Single Band}

Using {\em Spitzer} data, we have obtained accurate estimates of the
total infrared luminosities for 5MUSES. In the absence of
multi-wavelength data, single-band luminosities have often been used
to estimate L$_{\rm IR}$ \citep{Sajina07, Papovich07, Pope08, Rieke09,
  Bavouzet08, Symeonidis08}. However, the fractional contribution of
these photometric bands to the total infrared luminosity varies
substantially depending on the dominant energy source. Because the IRS
spectrum provides an unambiguous way to identify the energy source for
5MUSES galaxies, our sample is ideal for investigating the difference
in the fractional contributions of single band luminosities to L$_{\rm
  IR}$ in different types of objects.

\begin{figure}
  \epsscale{1.}
  \plotone{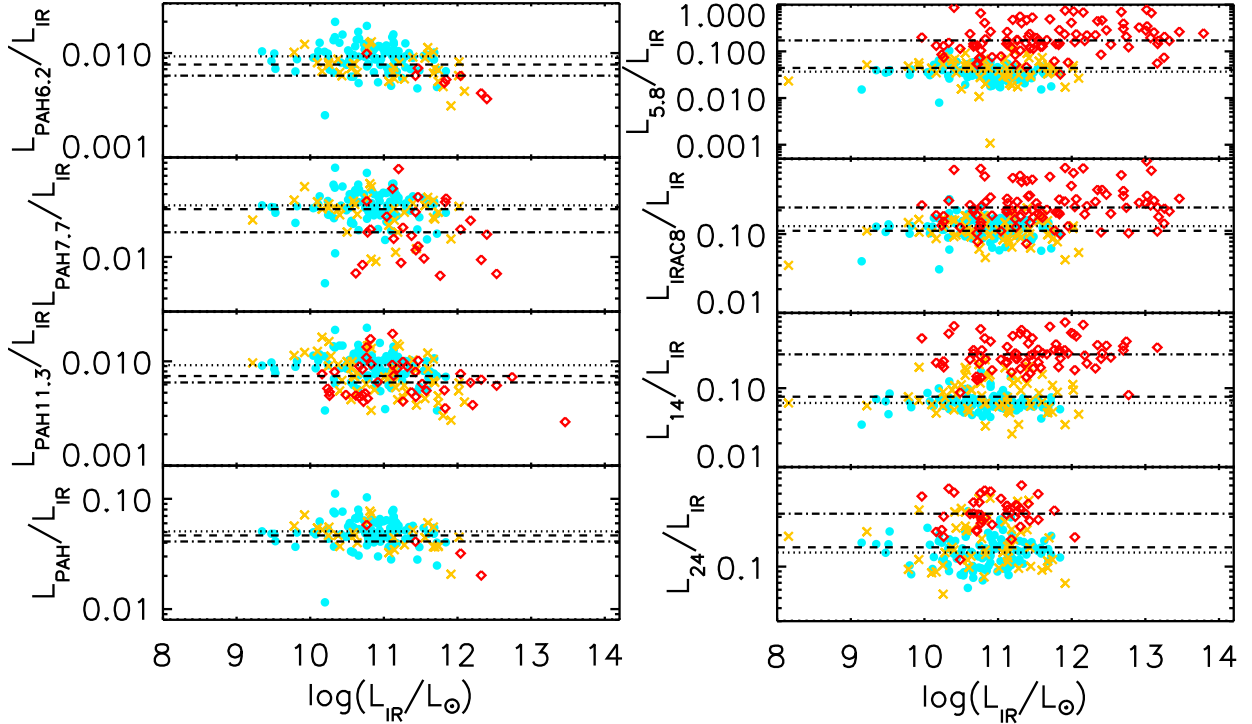}
  \caption{The luminosity ratio of several bands over L$_{\rm IR}$
    versus L$_{\rm IR}$. The blue circles, yellow crosses and red
    diamonds represent the SB, composite and AGN dominated sources in
    5MUSES. The dotted, dash, and dash-dotted lines stand for the
    median ratios for the SB, composite and AGN sources,
    respectively. The PAH luminosities are derived from PAHFIT
    measurements. The 5.8, 14 and 24\,$\mu$m luminosities are
    monochromatic luminosities calculated from the continua at these
    wavelengths. The IRAC 8\,$\mu$m luminosities are derived by
    convolving the rest-frame IRS spectra with the filter response
    curve of the IRAC 8\,$\mu$m band. The ratios and the associated
    uncertainties are also listed in Table \ref{l_ir_tab}.}
  \label{fig:lir_pah_con}
\end{figure}

In Figure \ref{fig:lir_pah_con}, we plot the ratio of several
luminosity bands to L$_{\rm IR}$. The PAH luminosities are plotted on
the left panel and continuum luminosities are on the right. The
dotted, dashed and dash-dotted lines respectively stand for the median
ratios for SB, composite and AGN dominated sources. Clearly the
fractional contribution of a certain band to L$_{\rm IR}$ is highly
dependent on the type of the object, e.g. the monochromatic 24\,$\mu$m
continuum luminosity $\nu$L$_{\nu}$ accounts for $\sim$13\% of the
total IR luminosity for SB galaxies, while it can contribute on
average 30\% of L$_{\rm IR}$ in AGN. The difference in the ratio of
PAH luminosity to L$_{\rm IR}$ is less significant for different types
of objects, because in order to be included on the left panel of this
plot, the AGN dominated sources also need to have a solid detection of
PAH feature that could be measured by PAHFIT, i.e. strong AGN sources
are excluded. The mean ratios of L$_{\rm singleband}$/L$_{\rm IR}$ are
summarized in Table \ref{l_ir_tab}.  Finally, we also provide our
calibration of using single band luminosity to estimate L$_{\rm IR}$
in the Appendix.

\section{Aromatic Feature Diagnostics}
\subsection{The Average Spectra}

We derive the stacked SEDs for the SB, composite and AGN dominated
sources in the 5MUSES sample, combining the low resolution IRS spectra
in the mid-IR with the MIPS photometry at FIR. Although we do not have
optical spectroscopy to classify the object types with the BPT diagram
\citep{Baldwin81, Kewley01}, the equivalent widths of PAH features can
be used as indicators of star formation activity. The 6.2 and
11.3\,$\mu$m PAH bands are relatively isolated with little
contamination from nearby features, which is important for
unambiguously defining the local continuum. However, the 11.3\,$\mu$m
band is located on the shoulder of 9.7\,$\mu$m silicate feature. Thus
its integrated flux and underlying continuum are likely to be affected
by dust extinction effects. As a result, in our discussion, we use the
6.2\,$\mu$m PAH EWs to classify objects. To be consistent with the
studies in the literature, we have adopted the following criteria for
our spectral classification: sources with EW$>$0.5\,$\mu$m are
SB-dominated; sources with 0.2$<$EW$\le$0.5\,$\mu$m are AGN-SB
composite and sources with EWs$\le$0.2\,$\mu$m are
AGN-dominated\footnote{Sources with a significant old stellar
  population could also have a reduced 6.2\,$\mu$m PAH EW. As will be
  shown in Shi et al. 2010 (in preparation), the stellar emission
  contributes less than $\sim$20\% to the 6\,$\mu$m continuum for our
  IR selected sample of 5MUSES.}.  \citep{Armus07}. The PAH EWs for
the sample are tabulated in Table \ref{tab:data}. Out of the 280
sources for which redshifts have been obtained from optical or
infrared spectroscopy, there are 123 SB galaxies (44\%), 62 composite
sources (22\%) and 95 AGN dominated sources (34\%).

\begin{figure}
  \epsscale{1.1}
  \plottwo{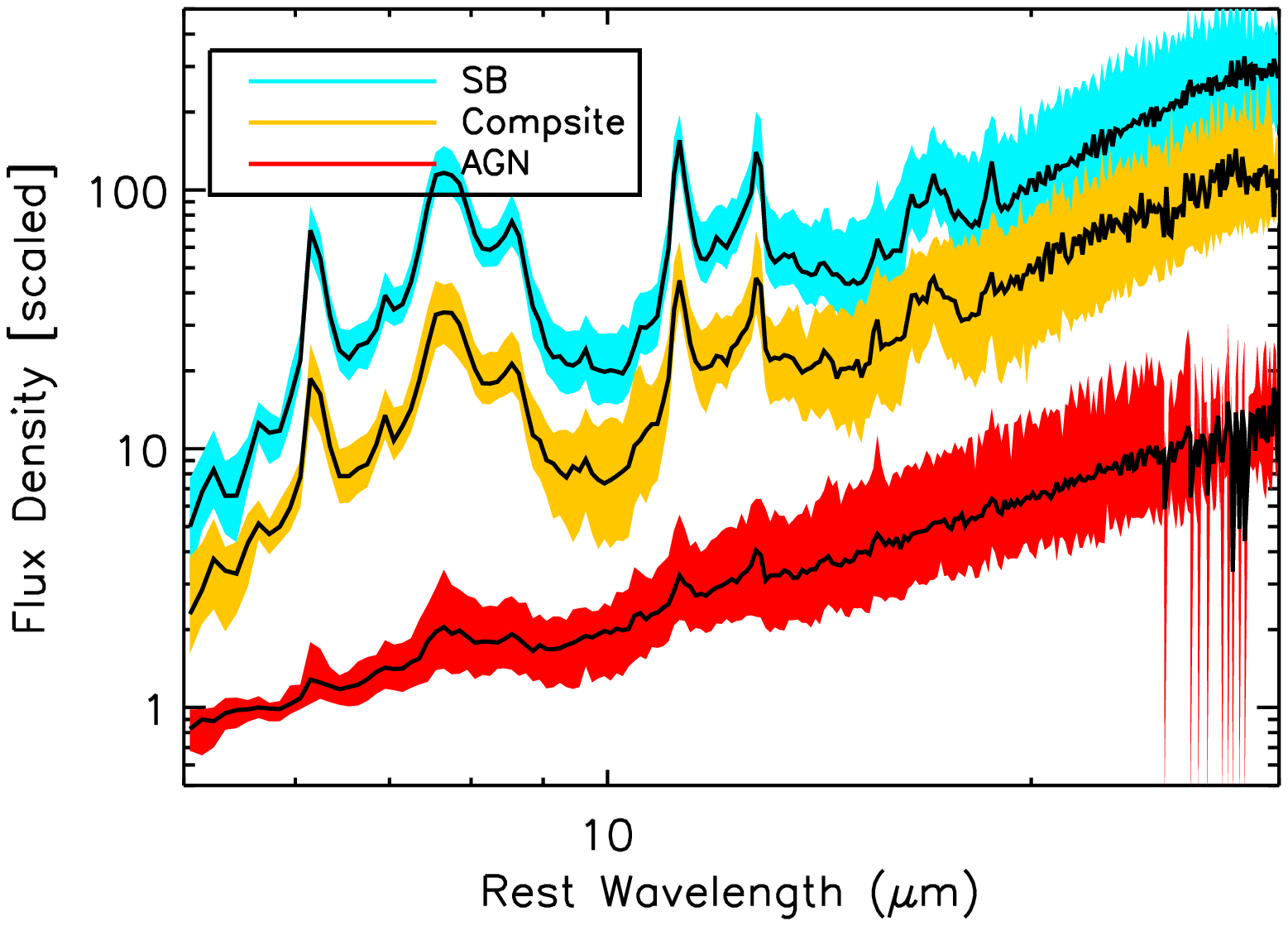}{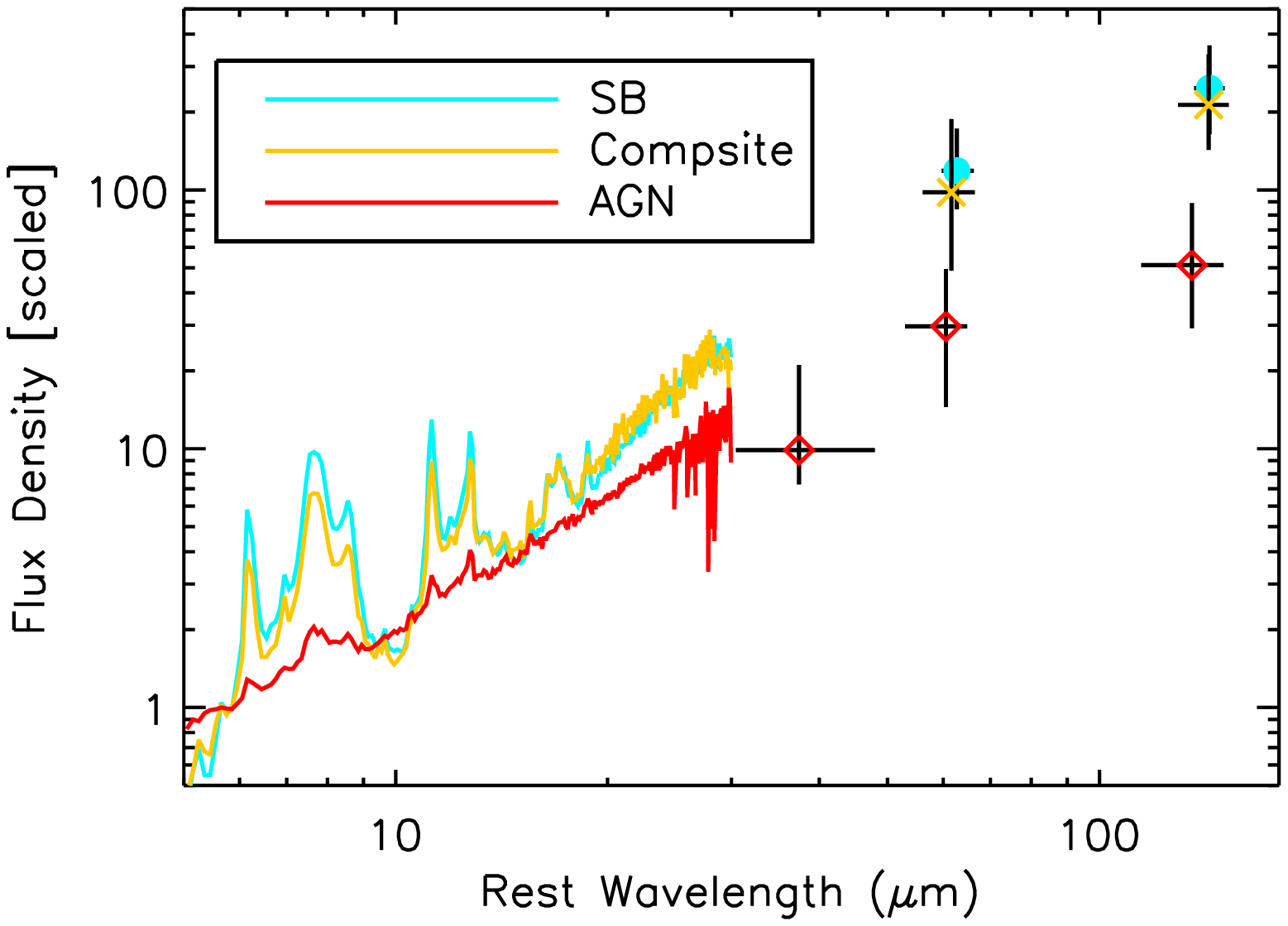}
  \caption{a) Left panel: The median IRS spectra for SB, composite and
    AGN dominated sources from 5MUSES after normalizing at
    5.8\,$\mu$m. The SEDs have been offset vertically. The shaded
    regions represent 1$\sigma$ uncertainties. b) Right panel: The
    median SEDs for SB, composite and AGN sources from mid-IR to FIR,
    normalized at 5.8\,$\mu$m.}
  \label{fig:5muses_avespec}
\end{figure}

The 5-30\,$\mu$m composite spectra are derived by first normalizing
individual spectra at rest-frame 5.8\,$\mu$m, and then taking the
median in each wavelength bin. In Figure \ref{fig:5muses_avespec} a,
we show the typical SED for SB galaxy in blue and AGN in red, while
yellow line represents the median SED for SB-AGN composite sources in
5MUSES. The average SEDs have been offset vertically. The shaded
regions represent the 16th and 84th percentile of the flux densities
at each wavelength.

The MIPS 70 and 160\,$\mu$m photometry is crucial for constraining the
SED shape of a galaxy and we have also included these data in the
final typical SED. Because of the difference in redshift range for
sources of different spectral types, we have divided the MIPS 70 and
160\,$\mu$m data into several rest-wavelength bins before we take the
median. For SB and composite sources, we take 2 bins: 40-70\,$\mu$m
and 70-160\,$\mu$m; For AGN, we choose to have 3 bins due to their
larger redshift range: 30-50, 50-100, and 100-160\,$\mu$m. We take the
median flux in each bin and assign the 16th and 84th percentile of the
data points in the same bin as the uncertainties.  The final median
SEDs are presented in Figure \ref{fig:5muses_avespec}\,b. We can
clearly see that besides having much less PAH emission in the mid-IR,
the continuum in the AGN also rises much more slowly than in the SB. The
SED of the composite source is between the SB and AGN and its shape is
dependent on how we define a composite source. As can be seen in
Figure \ref{fig:5muses_avespec} and \ref{fig:pahew_hist}, our
definition of composite sources with 0.2\,$\mu$m$<$6.2\,$\mu$m PAH
EW$<$0.5\,$\mu$m is likely biased towards star formation dominated
sources (see Section 4.2).

\subsection{The Distribution of PAH EWs}

\begin{figure}
  \epsscale{1.0}
  \plotone{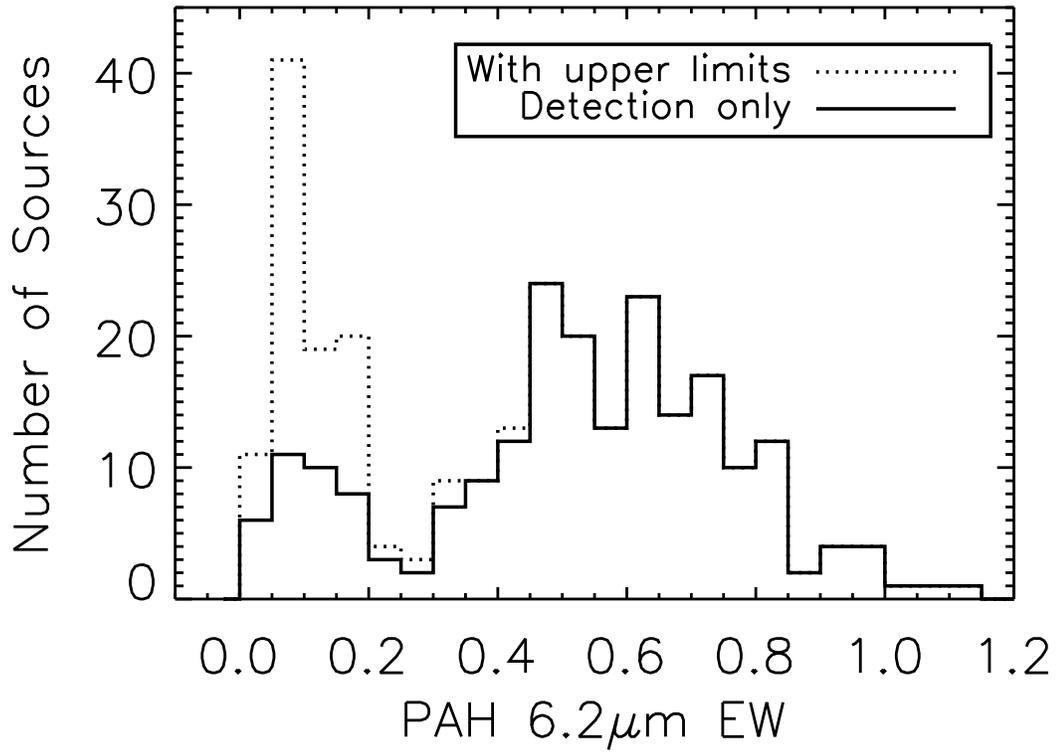}
  \caption{The histogram shows the distribution of the 6.2\,$\mu$m PAH
    EW for the 280 known-z sources from the 5MUSES sample. The solid
    line represents sources which have detection for the 6.2\,$\mu$m
    feature, while the dotted line includes upper limits. It is clear
    that both the solid line and dotted line show a dip in the PAH EW
    distribution at 0.2$\sim$0.3\,$\mu$m. See the text for detailed
    discussion on the bi-modality of the distribution.}
  \label{fig:pahew_hist}
\end{figure}

With the superb sensitivity and spectral coverage of the IRS, we are
able to quantify the strength of the PAH emission over nearly two
orders of magnitude in its EW. The distribution of the 6.2\,$\mu$m PAH
EWs for the 280 known-redshift galaxies in 5MUSES is shown in Figure
\ref{fig:pahew_hist}. The solid line represents the distribution for
sources with detection of the 6.2\,$\mu$m feature, while the dotted
line also includes upper limits. We clearly observe a bimodal
distribution in Figure \ref{fig:pahew_hist}, with two local peaks at
$\sim$0.1 and $\sim$0.6\,$\mu$m. This is somewhat surprising, because
5MUSES provides a representative sample completely selected based on
IR flux densities, and one would have expected a more continuous
distribution. Although we still lack redshift information for 50
sources in our sample, the featureless power-law shape of their IRS
spectra (except for a few cases where silicate absorption or very weak
PAH feature is present) indicate that these are likely to be
AGN-dominated. Thus if they were included in Figure
\ref{fig:pahew_hist}, they would most likely be located in the range
between 0$-$0.2\,$\mu$m, and the bimodal distribution would not be
affected. A similar bi-modality is also observed in the distribution
of the 11.3\,$\mu$m PAH EWs (not shown here). The observed bimodal
distribution of the PAH EWs may be a result of the selection effect
for this flux limited sample: objects at higher redshifts are more
likely to be AGN and thus pile up at the low EW end. However, if we
divide our sample into sources with z$>$0.5 and z$<$0.5, the bimodal
distribution is again observed in the z$<$0.5 population, although all
the z$>$0.5 objects are located at the low EW end. Detailed population
modelling is being performed and this issue will be addressed in a
later paper.

\subsection{PAH Properties versus mid-IR and FIR slopes}
Another important physical parameter that is often used to quantify
the dominant energy source of a galaxy is the ratio of warm to cold
dust. It has been shown in previous studies \citep{Desai07, Wu09} that
the 6.2 and 11.3 \,$\mu$m PAH EWs of galaxies are usually suppressed
in warmer systems dominated by AGN, as indicated by the low flux
ratios of {\em IRAS} $f_{\rm 60}/f_{\rm 25}$.  For the 5MUSES sample,
we have examined the correlation of the 6.2\,$\mu$m PAH EWs with
various continuum slopes, e.g. f$_{15}$/f$_{5.8}$, f$_{30}$/f$_{5.8}$,
f$_{30}$/f$_{15}$ and f$_{70}$/f$_{24}$. The rest-frame continuum
fluxes are estimated from the final SED obtained from the fits in
Section 3. We find that the continuum ratios of f$_{30}$/f$_{15}$ and
f$_{70}$/f$_{24}$ have the strongest global correlation with the
6.2\,$\mu$m PAH EWs and the correlation coefficients are both
$\sim$0.7.

\begin{figure}
  \epsscale{1.2}
  \plottwo{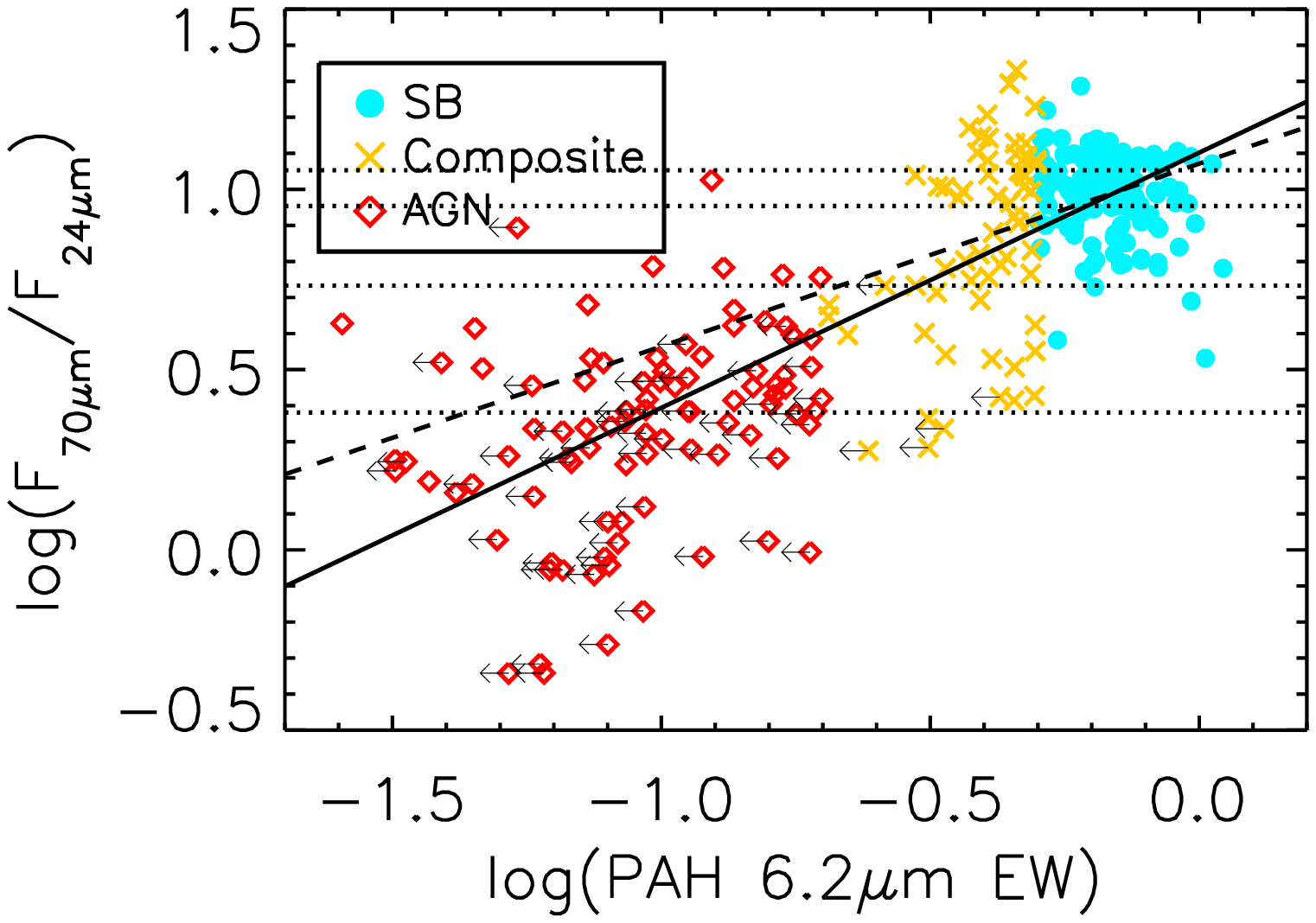}{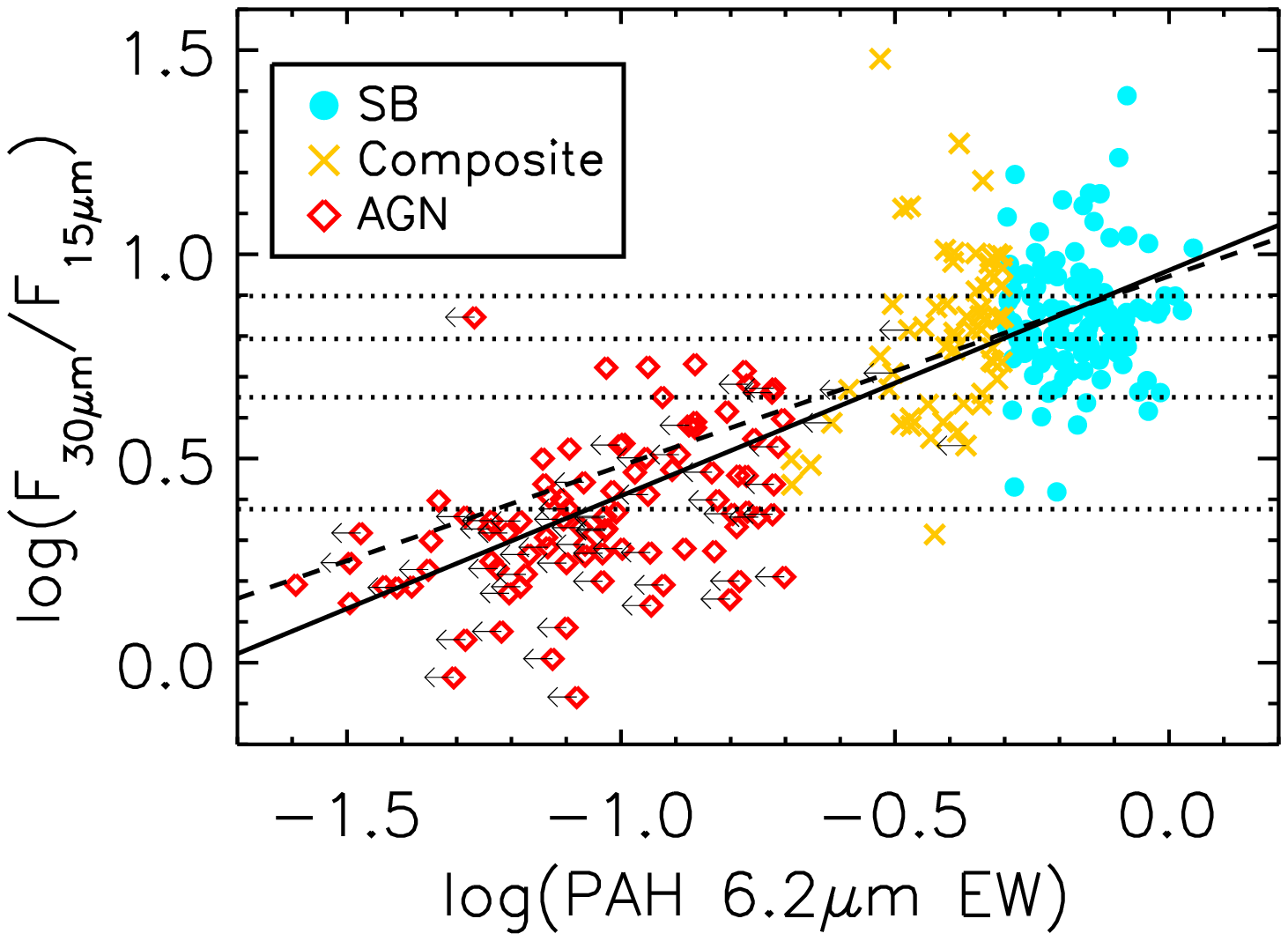}
  \caption{a) Left panel: The continuum flux ratio of f$_{\rm 70\mu
      m}$/f$_{\rm 24\mu m}$ versus the 6.2\,$\mu$m PAH EW. b) Right
    panel: The continuum flux ratio of f$_{\rm 30\mu m}$/f$_{\rm 15\mu
      m}$ versus the 6.2\,$\mu$m PAH EWs. The solid line is a fit to
    all the data points while the dashed line is a fit excluding
    sources with 6.2\,$\mu$m PAH EW upper limits. We bin the objects
    according to their continuum slopes and have equal numbers of
    objects in each bin. The dotted lines indicate the boundaries of
    those bins.}
  \label{fig:pah_color}
\end{figure}

In Figure \ref{fig:pah_color} a and b, we plot the 6.2\,$\mu$m PAH
EW against $f_{\rm 30}/f_{\rm 15}$ and f$_{70}$/f$_{24}$.  The 5MUSES
populations separate into two groups, one with steep spectra and high
aromatic content, and the other with slow rising spectra and low
aromatic content. The gap between SB and AGN dominated sources is
likely due to the selection effect of this sample. We should note that
within each group, there is little if any correlation between the
slope and the PAH EW, but it is the contrast between the two groups
that gives the overall impression of a correlation. This is consistent
with the studies of \citet{Veilleux09}, who have showed the power of
using the 7.7\,$\mu$m PAH EWs and $f_{\rm 30}/f_{\rm 15}$ ratios as
indicators of AGN activity, despite the large scatter associated with
each parameter.  To understand the variation in the PAH EWs and
continuum slopes, we further divide our sample into smaller bins and
estimate the average values in each bin. The sources are divided
according to their f$_{70}$/f$_{24}$ ratios or f$_{30}$/f$_{15}$
ratios and we assign an equal number of objects to each bin. We find
that sources in the first three bins with
log(f$_{30}$/f$_{15}$)$>$0.65 or log(f$_{70}$/f$_{24}$)$>$
0.73 \footnote{If the spectral index is defined as
  $\alpha$=log(f1/f2)/log($\nu1/\nu2$), then the continuum slope
  ratios of log(f$_{30}$/f$_{15}$)$>$0.65 can be translated to
  $\alpha_{30-15}<$-2.17 and log(f$_{70}$/f$_{24}$)$>$ 0.73 can be
  converted to $\alpha_{70-24}<$-1.57.} all have median 6.2\,$\mu$m
PAH EWs of $\sim$0.60\,$\mu$m and dispersion of $\sim$0.2\,dex, which
again confirms our observation that within the group of starburst
galaxies, there is little correlation between the slope and aromatic
content. Sources with log(f$_{30}$/f$_{15}$)$<$0.38 or
log(f$_{70}$/f$_{24}$)$<$0.38 are clearly AGN-dominated with very low
PAH EW. The median values and the associated uncertainties of the
6.2\,$\mu$m PAH EWs and continuum ratios are summarized in Table
\ref{pahew_slope}.

We also investigate the variation in the ratio of L$_{\rm
  PAH}$/L$_{\rm IR}$ when the galaxy color indicated by the continuum
slope changes. We use the sum of the PAH luminosity from the
6.2\,$\mu$m, 7.7\,$\mu$m complex and 11.3\,$\mu$m complex to represent
L$_{\rm PAH}$, measured from the composite spectra derived for each
bin. It is clear that the PAH fraction stays nearly constant for
starburst dominated systems, while its contribution drops significantly
when AGN becomes more dominant (see Table \ref{pahew_slope}).  It has
been shown that the PAH luminosity can contribute $\sim$10\% in
star-forming galaxies \citep{Smith07}. For our sample, we find that
L$_{\rm PAH}$ contributes $\sim$5\% to L$_{\rm IR}$. This ratio is
lower than the SINGS results.  We have only taken the 6.2, 7.7 and
11.3\,$\mu$m bands into account\footnote{The S/N ratios of the 5MUSES
  spectra are much lower than SINGS, thus we only include the
  strongest PAH bands.}, while the SINGS studies include all the PAH
emitting bands in the mid-IR.  Since the 6.2, 7.7 and 11.3\,$\mu$m
bands accounts for $\sim$68\% of the power in PAH emission
\citep{Smith07}, our L$_{\rm PAH}$/L$_{\rm IR}$ ratio can be converted
to $\sim$7.5\% for the total PAH contribution to L$_{\rm IR}$. This is
still slightly lower than the SINGS results, but consistent within
uncertainties.

\begin{figure}
  \epsscale{1.2}
  \plottwo{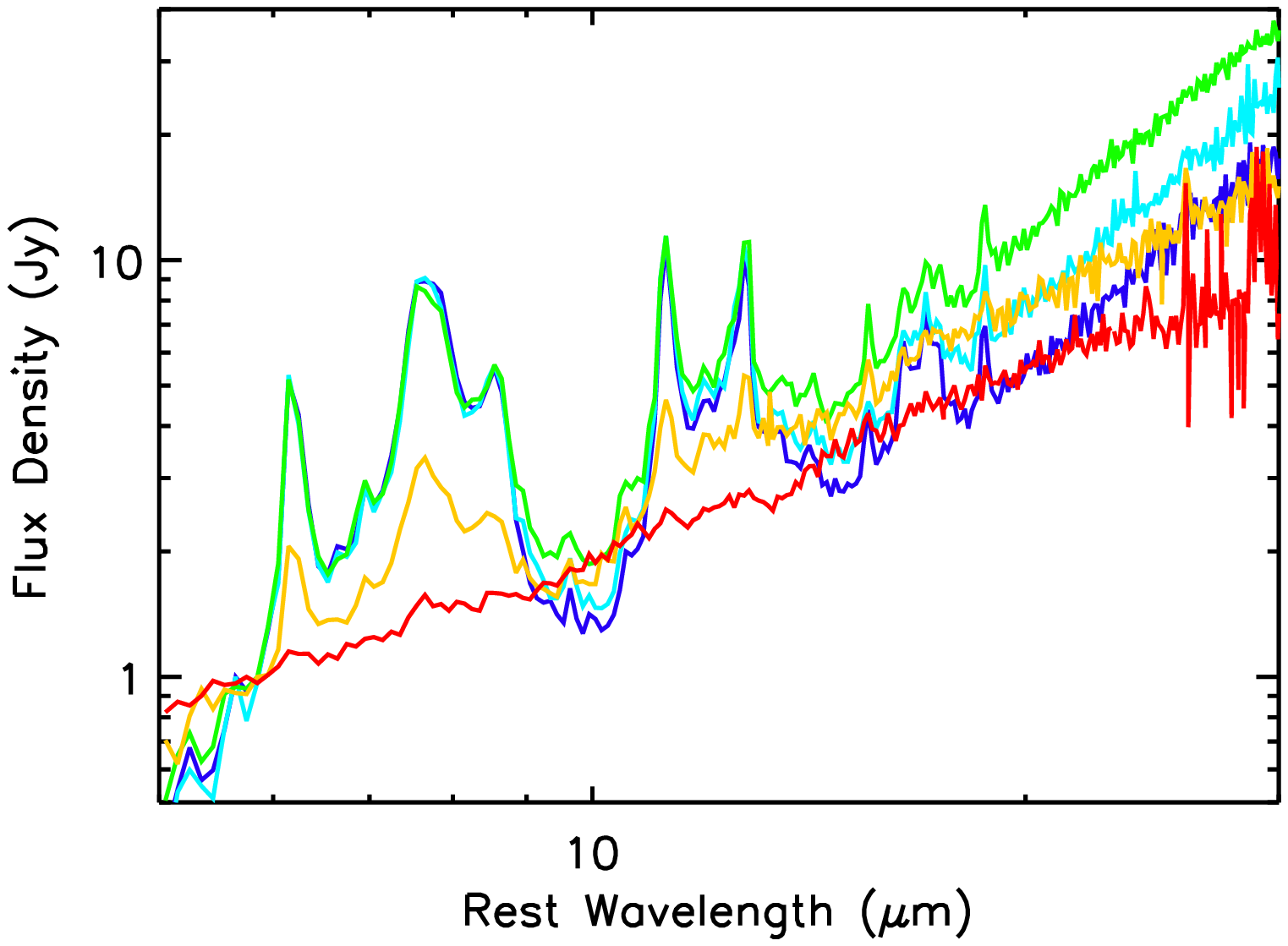}{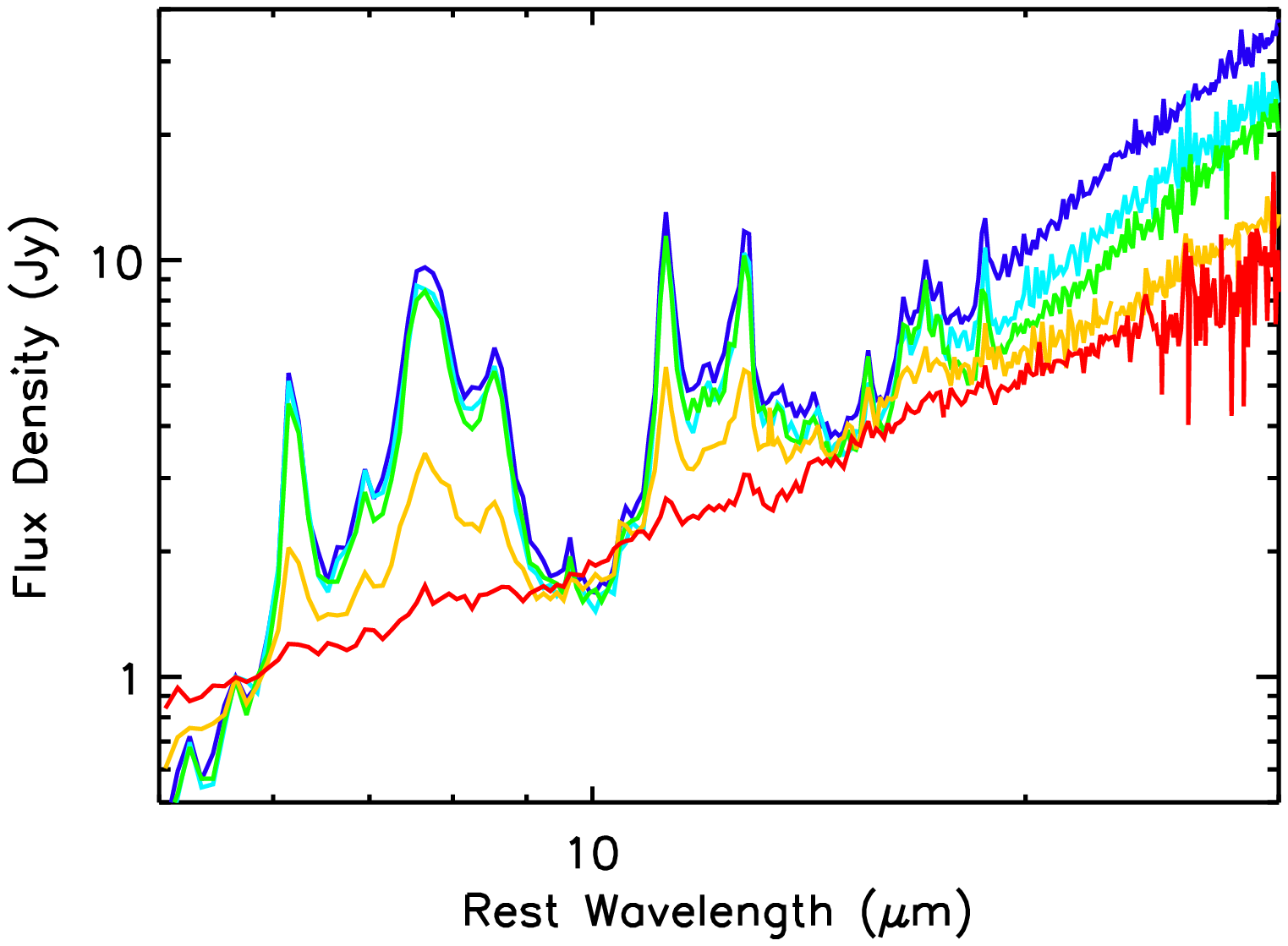}
  \caption{a) Left panel: The typical mid-IR SEDs in each bin of
    different f$_{\rm 70\mu m}$/f$_{\rm 24\mu m}$ ratios. b) Right
    panel: The typical SEDs in each bin of different f$_{\rm 30\mu
      m}$/f$_{\rm 15\mu m}$ ratios. All the SEDs have been normalized
    at 5.8\,$\mu$m. The colors represent the average spectra derived
    in each f$_{\rm 70\mu m}$/f$_{\rm 24\mu m}$ (or f$_{\rm 30\mu
      m}$/f$_{\rm 15\mu m}$) color bins listed in Table
    \ref{pahew_slope}.}
  \label{fig:slope_ave_spec}
\end{figure}

Finally, for each group of continuum slope sorted spectra, we derive
typical 5-30\,$\mu$m SEDs by taking the median flux densities in every
wavelength bin after normalizing at rest-frame 5.8\,$\mu$m. This will
be useful for SED studies when only galaxy colors estimated from broad
band photometry are available. These composite SEDs are shown in
Figure \ref{fig:slope_ave_spec}. Then we explore whether the
derivation of total IR luminosity from broadband photometry varies
with galaxy color. We assume L$_{\rm IR}$ is correlated with L$_{\rm
  24\mu m}$ and L$_{\rm 70\mu m}$ in the following manner and derive
the a and b coefficients in each f$_{70}$/f$_{24}$ continuum slope bin
(all in rest-frame):
\begin{equation}
  {\mathrm{log}}L_{\mathrm{IR}}=a~{\mathrm{log}}L_{\mathrm{24\mu m}}+b~{\mathrm{log}}L_{\mathrm{70\mu m}}
\end{equation}

\begin{figure}
  \epsscale{1.2}
  \plottwo{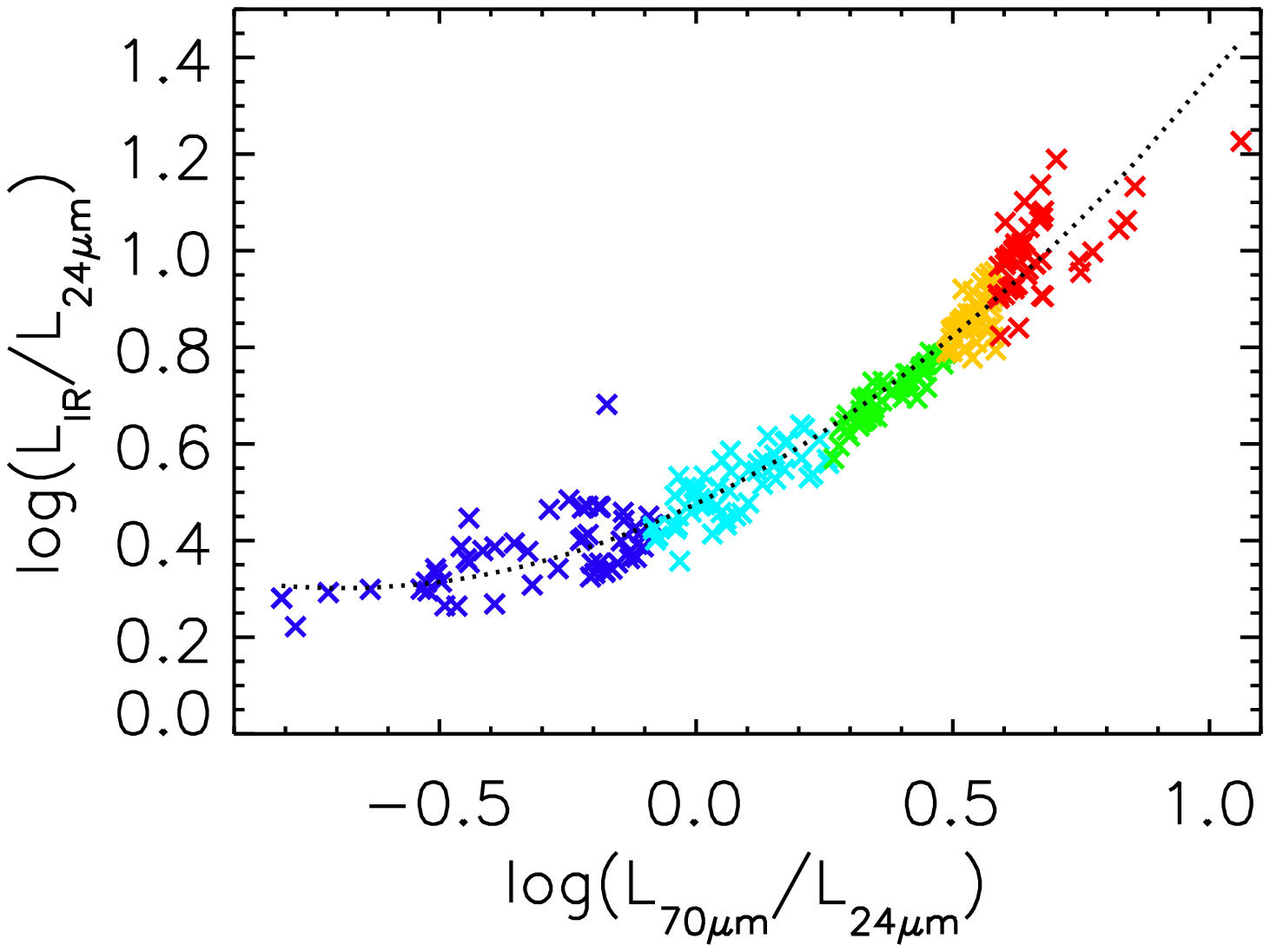}{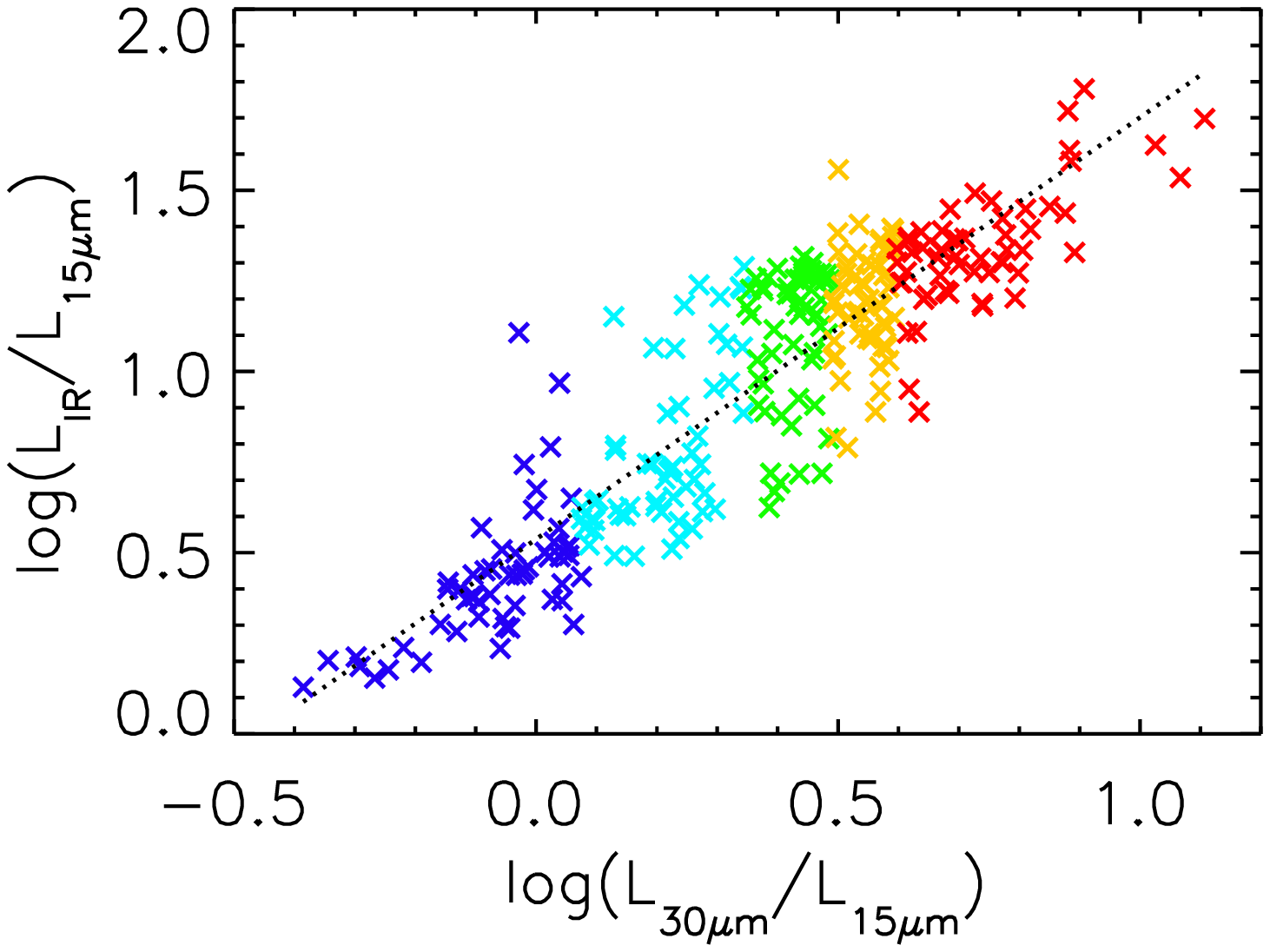}
  \caption{a) Left panel: The ratios of L$_{\rm IR}$/L$_{\rm 24\mu m}$
    versus L$_{\rm 70\mu m}$/L$_{\rm 24\mu m}$. The sources are
    colored according to their f$_{\rm 70}$/f$_{\rm 24}$ ratios. The
    dotted line is a 2nd-order polynomial fit to the data. b) Right
    panel: The ratios of L$_{\rm IR}$/L$_{\rm 15\mu m}$ versus L$_{\rm
      30\mu m}$/L$_{\rm 15\mu m}$. The sources are colored according
    to their f$_{\rm 30}$/f$_{\rm 15}$ ratios. The dotted line is a
    linear fit to the data.}
  \label{fig:LIR_slope}
\end{figure}

The values of a and b coefficients are summarized in Table
\ref{pahew_slope}. To illustrate the variations in each slope bin, we
plot the ratio of L$_{\rm IR}$/L$_{\rm 24\mu m}$ versus L$_{\rm 70\mu
  m}$/L$_{\rm 24\mu m}$ in Figure \ref{fig:LIR_slope} a. The sources
are colored according to their f$_{70}$/f$_{24}$ ratios. We clearly
observe that when normalized by the monochromatic 24\,$\mu$m
luminosity, L$_{\rm IR}$ is strongly correlated with L$_{\rm 70\mu m}$
and the slopes in each continuum ratio bin become steeper when L$_{\rm
  70\mu m}$/L$_{\rm 24\mu m}$ increase, except in the last slope bin
(see also the b coefficients).  We fit a 2nd-order polynomial to the
data and find the correlation to be:
\begin{equation}
  {\mathrm{log}}\frac{L_{\mathrm{IR}}}{L_{\mathrm{24\mu m}}}=(0.476\pm0.005)+(0.509\pm0.010)~{\mathrm{log}}\frac{L_{\mathrm{70\mu m}}}{L_{\mathrm{24\mu m}}}+(0.370\pm0.022)~({\mathrm{log}}\frac{L_{\mathrm{70\mu m}}}{L_{\mathrm{24\mu m}}})^2
\end{equation}

The above equation is derived based on the 5MUSES data. The majority
(90\%) of the 24\,$\mu$m luminosities of these 280 galaxies are
between 10$^{9.0}$L$_\odot$ and 10$^{12.0}$L$_\odot$. The ratio of
f$_{\rm 70\mu m}$/f$_{\rm 24\mu m}$ ranges from 0.45 to 34. Our result
is consistent with a similar correlation derived by
\citet{Papovich02}, while it diverges for sources with low f$_{\rm
  70\mu m}$/f$_{\rm 24\mu m}$ ratios, since their modelling work has
focused on star-forming galaxies only.

We repeat the same exercise for our sample binned with the
f$_{30}$/f$_{15}$ ratios. In Figure \ref{fig:LIR_slope} b, we find
that L$_{\rm IR}$ is correlated with L$_{\rm 30\mu m}$ when both
quantities are normalized by L$_{\rm 15\mu m}$, although with very
large scatter. The dotted line is a linear fit to the data. For a
given L$_{\rm 30\mu m}$/L$_{\rm 15\mu m}$ ratio, L$_{\rm IR}$/L$_{\rm
  15\mu m}$ can span as much as a factor of five. The median values in
each group binned by the f$_{30}$/f$_{15}$ ratios are also summarized
in Table \ref{pahew_slope}.

\subsection{The Variation in PAH Band-to-Band Strength Ratios}

The luminosity ratio of different PAH bands is thought to be a
function of the grain size and ionization state
\citep{Tielens08}. 
luminosity ratios of L$_{\rm PAH 7.7\mu m}$/L$_{\rm PAH 11.3\mu
  m}$\footnote{ We choose to use the L$_{\rm PAH 7.7\mu m}$/L$_{\rm
    PAH 11.3\mu m}$ ratio in this study for easier comparison with
  literature results, such as \citet{Smith07, Odowd09}.}  with the
6.2\,$\mu$m PAH EWs for the 5MUSES sample. Only sources with S/N$>$3
from PAHFIT measurements for the 7.7 and 11.3\,$\mu$m bands are
included in this plot. We find that the AGN-dominated sources on
average have lower L$_{\rm PAH 7.7\mu m}$/L$_{\rm PAH 11.3\mu m}$
ratios than the composite or SB-dominated sources. The mean
log(L$_{\rm PAH 7.7\mu m}$/L$_{\rm PAH 11.3\mu m}$) ratios for AGN,
composite and SB galaxies in 5MUSES are 0.32$\pm$0.18, 0.53$\pm$0.15
and 0.53$\pm$0.08, respectively. This is consistent with the studies
on the nuclear spectra of low luminosity star-forming galaxies from
SINGS \citep{Smith07}, which also show decreased L$_{\rm PAH 7.7\mu
  m}$/L$_{\rm PAH 11.3\mu m}$ ratios in spectra with AGN
signals. \citet{Smith07} suggest that this change in the ratio of
L$_{\rm PAH 7.7\mu m}$/L$_{\rm PAH 11.3\mu m}$ is likely due to the
destruction of the smallest PAHs by hard photons from the AGN. On the
other hand, AGN are less extinguished than SB or composite sources,
thus if PAHFIT underestimates the extinction correction, it will
preferentially underestimate the 11.3\,$\mu$m fluxes more than the
7.7\,$\mu$m feature in SB/Composite sources, thus resulting in the
elevated ratios of L$_{\rm PAH 7.7\mu m}$/L$_{\rm PAH 11.3\mu m}$ in
SB/composite systems.

\begin{figure}
  \epsscale{1.5}
  \plottwo{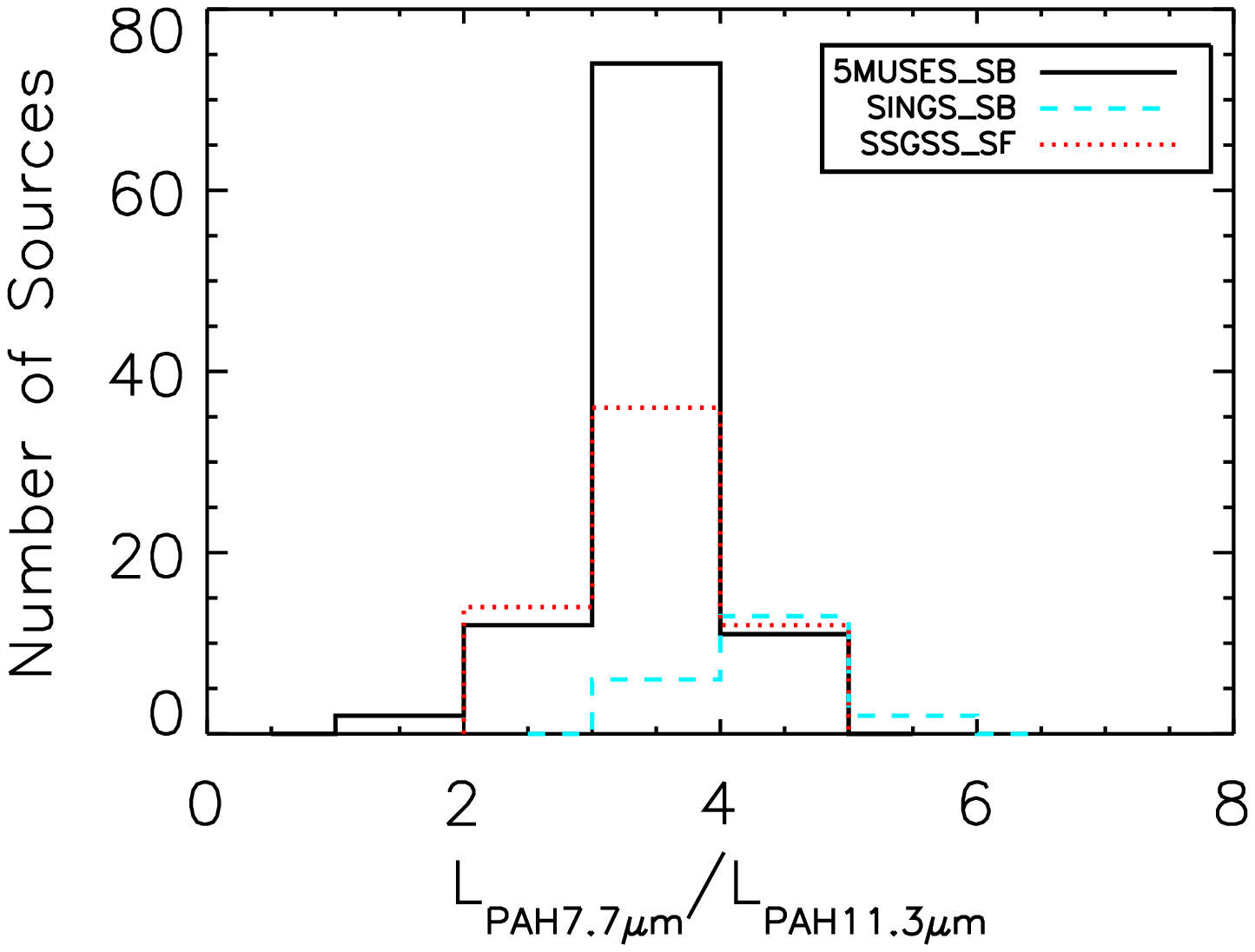}{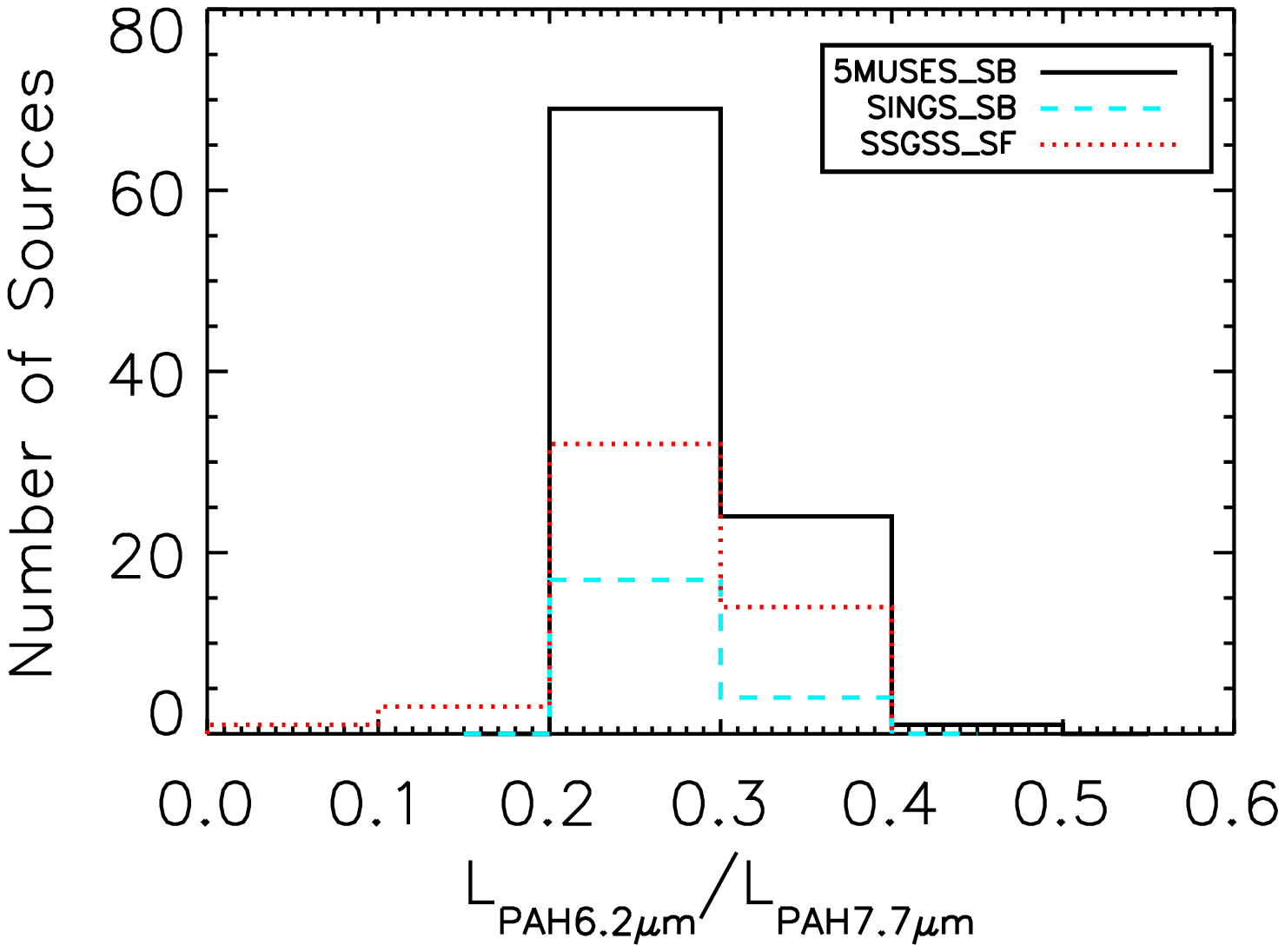}
  \caption{a) Top panel: A comparison of the distribution of PAH
    luminosity ratio of L$_{\rm PAH 7.7\mu m}$/L$_{\rm PAH 11.3\mu m}$
    for the IR classified SB-dominated sources from 5MUSES, SINGS and
    the optically classified SF-dominated sources in SSGSS. b) Bottom
    panel: Same as a), but for the PAH luminosity ratio of L$_{\rm PAH
      6.2\mu m}$/L$_{\rm PAH 7.7\mu m}$. The SB and SF galaxies in
    5MUSES and SSGSS appear to have a similar distribution for both
    the L$_{\rm PAH 7.7\mu m}$/L$_{\rm PAH 11.3\mu m}$ and L$_{\rm PAH
      6.2\mu m}$/L$_{\rm PAH 7.7\mu m}$ ratios, while the SINGS
    nuclear spectra appear to show higher L$_{\rm PAH 7.7\mu
      m}$/L$_{\rm PAH 11.3\mu m}$ ratios for the SB-dominated
    galaxies.}
  \label{fig:5muses_sings_ssgss}
\end{figure}

In Figure \ref{fig:5muses_sings_ssgss} a, we show the histogram of
the L$_{\rm PAH 7.7\mu m}$/L$_{\rm PAH 11.3\mu m}$ ratios for the
SB-dominated sources in 5MUSES. We have overplotted the values from
the SINGS sample. To make a fair comparison, we remeasure the PAH
luminosity and EWs for the SINGS nuclear spectra using the same method
as 5MUSES and classify the sources with 6.2\,$\mu$m PAH EWs larger
than 0.5\,$\mu$m as SB-dominated. We have also included the
distribution of the L$_{\rm PAH 7.7\mu m}$/L$_{\rm PAH 11.3\mu m}$
ratios from the UV/SDSS selected star-forming galaxies sample of SSGSS
\citep{Odowd09}.  For this last sample, the star-forming galaxies are
classified from optical spectroscopy using the BPT diagram method
\citep{Baldwin81, Kewley01}. We find that the distribution for SB
galaxies in 5MUSES and SSGSS is similar, while both samples appear to
have lower L$_{\rm PAH 7.7\mu m}$/L$_{\rm PAH 11.3\mu m}$ ratios than
the nuclear spectra of SINGS SB galaxies. The mean log(L$_{\rm PAH
  7.7\mu m}$/L$_{\rm PAH 11.3\mu m}$) ratio for SINGS starbursts is
0.63$\pm$0.06 while it is 0.53$\pm$0.08 for 5MUSES
starbursts. This might be a resolution effect: If the physical
conditions at the nuclear region of a galaxy indeed modifies the
distribution of the L$_{\rm PAH 7.7\mu m}$/L$_{\rm PAH 11.3\mu m}$
ratios, it might be visible only in the spectra taken through
apertures with small projected sizes. The median redshift for the
SB-dominated sources in the 5MUSES sample is 0.12, while the median
redshift for the SSGSS sample is 0.08. At the redshift of 0.08,
1$\arcsec$ corresponds to 1.53\,kpc. The IRS spectra (for SL, the slit
width is$\sim$3.6$\arcsec$) of 5MUSES and SSGSS sources are integrated
from the whole galaxy, thus diluting the signature of the nuclear
regions. \citet{Smith07} have shown the changes in the L$_{\rm PAH
  7.7\mu m}$/L$_{\rm PAH 11.3\mu m}$ ratios in spectra extracted from
bigger to smaller apertures in two star-forming galaxies: The L$_{\rm
  PAH 7.7\mu m}$/L$_{\rm PAH 11.3\mu m}$ ratios measured from {\em
  star-forming galaxy} spectra extracted with smaller apertures are
higher than those measured from larger apertures, consistent with our
results. More recently, \citet{Pereira-Santaella10} have suggested
that the 11.3\,$\mu$m PAH feature is more extended than the 6.2 or
7.7\,$\mu$m PAH from a spatially resolved mapping study of local
luminous infrared galaxies. They have observed lower L$_{\rm PAH
  6.2\mu m}$/L$_{\rm PAH 11.3\mu m}$ ratios in the nucleus, consistent
with our results.  We also show the distribution of L$_{\rm PAH 6.2\mu
  m}$/L$_{\rm PAH 7.7\mu m}$ ratios in Figure
\ref{fig:5muses_sings_ssgss} b.  No significant difference has been
observed between the 5MUSES, SINGS and SSGSS samples.

Finally in Figure \ref{fig:pah627711}, we present the variation in PAH
band-to-band ratios for the three strongest bands at 6.2, 7.7 and
11.3\,$\mu$m of the 5MUSES sample. Only sources with S/N$>$3 from
PAHFIT measurements for all three PAH bands are included in this
figure. The two dark lines represent the traces for fully neutral or
fully ionized PAH molecules with different numbers of carbon atoms
predicted from modelling work \citep{Draine01}. The L$_{\rm PAH 7.7\mu
  m}$/L$_{\rm PAH 11.3\mu m}$ ratios span a range of a factor of 5
while the L$_{\rm PAH 6.2\mu m}$/L$_{\rm PAH 7.7\mu m}$ ratios only
vary by a factor of 2. The uncertainty in the L$_{\rm PAH 7.7\mu
  m}$/L$_{\rm PAH 11.3\mu m}$ ratios is 0.09\,dex and it is 0.05\,dex
for the L$_{\rm PAH 6.2\mu m}$/L$_{\rm PAH 7.7\mu m}$ ratios. This
narrow range of L$_{\rm PAH 6.2\mu m}$/L$_{\rm PAH 7.7\mu m}$ ratios
is consistent with the values for the SINGS nuclear sample
\citep{Smith07}, while we have not observed any sources with extremely
low L$_{\rm PAH 6.2\mu m}$/L$_{\rm PAH 7.7\mu m}$ ratios ($<$0.2) as
has been found in the SSGSS sample \citep{Odowd09}.

\subsection{PAH band ratio versus [NeIII]/[NeII]}

Because of the large difference in ionization potentials of the
Ne$^{++}$ (41eV) and Ne$^{+}$ (21.6eV) ions, the ratio of
[NeIII]/[NeII] is often used as a tracer of the hardness of the
radiation field. The [NeIII]\,15.55\,$\mu$m and [NeII]\,12.81\,$\mu$m
lines are among the strongest lines emitted in the mid-IR and because
differential extinction effects between their wavelengths are small,
they are particularly valuable. We use the IRS low-res spectra to
identify and measure these lines\footnote{For the 5MUSES sample, only
  21 out of 330 sources have IRS high-resolution spectra, which limits
  our ability to probe the full dynamic range covered by the whole
  sample. Thus we use the low-resolution spectra to measure the [NeII]
  and [NeIII] fluxes to compare with the PAH band-to-band
  ratios.}. The line fluxes measured from low-resolution spectra have
on average an uncertainty of $\sim$20\%.

\begin{figure}
  \epsscale{1.}
  \plotone{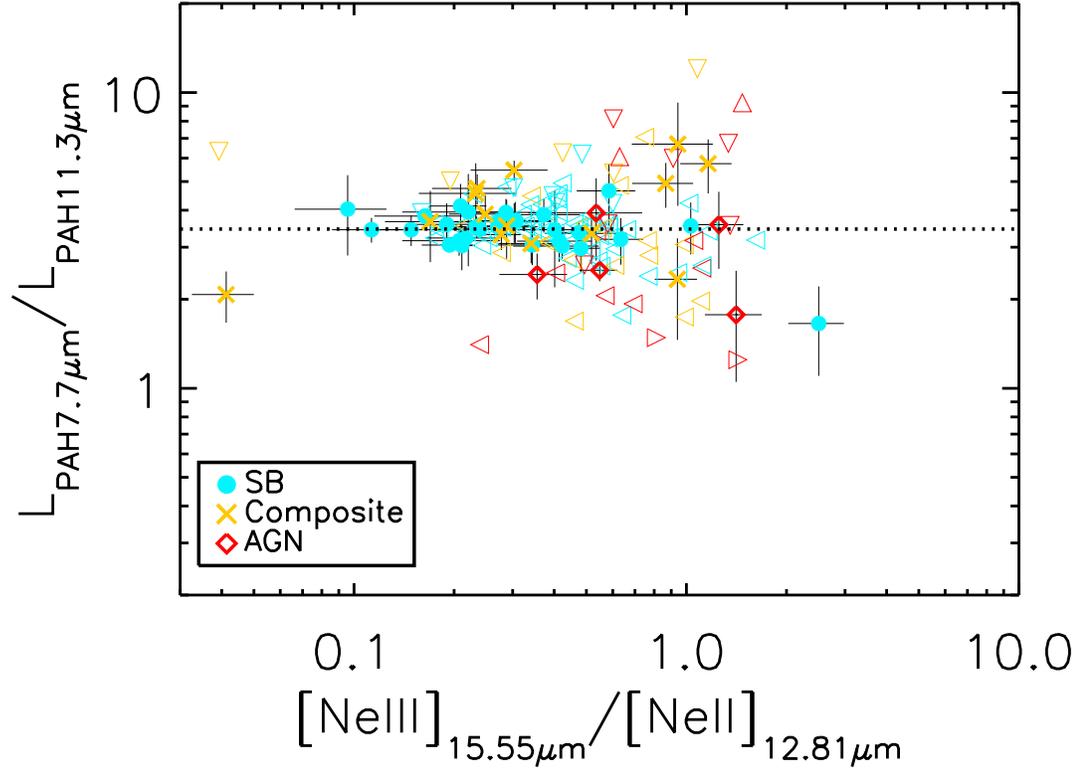}
  \caption{The flux ratios of [NeIII]/[NeII] versus the PAH luminosity
    ratios of L$_{\rm PAH 7.7\mu m}$/L$_{\rm PAH 11.3\mu m}$.  The
    blue, yellow and red open triangles represent upper/lower limits
    for the SB, composite and AGN in 5MUSES and the directions the
    triangles face are consistent with the directions of the
    limits. The dotted line is the median L$_{\rm PAH 7.7\mu
      m}$/L$_{\rm PAH 11.3\mu m}$ ratio for the SB-dominated sources
    on this plot. }
  \label{fig:pahratio_ne}
\end{figure}

In Figure \ref{fig:pahratio_ne}, we show the flux ratios of L$_{\rm
  PAH 7.7\mu m}$/L$_{\rm PAH 11.3\mu m}$ versus [NeIII]/[NeII]. The
solid symbols denote detections while the open triangles represent
upper/lower limits. We overplot the median L$_{\rm PAH 7.7\mu
  m}$/L$_{\rm PAH 11.3\mu m}$ for SB-dominated sources in 5MUSES as
the dotted line. We find that the SB, composite and AGN-dominated
sources (including sources with upper/lower limits) are almost evenly
distributed on the two sides of the dotted line. However, the AGN with
solid detections on both axes do appear to have lower L$_{\rm PAH
  7.7\mu m}$/L$_{\rm PAH 11.3\mu m}$ ratios in general.  As has been
discussed in Section 4.4, this is consistent with the studies of
\citet{Smith07} using the SINGS nuclear spectra. We note that the
5MUSES sample do not have sources with extreme L$_{\rm PAH 7.7\mu
  m}$/L$_{\rm PAH 11.3\mu m}$ ratios comparable to the lowest ones
reached by SINGS. This is probably because the SINGS spectra probe
smaller, more central and thus more AGN-dominated regions. It should
also be noted that the AGN luminosities in 5MUSES are substantially
higher than SINGS. Our results are consistent with \citet{Odowd09},
who have studied a UV-SDSS selected sample at z$\sim$0.1 and do not
observe extreme L$_{\rm PAH 7.7\mu m}$/L$_{\rm PAH 11.3\mu m}$ ratios
either. We also notice that the range of [NeIII]/[NeII] ratios are
similar for all three groups of objects that we have classified based
on their 6.2\,$\mu$m PAH EWs. This is consistent with the study of
\citet{Bernard-Salas09}, who found no correlation between the PAH EWs
and the [NeIII]/[NeII] ratios in a sample of starburst
galaxies. However, in more extreme radiation field conditions, such as
low-metallicity environment, PAH EWs have been observed to
anti-correlate with the radiation field hardness indicated by
[NeIII]/[NeII] ratios \citep{Wu06}.

\section{Conclusions}
We have studied a flux limited (f$_{24\mu m}>$5\,mJy) representative sample
of 330 galaxies surveyed with the Infrared Spectrograph on board the
{\em Spitzer} Space Telescope. Secure redshifts of 280 objects have
been obtained from optical or infrared spectroscopy. The redshifts of
the 5MUSES sample ranges from 0.08 to 4.27, with a median value of
0.144. This places the 5MUSES sample at intermediate redshift, which
bridges the gap between the nearby bright sources known from previous
studies and the z$\sim$2 objects pursued in most of the IRS follow up
observations of deep 24\,$\mu$m surveys. The simple selection criteria
ensures that our sample provides a complete census of galaxies with
crucial information on understanding the galaxy evolution processes.

Using mid-IR spectroscopy and mid-to-far IR photometry, we have
obtained accurate estimates on the total infrared luminosities of
5MUSES galaxies. This is achieved by minimizing the $\chi^2$ to find
the best fit template from our newly constructed empirical SED library
built upon recent {\em Spitzer} observations. The availability of
longer wavelength data also greatly reduces the uncertainties in
L$_{\rm IR}$. When only one IRS spectrum is available, one can still
predict the shape of the FIR SED from the mid-IR and estimate L$_{\rm
  IR}$, albeit with substantially larger uncertainties (0.2\,dex). The
IRS-only method does not introduce a systematic bias when estimating
L$_{\rm IR}$ for warm sources, but could underestimate the L$_{\rm
  IR}$ by $\sim$17\% for cold sources, due to the lack of information
sampling the peak of the SED. The fractional contribution of single
band luminosity to L$_{\rm IR}$ varies depending on the dominant
energy source and the average values have been calculated for the SB,
composite and AGN dominated sources, as well as the whole sample.

We analyze the properties of the PAH emission in our sample using the
IRS spectra. The PAH EWs show a bimodal distribution, which might be
related to the selection effect of the sample. The starburst and AGN
dominated sources form two clumps when comparing the continuum slopes
and PAH EWs, while there is little discernible correlation within each
group.  Average spectra binned with the 6.2\,$\mu$m PAH EWs, the
continuum slopes of log(f$_{30}$/f$_{15}$) and log(f$_{70}$/f$_{24}$)
have been derived to show the typical SED shapes. The variation in PAH
EW and L$_{\rm PAH}$/L$_{\rm IR}$ ratios when galaxy color changes
have also been inspected. The galaxy color provides essential
constraint on estimating the total infrared luminosity from broadband
photometry.

We have also inspected the band-to-band PAH intensity ratios with
regard to different spectral types. The L$_{\rm PAH7.7\mu m}$/L$_{\rm
  PAH11.3\mu m}$ ratios in AGN dominated sources in 5MUSES are on
average lower than the SB or composite sources. The SB, composite and
AGN dominated sources have mean log(L$_{\rm PAH7.7\mu m}$/L$_{\rm
  PAH11.3\mu m}$) ratios of 0.53$\pm$0.08, 0.54$\pm$0.15 and
0.32$\pm$0.18, respectively. The mean log(L$_{\rm PAH7.7\mu
  m}$/L$_{\rm PAH11.3\mu m}$) ratio for the SB dominated sources in
5MUSES is lower than the mean ratio derived from the nuclear spectra
of SB galaxies in SINGS (0.63$\pm$0.06), which might indicate a
difference in the physical conditions near the nucleus versus over the
entire galaxy. At the median redshift of our sample, the IRS SL slit
width corresponds to a few kpc, thus even if the ionization state or
grain size distribution is different at the nuclear level, the signal
might get diluted when we study the integrated spectrum and would
result in the different log(L$_{\rm PAH7.7\mu m}$/L$_{\rm PAH11.3\mu
  m}$) ratio distribution.

Finally, we provide our calibration of using PAH luminosity or mid-IR
continuum luminosity to estimate L$_{\rm IR}$ in the Appendix. We have
shown that single band luminosities trace the L$_{\rm IR}$ differently
in SB or AGN dominated sources and we provide calibrations for each
object type. This technique will be useful for luminosity estimates
when no multi-wavelength data are available. 

\acknowledgments 

We thank the anonymous referee whose comments have helped to improve
this manuscript. This work was based on observations made with the
{\em Spitzer Space Telescope}, which is operated by JPL/Caltech under
a contract with NASA. The observations are associated with the {\em
  Spitzer} Legacy Program 40539. The authors acknowledge support by
NASA through awards issued by JPL/Caltech. This research has made use
of the NASA/IPAC Extragalactic Database (NED) which is operated by the
Jet Propulsion Laboratory, California Institute of Technology, under
contract with the National Aeronautics and Space Administration.

\appendix
\section{Estimating the Total Infrared Luminosity from PAH or Monochromatic Continuum Luminosities}

\begin{figure}
  \epsscale{1.0}
  \plotone{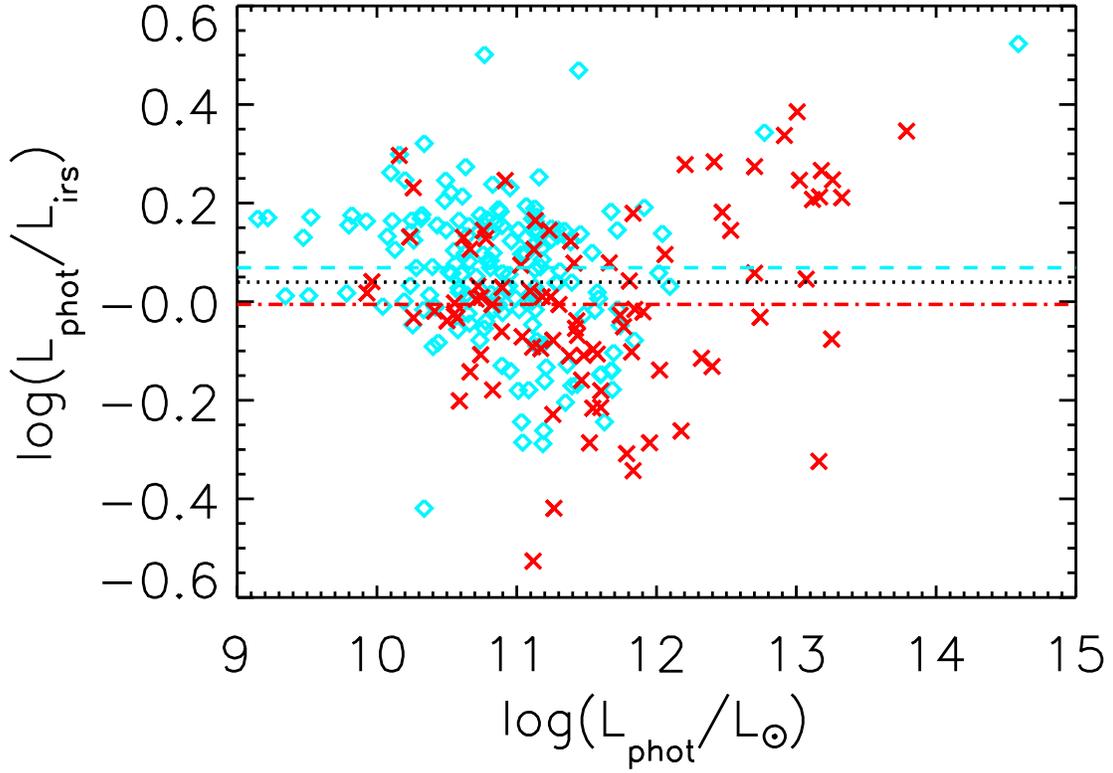}
  \caption{ The ratio of the L$_{\rm IR}$ derived from using IRAC 5.8,
    8.0 and MIPS 24, 70 and 160\,$\mu$m photometry over L$_{\rm IR}$
    predicted by IRS spectra versus L$_{\rm IR}$. The diamonds
    represent cold sources with $f_{\rm 24\mu m}/f_{\rm 70\mu m}<$0.2
    and the crosses represent the warm sources with $f_{\rm 24\mu
      m}/f_{\rm 70\mu m}>$0.2. The dotted line indicate the median of
    the luminosity ratios ($L_{\rm phot}/L_{\rm IR}=$1.10,
    1$\sigma$=0.16\,dex). The dashed line indicates the median ratio
    for the cold sources ($L_{\rm phot}/L_{\rm IR}=$1.17,
    1$\sigma$=0.14\,dex) and the dot-dashed line indicates the median
    ratio for the warm sources ($L_{\rm phot}/L_{\rm IR}=$0.99,
    1$\sigma$=0.18\,dex). }
  \label{fig:comp_irs_phot}
\end{figure}

In Section 3, we have discussed in detail our method to estimate the
total infrared luminosities for the 5MUSES sources. The empirical
library of SED templates built from {\em Spitzer} observations, as
well as the availability of photometric and spectroscopic data from
mid-IR to FIR for 5MUSES, allow us to have precise estimates on their
L$_{\rm IR}$. We have shown in Figure \ref{fig:comp_irs_phot} the
importance of having FIR data in determining the total energy output
in the infrared.  However, for high redshift galaxies, FIR
observations are not always available. {\em Herschel} Space
Observatory will provide FIR measurements from 70 to 500\,$\mu$m to
reveal the properties of cold dust in many systems. For now, we
provide our calibration of estimating L$_{\rm IR}$ from several bands
in the mid-IR and discuss its applications. The following correlations
are derived by performing a linear fit to the 5MUSES data with equal
weight on each object because the dispersion of the data point in the
x-y plane is larger than the measurement errors.

As has been shown in many studies, the infrared SED of a starburst
galaxy is drastically different from that of an AGN \citep{Brandl06,
  Hao07, Armus07}. Because of these substantial variations in the SED
shapes, it is crucial to calibrate the luminosity estimates for each
spectral type.  Here we provide our luminosity calibrations based on
the three spectral types : starburst, composite and AGN. The following
PAH luminosities are derived from the PAHFIT method.

1. 6.2\,$\mu$m PAH: With a wavelength cut at 28\,$\mu$m for the James
Webb Space Telescope (JWST), the 6.2\,$\mu$m PAH feature might be the
only PAH band that could be observed to quantify star formation
activities in z$\sim$3 sources when JWST is launched.\\
For SB sources:
\begin{equation}
  {\mathrm{log}}L_{\mathrm{IR}}=(2.40\pm0.22)+(0.96\pm0.03){\mathrm{log}}L_{\mathrm{PAH6.2\mu m}}
\end{equation}
For composite sources:
\begin{equation}
  {\mathrm{log}}L_{\mathrm{IR}}=(1.76\pm0.32)+(1.04\pm0.04){\mathrm{log}}L_{\mathrm{PAH6.2\mu m}}
\end{equation}
For AGN sources:
\begin{equation}
  {\mathrm{log}}L_{\mathrm{IR}}=(-0.58\pm0.58)+(1.30\pm0.06){\mathrm{log}}L_{\mathrm{PAH6.2\mu m}}
\end{equation}

2. 7.7\,$\mu$m PAH: The 7.7\,$\mu$m PAH complex is the strongest band
among the various PAH features. It is often used to estimate the total
infrared luminosities for the z$\sim$1-2 sources pursued in IRS
observations of 24\,$\mu$m selected sources. \\ 
For SB sources:
\begin{equation}
  {\mathrm{log}}L_{\mathrm{IR}}=(2.30\pm0.26)+(0.91\pm0.003){\mathrm{log}}L_{\mathrm{PAH7.7\mu m}}
\end{equation}
For composite sources:
\begin{equation}
  {\mathrm{log}}L_{\mathrm{IR}}=(1.80\pm0.44)+(0.98\pm0.05){\mathrm{log}}L_{\mathrm{PAH7.7\mu m}}
\end{equation}
For AGN sources:
\begin{equation}
  {\mathrm{log}}L_{\mathrm{IR}}=(3.45\pm0.90)+(0.83\pm0.09){\mathrm{log}}L_{\mathrm{PAH7.7\mu m}}
\end{equation}

3. 11.3\,$\mu$m PAH: The 11.3\,$\mu$m band is another strong PAH band
in the mid-IR that is relatively isolated from other PAH bands.
However, the integrated fluxes from this band might be affected by the
9.7\,$\mu$m silicate feature. \\ 
For SB sources:
\begin{equation}
  {\mathrm{log}}L_{\mathrm{IR}}=(2.18\pm0.22)+(0.98\pm0.03){\mathrm{log}}L_{\mathrm{PAH11.3\mu m}}
\end{equation}
For composite sources:
\begin{equation}
  {\mathrm{log}}L_{\mathrm{IR}}=(1.49\pm0.43)+(1.07\pm0.05){\mathrm{log}}L_{\mathrm{PAH11.3\mu m}}
\end{equation}
For AGN sources:
\begin{equation}
  {\mathrm{log}}L_{\mathrm{IR}}=(2.22\pm0.36)+(1.00\pm0.04){\mathrm{log}}L_{\mathrm{PAH11.3\mu m}}
\end{equation}

4. 6.2+7.7+11.3\,$\mu$m PAH: In normal star-forming galaxies, the PAH
emission accounts for $\sim$10\%-15\% of the total infrared
luminosities \citep{Smith07}, while this fraction is smaller for local
ULIRGs \citep{Armus07}. Here we use the sum of the three strongest PAH
bands, the 6.2, 7.7 and 11.3\,$\mu$m PAH luminosities to represent the
total PAH luminosities. However, when using the correlation provided
here, one needs to keep in mind that the properties of PAHs studied in
the local universe might be different at high z, as has already been
revealed in the study of several z$\sim$2 luminous infrared galaxies
\citep{Sajina07, Pope08}. Understanding the PAH contribution in our
intermediate redshift sample would also be instrumental for tackling
the problem of whether and how PAH emission evolves with redshift in
future studies.\\ For SB sources:
\begin{equation}
  {\mathrm{log}}L_{\mathrm{IR}}=(1.82\pm0.25)+(0.95\pm0.03){\mathrm{log}}L_{\mathrm{PAH6.2+7.7+11.3\mu m}}
\end{equation}
For composite sources:
\begin{equation}
  {\mathrm{log}}L_{\mathrm{IR}}=(0.73\pm0.36)+(1.06\pm0.04){\mathrm{log}}L_{\mathrm{PAH6.2+7.7+11.3\mu m}}
\end{equation}
For AGN sources:
\begin{equation}
  {\mathrm{log}}L_{\mathrm{IR}}=(-2.13\pm1.01)+(1.35\pm0.10){\mathrm{log}}L_{\mathrm{PAH6.2+7.7+11.3\mu m}}
\end{equation}

5. 5.8\,$\mu$m monochromatic continuum luminosity: The 5.8\,$\mu$m
continuum luminosity provides a crude estimate of L$_{\rm IR}$. In AGN
dominated sources, the 5.8\,$\mu$m continuum will be elevated due to
the presence of very hot dust component. This is also a band that is
available for most of the high redshift samples observed by {\em
  Spitzer}, and for JWST when it is launched. \\ 
For SB sources:
\begin{equation}
  {\mathrm{log}}L_{\mathrm{IR}}=(1.94\pm0.25)+(0.95\pm0.03){\mathrm{log}}L_{\mathrm{5.8\mu m}}
\end{equation}
For composite sources:
\begin{equation}
  {\mathrm{log}}L_{\mathrm{IR}}=(2.68\pm0.45)+(0.87\pm0.05){\mathrm{log}}L_{\mathrm{5.8\mu m}}
\end{equation}
For AGN sources:
\begin{equation}
  {\mathrm{log}}L_{\mathrm{IR}}=(2.34\pm0.29)+(0.85\pm0.03){\mathrm{log}}L_{\mathrm{5.8\mu m}}
\end{equation}

6. IRAC 8\,$\mu$m: The rest-frame IRAC 8.0\,$\mu$m band has included
both dust continuum emission and PAH emission from the 7.7, 8.3 and
8.6\,$\mu$m PAH band (if present).  It provides a useful channel for
estimating L$_{\rm IR}$ from PAH features when no spectroscopy is available.
\\ For SB sources:
\begin{equation}
  {\mathrm{log}}L_{\mathrm{IR}}=(1.60\pm0.21)+(0.93\pm0.02){\mathrm{log}}L_{\mathrm{IRAC 8\mu m}}
\end{equation}
For composite sources:
\begin{equation}
  {\mathrm{log}}L_{\mathrm{IR}}=(1.45\pm0.26)+(0.95\pm0.03){\mathrm{log}}L_{\mathrm{IRAC 8\mu m}}
\end{equation}
For AGN sources:
\begin{equation}
  {\mathrm{log}}L_{\mathrm{IR}}=(1.70\pm0.26)+(0.90\pm0.02){\mathrm{log}}L_{\mathrm{IRAC 8\mu m}}
\end{equation}

7. 14\,$\mu$m monochromatic continuum luminosity: The 14\,$\mu$m is an
important band in the mid-IR that is still sensitive to the AGN
emission.\\

For SB sources:
\begin{equation}
  {\mathrm{log}}L_{\mathrm{IR}}=(1.44\pm0.16)+(0.97\pm0.02){\mathrm{log}}L_{\mathrm{14\mu m}}
\end{equation}
For composite sources:
\begin{equation}
  {\mathrm{log}}L_{\mathrm{IR}}=(1.76\pm0.37)+(0.93\pm0.04){\mathrm{log}}L_{\mathrm{14\mu m}}
\end{equation}
For AGN sources:
\begin{equation}
  {\mathrm{log}}L_{\mathrm{IR}}=(1.62\pm0.29)+(0.90\pm0.03){\mathrm{log}}L_{\mathrm{14\mu m}}
\end{equation}

8. 24\,$\mu$m monochromatic continuum luminosity: Here we refer to the
24\,$\mu$m continuum luminosity averaged in one micron range, instead
of the rest-frame MIPS 24\,$\mu$m band. This is because if we use the
MIPS 24\,$\mu$m band, sources at z$>\sim$0.3 will be eliminated from
this study due to the limited wavelength coverage of its rest-frame
mid-IR spectra. The sources we use in the calibration mostly have
10$^{10}$L$_\odot$$<$L$_{\rm IR}$ $<$10$^{12}$L$_\odot$ and no quasars
have been included in this calibration because of the wavelength
cut. Since our sample is selected at 24\,$\mu$m, it tends to favor
warmer sources, which also needs to be kept in mind when using these
relations. \\
For SB sources:
\begin{equation}
  {\mathrm{log}}L_{\mathrm{IR}}=(1.56\pm0.23)+(0.93\pm0.02){\mathrm{log}}L_{\mathrm{24\mu m}}
\end{equation}
For composite sources:
\begin{equation}
  {\mathrm{log}}L_{\mathrm{IR}}=(1.64\pm0.44)+(0.91\pm0.04){\mathrm{log}}L_{\mathrm{24\mu m}}
\end{equation}
For AGN sources:
\begin{equation}
  {\mathrm{log}}L_{\mathrm{IR}}=(1.67\pm0.47)+(0.89\pm0.05){\mathrm{log}}L_{\mathrm{24\mu m}}
\end{equation}

\begin{deluxetable}{ccccc}
  \setlength{\tabcolsep}{0.2in}
  \tablecaption{On-Source Integration Time of the Sample\label{integration_time}}
  \tablewidth{0pc}
  \tablehead{
    \colhead{f$_{\rm 24\mu m}$ (mJy)} & \colhead{SL2 (second)} & \colhead{SL1 (second)} & \colhead{LL2 (second)} & \colhead{LL1 (second)}
    }
    \startdata
    5$\sim$7       &  480  &  480  &  480  &  480   \\
    7$\sim$10      &  480  &  240  &  240  &  240   \\
    10$\sim$15     &  480  &  480  &  180  &  180   \\
    15$\sim$25     &  240  &  120  &  120  &  120   \\
    $>$25          &  120  &  120  &  60   &   60   \\
    \enddata
\end{deluxetable}

\begin{deluxetable}{llllcrlr}
  \tabletypesize{\scriptsize}
  \setlength{\tabcolsep}{0.05in}
  \tablecaption{General Properties of the Sample\label{tab:data}}
  \tablewidth{0pc}
  \tablehead{
    \colhead{ID} & \colhead{Name} & \colhead{RA (J2000)} & \colhead {Dec (J2000)} & \colhead{Redshift\tablenotemark{a}} & \colhead{f$_{24\mu m}$ (mJy)} & \colhead{6.2$\mu$m EW}  & \colhead{log(L$_{\rm IR}$/L$_\odot$)}
    }
  \startdata
5MUSES-002 & 5MUSES$\_$J021503.52-042421.6 & 02h15m03.5s & -04d24m21.7s & 0.137(2) & 5.2~~~~~~ & 0.776$\pm$0.009 & 10.89$\pm$0.02 \\
5MUSES-004 & 5MUSES$\_$J021557.11-033729.0 & 02h15m57.1s & -03d37m29.1s & 0.032(2) & 8.8~~~~~~ & 0.504$\pm$0.048 & 9.80$\pm$0.03 \\
5MUSES-005 & 5MUSES$\_$J021638.21-042250.8 & 02h16m38.2s & -04d22m50.9s & 0.304(2) & 14.4~~~~~~ &  $<$0.094 & 11.54$\pm$0.02 \\
5MUSES-006 & 5MUSES$\_$J021640.72-044405.1 & 02h16m40.7s & -04d44m05.1s & 0.870(1) & 14.7~~~~~~ &  $<$0.045 & 12.70$\pm$0.01 \\
5MUSES-008 & 5MUSES$\_$J021649.71-042554.8 & 02h16m49.7s & -04d25m54.8s & 0.143(2) & 10.1~~~~~~ & 1.107$\pm$0.057 & 11.01$\pm$0.07 \\
5MUSES-009 & 5MUSES$\_$J021657.77-032459.7 & 02h16m57.8s & -03d24m59.8s & 0.137(1) & 23.8~~~~~~ &  $<$0.062 & 10.90$\pm$0.03 \\
5MUSES-010 & 5MUSES$\_$J021729.06-041937.8 & 02h17m29.1s & -04d19m37.8s & 1.146(1) & 8.8~~~~~~ &  $<$0.113 & 12.74$\pm$0.06 \\
5MUSES-011 & 5MUSES$\_$J021743.01-043625.1 & 02h17m43.0s & -04d36m25.2s & 0.784(2) & 5.5~~~~~~ &  $<$0.080 & 12.00$\pm$0.06 \\
5MUSES-012 & 5MUSES$\_$J021743.82-051751.7 & 02h17m43.8s & -05d17m51.8s & 0.031(1) & 17.1~~~~~~ & 0.645$\pm$0.080 & 10.11$\pm$0.03 \\
5MUSES-013 & 5MUSES$\_$J021754.88-035826.4 & 02h17m54.9s & -03d58m26.5s & 0.226(1) & 10.3~~~~~~ & 0.530$\pm$0.044 & 11.72$\pm$0.04 \\
5MUSES-014 & 5MUSES$\_$J021808.22-045845.3 & 02h18m08.2s & -04d58m45.3s & 0.712(1) & 9.1~~~~~~ &  $<$0.049 & 12.02$\pm$0.07 \\
5MUSES-016 & 5MUSES$\_$J021830.57-045622.9 & 02h18m30.6s & -04d56m23.0s & 1.401(1) & 8.4~~~~~~ &  $<$0.083 & 12.67$\pm$0.10 \\
5MUSES-018 & 5MUSES$\_$J021849.76-052158.2 & 02h18m49.8s & -05d21m58.2s & 0.292(1) & 5.3~~~~~~ & 0.571$\pm$0.058 & 11.63$\pm$0.03 \\
5MUSES-019 & 5MUSES$\_$J021859.74-040237.2 & 02h18m59.7s & -04d02m37.2s & 0.199(2) & 15.9~~~~~~ &  $<$0.160 & 11.23$\pm$0.06 \\
5MUSES-020 & 5MUSES$\_$J021909.60-052512.9 & 02h19m09.6s & -05d25m12.9s & 0.098(2) & 25.3~~~~~~ &  $<$0.194 & 10.74$\pm$0.02 \\
5MUSES-021 & 5MUSES$\_$J021912.71-050541.8 & 02h19m12.7s & -05d05m41.9s & 0.194(2) & 6.1~~~~~~ & 0.639$\pm$0.041 & 11.04$\pm$0.07 \\
5MUSES-022 & 5MUSES$\_$J021916.05-055726.9 & 02h19m16.1s & -05d57m27.0s & 0.103(2) & 11.0~~~~~~ & 0.198$\pm$0.027 & 10.71$\pm$0.05 \\
5MUSES-023 & 5MUSES$\_$J021928.33-042239.8 & 02h19m28.3s & -04d22m39.8s & 0.042(2) & 17.3~~~~~~ & 0.611$\pm$0.053 & 10.04$\pm$0.04 \\
5MUSES-025 & 5MUSES$\_$J021938.70-032508.2 & 02h19m38.7s & -03d25m08.3s & 0.435(2) & 6.8~~~~~~ &  $<$0.094 & 11.66$\pm$0.02 \\
5MUSES-026 & 5MUSES$\_$J021939.08-051133.8 & 02h19m39.1s & -05d11m33.9s & 0.151(2) & 32.5~~~~~~ & 0.101$\pm$0.010 & 11.38$\pm$0.06 \\
5MUSES-028 & 5MUSES$\_$J021953.04-051824.1 & 02h19m53.0s & -05d18m24.2s & 0.072(2) & 30.3~~~~~~ & 0.781$\pm$0.019 & 10.93$\pm$0.03 \\
5MUSES-029 & 5MUSES$\_$J021956.96-052440.4 & 02h19m57.0s & -05d24m40.5s & 0.081(2) & 5.6~~~~~~ & 0.699$\pm$0.079 & 10.44$\pm$0.04 \\
5MUSES-030 & 5MUSES$\_$J022000.22-043947.6 & 02h20m00.2s & -04d39m47.7s & 0.350(1) & 5.8~~~~~~ & 0.137$\pm$0.007 & 11.48$\pm$0.06 \\
5MUSES-031 & 5MUSES$\_$J022005.93-031545.7 & 02h20m05.9s & -03d15m45.8s & 1.560(2) & 6.9~~~~~~ &  $<$0.178 & 13.17$\pm$0.05 \\
5MUSES-032 & 5MUSES$\_$J022012.21-034111.8 & 02h20m12.2s & -03d41m11.8s & 0.166(2) & 6.7~~~~~~ &  $<$0.079 & 10.40$\pm$0.08 \\
5MUSES-034 & 5MUSES$\_$J022145.09-053207.4 & 02h21m45.1s & -05d32m07.4s & 0.008(2) & 6.2~~~~~~ & 0.391$\pm$0.049 & 8.16$\pm$0.05 \\
5MUSES-035 & 5MUSES$\_$J022147.82-025730.7 & 02h21m47.8s & -02d57m30.7s & 0.068(2) & 21.0~~~~~~ & 0.714$\pm$0.037 & 10.88$\pm$0.04 \\
5MUSES-036 & 5MUSES$\_$J022147.87-044613.5 & 02h21m47.9s & -04d46m13.5s & 0.025(2) & 5.1~~~~~~ & 0.809$\pm$0.035 & 9.15$\pm$0.02 \\
5MUSES-037 & 5MUSES$\_$J022151.54-032911.8 & 02h21m51.5s & -03d29m11.8s & 0.164(1) & 6.9~~~~~~ & 0.748$\pm$0.104 & 11.14$\pm$0.03 \\
5MUSES-038 & 5MUSES$\_$J022205.03-050537.0 & 02h22m05.0s & -05d05m37.0s & 0.258(2) & 6.3~~~~~~ & 0.696$\pm$0.035 & 11.68$\pm$0.04 \\
5MUSES-039 & 5MUSES$\_$J022223.26-044319.8 & 02h22m23.3s & -04d43m19.9s & 0.073(2) & 5.1~~~~~~ & 0.356$\pm$0.023 & 10.28$\pm$0.03 \\
5MUSES-040 & 5MUSES$\_$J022224.06-050550.3 & 02h22m24.1s & -05d05m50.4s & 0.149(2) & 5.7~~~~~~ & 0.602$\pm$0.022 & 10.95$\pm$0.02 \\
5MUSES-041 & 5MUSES$\_$J022241.34-045652.0 & 02h22m41.3s & -04d56m52.1s & 0.139(2) & 5.1~~~~~~ & 0.308$\pm$0.008 & 10.57$\pm$0.08 \\
5MUSES-043 & 5MUSES$\_$J022257.96-041840.8 & 02h22m58.0s & -04d18m40.8s & 0.239(2) & 5.3~~~~~~ & 0.205$\pm$0.013 & 11.18$\pm$0.05 \\
5MUSES-044 & 5MUSES$\_$J022301.97-052335.8 & 02h23m02.0s & -05d23m35.9s & 0.708(2) & 6.8~~~~~~ &  $<$0.054 & 12.77$\pm$0.04 \\
5MUSES-045 & 5MUSES$\_$J022309.31-052316.1 & 02h23m09.3s & -05d23m16.2s & 0.084(2) & 5.3~~~~~~ &  $<$0.426 & 9.93$\pm$0.06 \\
5MUSES-047 & 5MUSES$\_$J022315.58-040606.0 & 02h23m15.6s & -04d06m06.0s & 0.199(2) & 9.4~~~~~~ & 0.486$\pm$0.067 & 11.31$\pm$0.04 \\
5MUSES-048 & 5MUSES$\_$J022329.13-043209.5 & 02h23m29.1s & -04d32m09.6s & 0.144(2) & 7.6~~~~~~ & 0.585$\pm$0.075 & 10.95$\pm$0.05 \\
5MUSES-049 & 5MUSES$\_$J022334.65-035229.4 & 02h23m34.7s & -03d52m29.4s & 0.176(2) & 7.6~~~~~~ & 0.966$\pm$0.129 & 11.03$\pm$0.10 \\
5MUSES-050 & 5MUSES$\_$J022345.04-054234.4 & 02h23m45.0s & -05d42m34.5s & 0.143(2) & 9.1~~~~~~ & 0.689$\pm$0.003 & 11.17$\pm$0.02 \\
5MUSES-051 & 5MUSES$\_$J022356.49-025431.1 & 02h23m56.5s & -02d54m31.1s & 0.451(2) & 10.4~~~~~~ & 0.058$\pm$0.004 & 11.79$\pm$0.07 \\
5MUSES-052 & 5MUSES$\_$J022413.64-042227.8 & 02h24m13.6s & -04d22m27.8s & 0.116(2) & 9.2~~~~~~ & 0.626$\pm$0.062 & 10.96$\pm$0.04 \\
5MUSES-053 & 5MUSES$\_$J022422.48-040230.5 & 02h24m22.5s & -04d02m30.6s & 0.171(2) & 7.5~~~~~~ & 0.414$\pm$0.007 & 11.16$\pm$0.04 \\
5MUSES-054 & 5MUSES$\_$J022431.58-052818.8 & 02h24m31.6s & -05d28m18.8s & 2.068(2) & 9.4~~~~~~ &  \nodata  & 13.02$\pm$0.25 \\
5MUSES-055 & 5MUSES$\_$J022434.28-041531.2 & 02h24m34.3s & -04d15m31.2s & 0.259(2) & 6.3~~~~~~ & 0.584$\pm$0.019 & 11.58$\pm$0.02 \\
5MUSES-056 & 5MUSES$\_$J022438.97-042706.3 & 02h24m39.0s & -04d27m06.4s & 0.252(2) & 6.6~~~~~~ & 0.156$\pm$0.034 & 11.30$\pm$0.06 \\
5MUSES-057 & 5MUSES$\_$J022446.99-040851.3 & 02h24m47.0s & -04d08m51.4s & 0.096(2) & 5.3~~~~~~ & 0.456$\pm$0.012 & 10.81$\pm$0.01 \\
5MUSES-058 & 5MUSES$\_$J022457.64-041417.9 & 02h24m57.6s & -04d14m18.0s & 0.063(2) & 11.9~~~~~~ & 0.476$\pm$0.035 & 10.59$\pm$0.04 \\
5MUSES-060 & 5MUSES$\_$J022507.43-041835.7 & 02h25m07.4s & -04d18m35.8s & 0.105(2) & 6.8~~~~~~ & 0.632$\pm$0.062 & 10.56$\pm$0.05 \\
5MUSES-061 & 5MUSES$\_$J022508.33-053917.7 & 02h25m08.3s & -05d39m17.7s & 0.293(2) & 9.6~~~~~~ & 0.025$\pm$0.002 & 11.55$\pm$0.05 \\
5MUSES-062 & 5MUSES$\_$J022522.59-045452.2 & 02h25m22.6s & -04d54m52.2s & 0.144(2) & 10.1~~~~~~ & 0.719$\pm$0.007 & 11.25$\pm$0.02 \\
5MUSES-063 & 5MUSES$\_$J022536.44-050011.5 & 02h25m36.4s & -05d00m11.6s & 0.053(1) & 13.7~~~~~~ & 0.709$\pm$0.051 & 10.77$\pm$0.11 \\
5MUSES-064 & 5MUSES$\_$J022548.21-050051.5 & 02h25m48.2s & -05d00m51.5s & 0.150(1) & 8.0~~~~~~ & 0.297$\pm$0.051 & 11.19$\pm$0.04 \\
5MUSES-065 & 5MUSES$\_$J022549.78-040024.6 & 02h25m49.8s & -04d00m24.7s & 0.044(2) & 58.5~~~~~~ & 0.438$\pm$0.009 & 10.64$\pm$0.03 \\
5MUSES-066 & 5MUSES$\_$J022559.99-050145.3 & 02h26m00.0s & -05d01m45.3s & 0.205(2) & 5.7~~~~~~ & 0.916$\pm$0.027 & 11.39$\pm$0.04 \\
5MUSES-067 & 5MUSES$\_$J022602.92-045306.8 & 02h26m02.9s & -04d53m06.8s & 0.056(2) & 6.4~~~~~~ & 0.669$\pm$0.028 & 10.13$\pm$0.03 \\
5MUSES-068 & 5MUSES$\_$J022603.61-045903.8 & 02h26m03.6s & -04d59m03.8s & 0.055(2) & 31.4~~~~~~ & 0.634$\pm$0.047 & 10.59$\pm$0.04 \\
5MUSES-069 & 5MUSES$\_$J022617.43-050443.4 & 02h26m17.4s & -05d04m43.5s & 0.057(2) & 48.7~~~~~~ & 0.168$\pm$0.005 & 10.77$\pm$0.04 \\
5MUSES-070 & 5MUSES$\_$J022637.79-035841.6 & 02h26m37.8s & -03d58m41.7s & 0.070(2) & 13.5~~~~~~ & 0.377$\pm$0.019 & 10.43$\pm$0.04 \\
5MUSES-071 & 5MUSES$\_$J022655.87-040302.2 & 02h26m55.9s & -04d03m02.5s & 0.135(2) & 6.9~~~~~~ & 1.026$\pm$0.179 & 10.59$\pm$0.05 \\
5MUSES-073 & 5MUSES$\_$J022720.68-044537.1 & 02h27m20.7s & -04d45m37.2s & 0.055(2) & 73.1~~~~~~ & 0.625$\pm$0.032 & 11.06$\pm$0.04 \\
5MUSES-074 & 5MUSES$\_$J022738.53-044702.7 & 02h27m38.5s & -04d47m02.8s & 0.173(2) & 7.1~~~~~~ & 0.918$\pm$0.032 & 11.13$\pm$0.03 \\
5MUSES-075 & 5MUSES$\_$J022741.64-045650.5 & 02h27m41.6s & -04d56m50.6s & 0.055(2) & 11.4~~~~~~ & 0.627$\pm$0.004 & 10.53$\pm$0.03 \\
5MUSES-077 & 5MUSES$\_$J103237.44+580845.9 & 10h32m37.4s & +58d08m46.0s & 0.251(2) & 6.1~~~~~~ & 0.394$\pm$0.060 & 11.74$\pm$0.04 \\
5MUSES-079 & 5MUSES$\_$J103450.50+584418.2 & 10h34m50.5s & +58d44m18.2s & 0.091(1) & 20.1~~~~~~ & 0.643$\pm$0.047 & 10.90$\pm$0.06 \\
5MUSES-080 & 5MUSES$\_$J103513.72+573444.6 & 10h35m13.7s & +57d34m44.6s & 1.537(2) & 5.5~~~~~~ &  $<$0.171 & 13.25$\pm$0.08 \\
5MUSES-081 & 5MUSES$\_$J103527.20+583711.9 & 10h35m27.2s & +58d37m12.0s & 0.885(2) & 6.9~~~~~~ & 0.080$\pm$0.010 & 12.53$\pm$0.04 \\
5MUSES-082 & 5MUSES$\_$J103531.46+581234.2 & 10h35m31.5s & +58d12m34.2s & 0.176(2) & 5.0~~~~~~ & 0.574$\pm$0.018 & 11.25$\pm$0.04 \\
5MUSES-083 & 5MUSES$\_$J103542.76+583313.1 & 10h35m42.8s & +58d33m13.1s & 0.087(2) & 6.6~~~~~~ & 0.761$\pm$0.002 & 10.51$\pm$0.06 \\
5MUSES-084 & 5MUSES$\_$J103601.81+581836.2 & 10h36m01.8s & +58d18m36.2s & 0.100(1) & 6.0~~~~~~ & 0.421$\pm$0.012 & 10.63$\pm$0.04 \\
5MUSES-085 & 5MUSES$\_$J103606.45+581829.7 & 10h36m06.5s & +58d18m29.7s & 0.210(1) & 22.5~~~~~~ &  $<$0.068 & 11.41$\pm$0.02 \\
5MUSES-086 & 5MUSES$\_$J103646.42+584330.6 & 10h36m46.4s & +58d43m30.6s & 0.140(2) & 6.8~~~~~~ & 0.549$\pm$0.020 & 10.94$\pm$0.03 \\
5MUSES-087 & 5MUSES$\_$J103701.99+574414.8 & 10h37m02.0s & +57d44m14.8s & 0.577(2) & 12.8~~~~~~ &  $<$0.065 & 12.06$\pm$0.05 \\
5MUSES-088 & 5MUSES$\_$J103724.74+580512.9 & 10h37m24.7s & +58d05m12.9s & 1.517(1) & 8.6~~~~~~ &  $<$0.158 & 13.02$\pm$0.06 \\
5MUSES-089 & 5MUSES$\_$J103803.35+572701.5 & 10h38m03.4s & +57d27m01.5s & 1.285(2) & 15.4~~~~~~ &  $<$0.086 & 13.33$\pm$0.06 \\
5MUSES-090 & 5MUSES$\_$J103813.90+580047.3 & 10h38m13.9s & +58d00m47.4s & 0.205(2) & 6.2~~~~~~ &  $<$0.335 & 10.89$\pm$0.14 \\
5MUSES-091 & 5MUSES$\_$J103818.19+583556.5 & 10h38m18.2s & +58d35m56.5s & 0.129(2) & 7.8~~~~~~ & 0.312$\pm$0.018 & 10.50$\pm$0.03 \\
5MUSES-093 & 5MUSES$\_$J103856.16+570333.9 & 10h38m56.2s & +57d03m33.9s & 0.178(2) & 5.7~~~~~~ & 0.338$\pm$0.011 & 10.82$\pm$0.03 \\
5MUSES-097 & 5MUSES$\_$J104016.32+570846.0 & 10h40m16.3s & +57d08m46.1s & 0.118(2) & 5.2~~~~~~ & 0.661$\pm$0.003 & 10.87$\pm$0.02 \\
5MUSES-098 & 5MUSES$\_$J104058.79+581703.3 & 10h40m58.8s & +58d17m03.4s & 0.072(1) & 10.4~~~~~~ &  $<$0.119 & 9.96$\pm$0.06 \\
5MUSES-099 & 5MUSES$\_$J104131.79+592258.4 & 10h41m31.8s & +59d22m58.4s & 0.925(1) & 7.0~~~~~~ &  $<$0.061 & 12.16$\pm$0.09 \\
5MUSES-100 & 5MUSES$\_$J104132.49+565953.0 & 10h41m32.5s & +56d59m53.0s & 0.346(1) & 8.3~~~~~~ & 0.454$\pm$0.044 & 11.74$\pm$0.04 \\
5MUSES-101 & 5MUSES$\_$J104159.83+585856.4 & 10h41m59.8s & +58d58m56.4s & 0.360(2) & 21.7~~~~~~ &  $<$0.127 & 11.95$\pm$0.02 \\
5MUSES-102 & 5MUSES$\_$J104255.66+575549.7 & 10h42m55.7s & +57d55m49.8s & 1.468(1) & 6.4~~~~~~ &  $<$0.067 & 13.01$\pm$0.05 \\
5MUSES-103 & 5MUSES$\_$J104303.50+585718.1 & 10h43m03.5s & +58d57m18.1s & 0.595(1) & 5.4~~~~~~ &  $<$0.066 & 11.90$\pm$0.05 \\
5MUSES-105 & 5MUSES$\_$J104432.94+564041.6 & 10h44m32.9s & +56d40m41.6s & 0.067(1) & 28.7~~~~~~ & 0.637$\pm$0.117 & 10.92$\pm$0.03 \\
5MUSES-106 & 5MUSES$\_$J104438.21+562210.7 & 10h44m38.2s & +56d22m10.8s & 0.025(1) & 80.6~~~~~~ & 0.509$\pm$0.027 & 10.49$\pm$0.05 \\
5MUSES-107 & 5MUSES$\_$J104454.08+574425.7 & 10h44m54.1s & +57d44m25.8s & 0.118(1) & 6.5~~~~~~ & 0.585$\pm$0.096 & 10.99$\pm$0.02 \\
5MUSES-108 & 5MUSES$\_$J104501.73+571111.3 & 10h45m01.7s & +57d11m11.4s & 0.390(1) & 10.9~~~~~~ &  $<$0.164 & 11.60$\pm$0.05 \\
5MUSES-109 & 5MUSES$\_$J104516.02+592304.7 & 10h45m16.0s & +59d23m04.7s & 0.322(1) & 5.1~~~~~~ & 0.094$\pm$0.005 & 11.39$\pm$0.07 \\
5MUSES-110 & 5MUSES$\_$J104643.26+584715.1 & 10h46m43.3s & +58d47m15.1s & 0.140(1) & 5.4~~~~~~ & 0.522$\pm$0.017 & 10.90$\pm$0.03 \\
5MUSES-112 & 5MUSES$\_$J104705.07+590728.4 & 10h47m05.1s & +59d07m28.5s & 0.391(1) & 7.0~~~~~~ & 0.032$\pm$0.003 & 11.39$\pm$0.10 \\
5MUSES-114 & 5MUSES$\_$J104729.89+572842.9 & 10h47m29.9s & +57d28m42.9s & 0.230(2) & 6.2~~~~~~ & 0.477$\pm$0.052 & 11.60$\pm$0.01 \\
5MUSES-115 & 5MUSES$\_$J104837.81+582642.1 & 10h48m37.8s & +58d26m42.2s & 0.232(1) & 7.6~~~~~~ & 0.729$\pm$0.022 & 11.68$\pm$0.02 \\
5MUSES-116 & 5MUSES$\_$J104839.73+555356.4 & 10h48m39.7s & +55d53m56.5s & 2.043(1) & 9.8~~~~~~ &  \nodata  & 13.46$\pm$0.25 \\
5MUSES-117 & 5MUSES$\_$J104843.90+580341.2 & 10h48m43.9s & +58d03m41.3s & 0.162(2) & 7.1~~~~~~ & 0.838$\pm$0.029 & 11.04$\pm$0.05 \\
5MUSES-118 & 5MUSES$\_$J104907.15+565715.3 & 10h49m07.2s & +56d57m15.4s & 0.072(1) & 9.7~~~~~~ & 0.805$\pm$0.014 & 10.65$\pm$0.03 \\
5MUSES-119 & 5MUSES$\_$J104918.33+562512.9 & 10h49m18.3s & +56d25m13.0s & 0.330(1) & 7.1~~~~~~ & 0.037$\pm$0.001 & 11.20$\pm$0.08 \\
5MUSES-123 & 5MUSES$\_$J105005.97+561500.0 & 10h50m06.0s & +56d15m00.0s & 0.119(2) & 14.8~~~~~~ & 0.714$\pm$0.097 & 11.14$\pm$0.04 \\
5MUSES-124 & 5MUSES$\_$J105047.83+590348.3 & 10h50m47.8s & +59d03m48.4s & 0.131(2) & 5.2~~~~~~ & 0.623$\pm$0.015 & 10.90$\pm$0.04 \\
5MUSES-126 & 5MUSES$\_$J105058.76+560550.0 & 10h50m58.8s & +56d05m50.0s & 0.125(2) & 5.5~~~~~~ & 0.496$\pm$0.054 & 10.41$\pm$0.05 \\
5MUSES-127 & 5MUSES$\_$J105106.12+591625.3 & 10h51m06.1s & +59d16m25.3s & 0.768(1) & 5.4~~~~~~ & 0.078$\pm$0.003 & 12.32$\pm$0.06 \\
5MUSES-128 & 5MUSES$\_$J105128.05+573502.4 & 10h51m28.1s & +57d35m02.4s & 0.073(1) & 10.0~~~~~~ & 0.695$\pm$0.081 & 10.42$\pm$0.03 \\
5MUSES-130 & 5MUSES$\_$J105158.53+590652.0 & 10h51m58.5s & +59d06m52.1s & 1.814(2) & 5.4~~~~~~ &  $<$0.093 & 13.26$\pm$0.05 \\
5MUSES-131 & 5MUSES$\_$J105200.29+591933.7 & 10h52m00.3s & +59d19m33.8s & 0.115(1) & 11.4~~~~~~ & 0.297$\pm$0.036 & 10.76$\pm$0.03 \\
5MUSES-132 & 5MUSES$\_$J105206.56+580947.1 & 10h52m06.6s & +58d09m47.1s & 0.117(2) & 16.7~~~~~~ & 0.661$\pm$0.009 & 11.34$\pm$0.03 \\
5MUSES-133 & 5MUSES$\_$J105336.87+580350.7 & 10h53m36.9s & +58d03m50.7s & 0.460(1) & 5.9~~~~~~ & 0.368$\pm$0.001 & 12.02$\pm$0.04 \\
5MUSES-135 & 5MUSES$\_$J105404.11+574019.7 & 10h54m04.1s & +57d40m19.7s & 1.101(1) & 8.5~~~~~~ &  $<$0.084 & 12.70$\pm$0.03 \\
5MUSES-136 & 5MUSES$\_$J105421.65+582344.6 & 10h54m21.7s & +58d23m44.7s & 0.205(2) & 16.8~~~~~~ & 0.074$\pm$0.001 & 11.43$\pm$0.03 \\
5MUSES-138 & 5MUSES$\_$J105604.84+574229.9 & 10h56m04.8s & +57d42m30.0s & 1.211(1) & 11.2~~~~~~ &  $<$0.146 & 13.16$\pm$0.11 \\
5MUSES-139 & 5MUSES$\_$J105636.95+573449.3 & 10h56m37.0s & +57d34m49.4s & 0.047(1) & 6.4~~~~~~ & 0.444$\pm$0.060 & 10.16$\pm$0.04 \\
5MUSES-140 & 5MUSES$\_$J105641.81+580046.0 & 10h56m41.8s & +58d00m46.0s & 0.130(1) & 7.5~~~~~~ & 0.686$\pm$0.014 & 11.03$\pm$0.03 \\
5MUSES-141 & 5MUSES$\_$J105705.43+580437.4 & 10h57m05.4s & +58d04m37.4s & 0.140(2) & 16.5~~~~~~ & 0.097$\pm$0.001 & 11.18$\pm$0.03 \\
5MUSES-142 & 5MUSES$\_$J105733.53+565737.4 & 10h57m33.5s & +56d57m37.5s & 0.086(1) & 5.6~~~~~~ & 0.454$\pm$0.023 & 10.38$\pm$0.06 \\
5MUSES-143 & 5MUSES$\_$J105740.55+570616.4 & 10h57m40.6s & +57d06m16.5s & 0.073(1) & 6.1~~~~~~ & 0.503$\pm$0.058 & 10.24$\pm$0.03 \\
5MUSES-144 & 5MUSES$\_$J105829.28+580439.2 & 10h58m29.3s & +58d04m39.3s & 0.136(1) & 7.1~~~~~~ & 0.452$\pm$0.075 & 10.56$\pm$0.03 \\
5MUSES-145 & 5MUSES$\_$J105854.08+574130.0 & 10h58m54.1s & +57d41m30.0s & 0.232(1) & 6.1~~~~~~ & 0.222$\pm$0.031 & 11.10$\pm$0.07 \\
5MUSES-146 & 5MUSES$\_$J105903.47+572155.1 & 10h59m03.5s & +57d21m55.1s & 0.119(2) & 13.8~~~~~~ &  $<$0.261 & 10.87$\pm$0.05 \\
5MUSES-147 & 5MUSES$\_$J105951.71+581802.9 & 10h59m51.7s & +58d18m02.9s & 2.335(1) & 5.3~~~~~~ &  \nodata  & 13.08$\pm$0.17 \\
5MUSES-148 & 5MUSES$\_$J105959.95+574848.1 & 11h00m00.0s & +57d48m48.2s & 0.453(1) & 9.1~~~~~~ &  $<$0.052 & 11.83$\pm$0.02 \\
5MUSES-149 & 5MUSES$\_$J110002.06+573142.1 & 11h00m02.1s & +57d31m42.2s & 0.387(2) & 8.3~~~~~~ & 0.496$\pm$0.027 & 12.02$\pm$0.05 \\
5MUSES-151 & 5MUSES$\_$J110124.97+574315.8 & 11h01m25.0s & +57d43m15.9s & 0.243(1) & 6.1~~~~~~ & 0.545$\pm$0.058 & 11.17$\pm$0.06 \\
5MUSES-152 & 5MUSES$\_$J110133.80+575206.6 & 11h01m33.8s & +57d52m06.6s & 0.277(2) & 6.4~~~~~~ & 0.509$\pm$0.057 & 11.84$\pm$0.04 \\
5MUSES-153 & 5MUSES$\_$J110223.58+574436.2 & 11h02m23.6s & +57d44m36.2s & 0.226(1) & 10.2~~~~~~ &  $<$0.093 & 11.12$\pm$0.02 \\
5MUSES-154 & 5MUSES$\_$J110235.02+574655.7 & 11h02m35.0s & +57d46m55.7s & 0.226(2) & 6.2~~~~~~ & 0.523$\pm$0.066 & 11.48$\pm$0.04 \\
5MUSES-155 & 5MUSES$\_$J155833.00+544426.9 & 15h58m32.9s & +54d44m27.2s & 0.350(1) & 9.1~~~~~~ & 0.086$\pm$0.001 & 11.52$\pm$0.03 \\
5MUSES-156 & 5MUSES$\_$J155833.28+545937.1 & 15h58m33.3s & +54d59m37.2s & 0.340(2) & 6.3~~~~~~ & 0.327$\pm$0.012 & 12.10$\pm$0.03 \\
5MUSES-157 & 5MUSES$\_$J155936.12+544203.7 & 15h59m36.1s & +54d42m03.8s & 0.308(2) & 14.5~~~~~~ &  $<$0.060 & 11.32$\pm$0.06 \\
5MUSES-158 & 5MUSES$\_$J160038.82+551018.6 & 16h00m38.8s & +55d10m18.7s & 0.144(2) & 20.1~~~~~~ & 0.637$\pm$0.020 & 11.45$\pm$0.04 \\
5MUSES-160 & 5MUSES$\_$J160114.49+551304.1 & 16h01m14.5s & +55d13m04.1s & 0.220(2) & 7.9~~~~~~ &  $<$0.079 & 10.82$\pm$0.06 \\
5MUSES-162 & 5MUSES$\_$J160128.52+544521.3 & 16h01m28.5s & +54d45m21.4s & 0.728(1) & 12.8~~~~~~ &  $<$0.034 & 12.47$\pm$0.01 \\
5MUSES-163 & 5MUSES$\_$J160322.77+544237.3 & 16h03m22.8s & +54d42m37.3s & 0.215(1) & 5.7~~~~~~ & 0.687$\pm$0.070 & 11.35$\pm$0.03 \\
5MUSES-165 & 5MUSES$\_$J160341.30+552612.7 & 16h03m41.3s & +55d26m12.7s & 0.146(1) & 5.3~~~~~~ & 0.610$\pm$0.012 & 11.11$\pm$0.03 \\
5MUSES-166 & 5MUSES$\_$J160358.18+555504.4 & 16h03m58.2s & +55d55m04.4s & 0.322(2) & 5.0~~~~~~ & 0.406$\pm$0.030 & 11.56$\pm$0.06 \\
5MUSES-167 & 5MUSES$\_$J160401.21+551502.7 & 16h04m01.2s & +55d15m02.7s & 0.182(1) & 11.4~~~~~~ &  $<$0.112 & 11.11$\pm$0.05 \\
5MUSES-168 & 5MUSES$\_$J160408.18+542531.2 & 16h04m08.2s & +54d25m31.2s & 0.260(1) & 5.0~~~~~~ & 0.604$\pm$0.054 & 11.54$\pm$0.02 \\
5MUSES-169 & 5MUSES$\_$J160408.30+545813.0 & 16h04m08.3s & +54d58m13.1s & 0.064(1) & 26.2~~~~~~ & 0.602$\pm$0.009 & 10.83$\pm$0.03 \\
5MUSES-171 & 5MUSES$\_$J160440.64+553409.2 & 16h04m40.6s & +55d34m09.3s & 0.078(1) & 22.9~~~~~~ & 0.521$\pm$0.035 & 11.10$\pm$0.04 \\
5MUSES-173 & 5MUSES$\_$J160630.59+542007.4 & 16h06m30.6s & +54d20m07.4s & 0.820(1) & 5.5~~~~~~ &  $<$0.052 & 11.91$\pm$0.10 \\
5MUSES-174 & 5MUSES$\_$J160655.35+534016.9 & 16h06m55.4s & +53d40m16.9s & 0.214(1) & 14.6~~~~~~ &  $<$0.086 & 11.26$\pm$0.01 \\
5MUSES-176 & 5MUSES$\_$J160730.41+554905.5 & 16h07m30.4s & +55d49m05.6s & 0.118(1) & 6.2~~~~~~ & 0.835$\pm$0.100 & 10.81$\pm$0.03 \\
5MUSES-177 & 5MUSES$\_$J160743.09+554416.5 & 16h07m43.1s & +55d44m16.5s & 0.118(1) & 9.6~~~~~~ & 0.752$\pm$0.049 & 11.13$\pm$0.03 \\
5MUSES-178 & 5MUSES$\_$J160801.79+555359.7 & 16h08m01.8s & +55d53m59.7s & 0.062(1) & 6.2~~~~~~ & 1.057$\pm$0.011 & 10.24$\pm$0.04 \\
5MUSES-179 & 5MUSES$\_$J160803.71+545301.9 & 16h08m03.7s & +54d53m02.0s & 0.053(1) & 5.1~~~~~~ & 0.373$\pm$0.019 & 10.26$\pm$0.01 \\
5MUSES-180 & 5MUSES$\_$J160819.57+553314.2 & 16h08m19.6s & +55d33m14.3s & 0.115(1) & 7.2~~~~~~ & 0.337$\pm$0.005 & 10.83$\pm$0.02 \\
5MUSES-181 & 5MUSES$\_$J160832.59+552926.9 & 16h08m32.6s & +55d29m27.0s & 0.065(1) & 5.9~~~~~~ & 0.844$\pm$0.121 & 10.32$\pm$0.03 \\
5MUSES-183 & 5MUSES$\_$J160839.73+552330.6 & 16h08m39.7s & +55d23m30.7s & 0.064(1) & 5.8~~~~~~ & 0.964$\pm$0.029 & 10.33$\pm$0.02 \\
5MUSES-184 & 5MUSES$\_$J160847.02+563702.2 & 16h08m47.0s & +56d37m02.2s & 0.590(1) & 8.3~~~~~~ & 0.045$\pm$0.001 & 12.21$\pm$0.02 \\
5MUSES-185 & 5MUSES$\_$J160858.38+553010.2 & 16h08m58.4s & +55d30m10.3s & 0.066(1) & 8.8~~~~~~ & 0.586$\pm$0.104 & 10.34$\pm$0.02 \\
5MUSES-186 & 5MUSES$\_$J160858.66+563635.6 & 16h08m58.7s & +56d36m35.7s & 0.117(1) & 5.0~~~~~~ & 0.566$\pm$0.050 & 10.77$\pm$0.04 \\
5MUSES-187 & 5MUSES$\_$J160907.56+552428.4 & 16h09m07.6s & +55d24m28.4s & 0.065(1) & 7.7~~~~~~ & 0.670$\pm$0.003 & 10.54$\pm$0.03 \\
5MUSES-188 & 5MUSES$\_$J160908.28+552241.4 & 16h09m08.3s & +55d22m41.5s & 0.084(1) & 6.6~~~~~~ & 0.824$\pm$0.056 & 10.65$\pm$0.02 \\
5MUSES-189 & 5MUSES$\_$J160926.69+551642.3 & 16h09m26.7s & +55d16m42.3s & 0.068(2) & 6.8~~~~~~ & 0.507$\pm$0.058 & 10.19$\pm$0.02 \\
5MUSES-190 & 5MUSES$\_$J160930.53+563509.0 & 16h09m30.5s & +56d35m09.1s & 0.030(1) & 5.1~~~~~~ & 0.428$\pm$0.028 & 9.22$\pm$0.06 \\
5MUSES-191 & 5MUSES$\_$J160931.55+541827.3 & 16h09m31.6s & +54d18m27.4s & 0.082(1) & 5.6~~~~~~ & 0.497$\pm$0.033 & 10.61$\pm$0.04 \\
5MUSES-192 & 5MUSES$\_$J160937.48+541259.2 & 16h09m37.5s & +54d12m59.3s & 0.086(1) & 5.7~~~~~~ & 0.681$\pm$0.018 & 10.66$\pm$0.02 \\
5MUSES-193 & 5MUSES$\_$J161103.73+544322.0 & 16h11m03.7s & +54d43m22.1s & 0.063(2) & 6.6~~~~~~ & 0.536$\pm$0.018 & 10.26$\pm$0.03 \\
5MUSES-194 & 5MUSES$\_$J161119.36+553355.4 & 16h11m19.4s & +55d33m55.4s & 0.227(1) & 35.4~~~~~~ &  $<$0.100 & 11.76$\pm$0.03 \\
5MUSES-195 & 5MUSES$\_$J161123.44+545158.2 & 16h11m23.4s & +54d51m58.2s & 0.078(2) & 5.5~~~~~~ & 0.516$\pm$0.002 & 10.40$\pm$0.03 \\
5MUSES-196 & 5MUSES$\_$J161223.39+540339.2 & 16h12m23.4s & +54d03m39.2s & 0.138(2) & 13.0~~~~~~ & 0.839$\pm$0.136 & 11.07$\pm$0.03 \\
5MUSES-197 & 5MUSES$\_$J161233.43+545630.4 & 16h12m33.4s & +54d56m30.5s & 0.083(1) & 8.3~~~~~~ & 0.560$\pm$0.083 & 10.66$\pm$0.04 \\
5MUSES-198 & 5MUSES$\_$J161241.05+543956.8 & 16h12m41.1s & +54d39m56.8s & 0.035(2) & 5.7~~~~~~ & 0.841$\pm$0.078 & 9.51$\pm$0.03 \\
5MUSES-199 & 5MUSES$\_$J161249.54+564232.7 & 16h12m49.5s & +56d42m32.8s & 0.336(1) & 8.0~~~~~~ & 0.411$\pm$0.036 & 11.60$\pm$0.08 \\
5MUSES-200 & 5MUSES$\_$J161250.85+532304.9 & 16h12m50.9s & +53d23m05.0s & 0.048(2) & 17.9~~~~~~ & 0.405$\pm$0.074 & 10.40$\pm$0.05 \\
5MUSES-202 & 5MUSES$\_$J161254.17+545525.4 & 16h12m54.2s & +54d55m25.4s & 0.065(2) & 8.0~~~~~~ & 0.624$\pm$0.015 & 10.59$\pm$0.01 \\
5MUSES-203 & 5MUSES$\_$J161301.82+552123.0 & 16h13m01.8s & +55d21m23.1s & 0.012(2) & 36.3~~~~~~ & 0.563$\pm$0.044 & 9.47$\pm$0.05 \\
5MUSES-204 & 5MUSES$\_$J161357.01+534105.3 & 16h13m57.0s & +53d41m05.3s & 0.180(2) & 6.5~~~~~~ & 0.106$\pm$0.004 & 10.83$\pm$0.03 \\
5MUSES-205 & 5MUSES$\_$J161402.98+560756.9 & 16h14m03.0s & +56d07m57.0s & 0.063(2) & 21.0~~~~~~ & 0.746$\pm$0.052 & 10.79$\pm$0.06 \\
5MUSES-207 & 5MUSES$\_$J161406.87+551451.9 & 16h14m06.9s & +55d14m52.0s & 0.564(2) & 9.2~~~~~~ & 0.047$\pm$0.010 & 12.18$\pm$0.03 \\
5MUSES-208 & 5MUSES$\_$J161411.52+540554.3 & 16h14m11.5s & +54d05m54.3s & 0.305(1) & 5.9~~~~~~ & 0.587$\pm$0.123 & 11.72$\pm$0.04 \\
5MUSES-209 & 5MUSES$\_$J161449.08+554512.9 & 16h14m49.1s & +55d45m12.9s & 0.064(1) & 15.0~~~~~~ & 0.148$\pm$0.007 & 10.26$\pm$0.03 \\
5MUSES-210 & 5MUSES$\_$J161521.78+543148.3 & 16h15m21.8s & +54d31m48.3s & 0.474(1) & 5.1~~~~~~ &  $<$0.058 & 11.47$\pm$0.08 \\
5MUSES-211 & 5MUSES$\_$J161528.07+534402.4 & 16h15m28.1s & +53d44m02.5s & 0.133(2) & 6.0~~~~~~ & 0.476$\pm$0.071 & 11.01$\pm$0.03 \\
5MUSES-212 & 5MUSES$\_$J161542.10+561814.7 & 16h15m42.1s & +56d18m14.7s & 0.109(1) & 13.7~~~~~~ &  $<$0.150 & 10.67$\pm$0.04 \\
5MUSES-214 & 5MUSES$\_$J161546.51+550330.9 & 16h15m46.5s & +55d03m31.0s & 0.087(1) & 8.9~~~~~~ & 0.169$\pm$0.003 & 10.26$\pm$0.03 \\
5MUSES-215 & 5MUSES$\_$J161548.31+534551.1 & 16h15m48.3s & +53d45m51.1s & 0.147(1) & 7.5~~~~~~ & 0.512$\pm$0.104 & 11.18$\pm$0.03 \\
5MUSES-216 & 5MUSES$\_$J161551.45+541535.9 & 16h15m51.5s & +54d15m36.0s & 0.215(2) & 6.3~~~~~~ & 0.445$\pm$0.049 & 11.43$\pm$0.04 \\
5MUSES-217 & 5MUSES$\_$J161644.45+533734.0 & 16h16m44.4s & +53d37m34.3s & 0.147(1) & 8.8~~~~~~ & 0.828$\pm$0.080 & 11.19$\pm$0.02 \\
5MUSES-219 & 5MUSES$\_$J161645.92+542554.4 & 16h16m45.9s & +54d25m54.4s & 0.223(1) & 12.4~~~~~~ & 0.162$\pm$0.002 & 11.26$\pm$0.02 \\
5MUSES-220 & 5MUSES$\_$J161655.96+545307.0 & 16h16m56.0s & +54d53m07.1s & 0.418(1) & 5.1~~~~~~ & 0.391$\pm$0.039 & 11.81$\pm$0.05 \\
5MUSES-221 & 5MUSES$\_$J161659.95+560027.2 & 16h17m00.0s & +56d00m27.2s & 0.063(1) & 10.8~~~~~~ & 0.517$\pm$0.049 & 10.66$\pm$0.02 \\
5MUSES-222 & 5MUSES$\_$J161712.27+551853.0 & 16h17m12.3s & +55d18m53.0s & 0.037(1) & 6.7~~~~~~ & 0.714$\pm$0.030 & 9.53$\pm$0.06 \\
5MUSES-223 & 5MUSES$\_$J161716.57+550920.3 & 16h17m16.6s & +55d09m20.3s & 0.092(2) & 7.3~~~~~~ & 0.728$\pm$0.022 & 10.65$\pm$0.04 \\
5MUSES-225 & 5MUSES$\_$J161748.06+551831.1 & 16h17m48.1s & +55d18m31.1s & 0.145(1) & 7.0~~~~~~ & 0.363$\pm$0.030 & 11.13$\pm$0.05 \\
5MUSES-227 & 5MUSES$\_$J161759.22+541501.3 & 16h17m59.2s & +54d15m01.3s & 0.135(1) & 22.7~~~~~~ & 0.137$\pm$0.006 & 11.12$\pm$0.06 \\
5MUSES-228 & 5MUSES$\_$J161809.36+551522.0 & 16h18m09.4s & +55d15m22.1s & 0.136(1) & 6.4~~~~~~ & 0.137$\pm$0.002 & 10.71$\pm$0.05 \\
5MUSES-229 & 5MUSES$\_$J161819.31+541859.0 & 16h18m19.3s & +54d18m59.1s & 0.083(1) & 28.3~~~~~~ & 0.472$\pm$0.005 & 11.14$\pm$0.04 \\
5MUSES-230 & 5MUSES$\_$J161823.11+552721.4 & 16h18m23.1s & +55d27m21.4s & 0.084(1) & 25.3~~~~~~ & 0.613$\pm$0.016 & 11.13$\pm$0.03 \\
5MUSES-231 & 5MUSES$\_$J161827.72+552208.6 & 16h18m27.7s & +55d22m08.6s & 0.083(1) & 9.9~~~~~~ & 0.673$\pm$0.082 & 10.74$\pm$0.05 \\
5MUSES-232 & 5MUSES$\_$J161843.35+554433.1 & 16h18m43.4s & +55d44m33.1s & 0.153(1) & 10.1~~~~~~ & 0.618$\pm$0.046 & 11.29$\pm$0.03 \\
5MUSES-233 & 5MUSES$\_$J161848.03+535837.5 & 16h18m48.0s & +53d58m37.6s & 0.079(1) & 7.2~~~~~~ & 0.124$\pm$0.005 & 10.49$\pm$0.09 \\
5MUSES-234 & 5MUSES$\_$J161929.57+541841.9 & 16h19m29.6s & +54d18m41.9s & 0.100(1) & 16.5~~~~~~ & 0.487$\pm$0.048 & 11.07$\pm$0.02 \\
5MUSES-235 & 5MUSES$\_$J161950.52+543715.3 & 16h19m50.5s & +54d37m15.4s & 0.146(1) & 7.0~~~~~~ & 0.761$\pm$0.041 & 11.14$\pm$0.03 \\
5MUSES-239 & 5MUSES$\_$J162033.98+542323.5 & 16h20m34.0s & +54d23m23.5s & 0.133(1) & 9.1~~~~~~ & 0.622$\pm$0.058 & 11.07$\pm$0.05 \\
5MUSES-240 & 5MUSES$\_$J162038.10+553521.4 & 16h20m38.1s & +55d35m21.5s & 0.191(1) & 8.6~~~~~~ & 0.716$\pm$0.099 & 11.39$\pm$0.03 \\
5MUSES-241 & 5MUSES$\_$J162058.82+542513.1 & 16h20m58.8s & +54d25m13.2s & 0.082(1) & 21.3~~~~~~ & 0.880$\pm$0.005 & 11.11$\pm$0.03 \\
5MUSES-242 & 5MUSES$\_$J162059.02+542601.5 & 16h20m59.0s & +54d26m01.5s & 0.046(1) & 17.2~~~~~~ & 0.732$\pm$0.068 & 10.20$\pm$0.07 \\
5MUSES-243 & 5MUSES$\_$J162110.51+544116.8 & 16h21m10.5s & +54d41m16.8s & 0.155(1) & 9.0~~~~~~ & 0.175$\pm$0.008 & 10.92$\pm$0.06 \\
5MUSES-244 & 5MUSES$\_$J162127.98+551452.9 & 16h21m28.0s & +55d14m52.9s & 0.100(1) & 5.6~~~~~~ & 0.707$\pm$0.091 & 10.74$\pm$0.02 \\
5MUSES-245 & 5MUSES$\_$J162133.00+551829.9 & 16h21m33.0s & +55d18m29.9s & 0.238(1) & 7.7~~~~~~ & 0.494$\pm$0.081 & 11.27$\pm$0.13 \\
5MUSES-247 & 5MUSES$\_$J162150.85+553008.8 & 16h21m50.9s & +55d30m08.9s & 0.099(1) & 6.6~~~~~~ & 0.911$\pm$0.009 & 10.82$\pm$0.02 \\
5MUSES-248 & 5MUSES$\_$J162210.87+550253.7 & 16h22m10.9s & +55d02m53.8s & 0.034(1) & 47.7~~~~~~ & 0.527$\pm$0.062 & 10.58$\pm$0.04 \\
5MUSES-249 & 5MUSES$\_$J162214.77+550614.1 & 16h22m14.8s & +55d06m14.2s & 0.237(1) & 7.4~~~~~~ & 0.470$\pm$0.021 & 11.67$\pm$0.02 \\
5MUSES-250 & 5MUSES$\_$J162313.11+551111.5 & 16h23m13.1s & +55d11m11.6s & 0.236(1) & 6.6~~~~~~ & 0.405$\pm$0.001 & 11.67$\pm$0.02 \\
5MUSES-251 & 5MUSES$\_$J163001.46+410952.9 & 16h30m01.5s & +41d09m52.9s & 0.121(1) & 7.3~~~~~~ & 0.697$\pm$0.127 & 10.84$\pm$0.01 \\
5MUSES-252 & 5MUSES$\_$J163111.27+404805.2 & 16h31m11.3s & +40d48m05.2s & 0.258(1) & 16.7~~~~~~ & 0.042$\pm$0.002 & 11.30$\pm$0.24 \\
5MUSES-253 & 5MUSES$\_$J163128.57+404536.0 & 16h31m28.6s & +40d45m36.0s & 0.181(1) & 14.8~~~~~~ & 0.170$\pm$0.009 & 11.13$\pm$0.04 \\
5MUSES-254 & 5MUSES$\_$J163220.40+402334.4 & 16h32m20.4s & +40d23m34.4s & 0.079(1) & 8.3~~~~~~ & 0.602$\pm$0.002 & 10.79$\pm$0.05 \\
5MUSES-255 & 5MUSES$\_$J163308.28+403321.5 & 16h33m08.3s & +40d33m21.6s & 0.404(1) & 8.3~~~~~~ & 0.164$\pm$0.013 & 11.82$\pm$0.05 \\
5MUSES-256 & 5MUSES$\_$J163310.92+405641.3 & 16h33m10.9s & +40d56m41.4s & 0.136(1) & 8.0~~~~~~ & 0.725$\pm$0.022 & 10.86$\pm$0.03 \\
5MUSES-258 & 5MUSES$\_$J163317.57+403443.6 & 16h33m17.6s & +40d34m43.6s & 0.378(1) & 7.2~~~~~~ &  $<$0.073 & 11.46$\pm$0.03 \\
5MUSES-260 & 5MUSES$\_$J163335.85+401529.1 & 16h33m35.9s & +40d15m29.1s & 0.028(1) & 30.3~~~~~~ & 0.954$\pm$0.037 & 10.07$\pm$0.04 \\
5MUSES-261 & 5MUSES$\_$J163359.12+405304.7 & 16h33m59.1s & +40d53m04.7s & 0.032(1) & 11.9~~~~~~ & 0.474$\pm$0.027 & 9.92$\pm$0.03 \\
5MUSES-262 & 5MUSES$\_$J163401.79+412052.5 & 16h34m01.8s & +41d20m52.6s & 0.028(1) & 47.0~~~~~~ & 0.739$\pm$0.030 & 10.32$\pm$0.04 \\
5MUSES-263 & 5MUSES$\_$J163506.06+411038.4 & 16h35m06.1s & +41d10m38.5s & 0.079(1) & 13.5~~~~~~ & 0.462$\pm$0.001 & 10.83$\pm$0.03 \\
5MUSES-264 & 5MUSES$\_$J163541.68+405900.6 & 16h35m41.7s & +40d59m00.7s & 0.188(1) & 10.4~~~~~~ &  $<$0.189 & 11.04$\pm$0.03 \\
5MUSES-265 & 5MUSES$\_$J163546.87+403903.6 & 16h35m46.9s & +40d39m03.6s & 0.122(1) & 8.3~~~~~~ & 0.613$\pm$0.007 & 11.09$\pm$0.03 \\
5MUSES-266 & 5MUSES$\_$J163608.13+410507.6 & 16h36m08.1s & +41d05m07.7s & 0.170(1) & 13.2~~~~~~ & 0.457$\pm$0.010 & 11.91$\pm$0.10 \\
5MUSES-267 & 5MUSES$\_$J163645.27+415133.6 & 16h36m45.3s & +41d51m33.7s & 0.081(1) & 7.8~~~~~~ &  $<$0.190 & 10.24$\pm$0.06 \\
5MUSES-268 & 5MUSES$\_$J163651.65+405600.1 & 16h36m51.7s & +40d56m00.2s & 0.476(1) & 9.6~~~~~~ &  $<$0.101 & 11.83$\pm$0.02 \\
5MUSES-269 & 5MUSES$\_$J163705.29+413155.8 & 16h37m05.3s & +41d31m55.9s & 0.122(2) & 10.6~~~~~~ & 0.704$\pm$0.001 & 11.20$\pm$0.03 \\
5MUSES-270 & 5MUSES$\_$J163709.31+414030.8 & 16h37m09.3s & +41d40m30.9s & 0.760(1) & 9.5~~~~~~ &  $<$0.032 & 12.41$\pm$0.05 \\
5MUSES-271 & 5MUSES$\_$J163715.58+414933.7 & 16h37m15.6s & +41d49m33.7s & 0.121(1) & 8.8~~~~~~ & 0.580$\pm$0.012 & 10.95$\pm$0.03 \\
5MUSES-272 & 5MUSES$\_$J163729.26+405248.5 & 16h37m29.3s & +40d52m48.5s & 0.026(2) & 19.1~~~~~~ & 0.406$\pm$0.020 & 10.10$\pm$0.04 \\
5MUSES-273 & 5MUSES$\_$J163731.41+405155.5 & 16h37m31.4s & +40d51m55.6s & 0.189(1) & 7.6~~~~~~ & 0.404$\pm$0.045 & 11.44$\pm$0.05 \\
5MUSES-274 & 5MUSES$\_$J163751.24+401439.9 & 16h37m51.2s & +40d14m39.9s & 0.072(2) & 11.8~~~~~~ & 0.880$\pm$0.043 & 10.63$\pm$0.03 \\
5MUSES-275 & 5MUSES$\_$J163751.35+413027.3 & 16h37m51.4s & +41d30m27.3s & 0.287(1) & 25.8~~~~~~ & 0.131$\pm$0.011 & 12.04$\pm$0.04 \\
5MUSES-276 & 5MUSES$\_$J163751.85+401503.9 & 16h37m51.9s & +40d15m04.0s & 0.070(2) & 8.6~~~~~~ & 0.771$\pm$0.021 & 10.61$\pm$0.03 \\
5MUSES-277 & 5MUSES$\_$J163802.24+404653.4 & 16h38m02.2s & +40d46m53.4s & 0.103(2) & 9.1~~~~~~ & 0.204$\pm$0.005 & 10.56$\pm$0.06 \\
5MUSES-278 & 5MUSES$\_$J163805.85+413508.1 & 16h38m05.9s & +41d35m08.2s & 0.119(2) & 10.6~~~~~~ & 0.573$\pm$0.134 & 11.02$\pm$0.03 \\
5MUSES-279 & 5MUSES$\_$J163808.47+403213.7 & 16h38m08.5s & +40d32m13.8s & 0.220(2) & 11.9~~~~~~ & 0.486$\pm$0.015 & 11.69$\pm$0.05 \\
5MUSES-280 & 5MUSES$\_$J163809.65+402844.7 & 16h38m09.6s & +40d28m44.8s & 0.072(2) & 17.3~~~~~~ & 0.696$\pm$0.067 & 10.55$\pm$0.04 \\
5MUSES-281 & 5MUSES$\_$J163906.16+404003.2 & 16h39m06.2s & +40d40m03.3s & 0.035(1) & 6.7~~~~~~ & 0.719$\pm$0.007 & 9.82$\pm$0.03 \\
5MUSES-282 & 5MUSES$\_$J164019.68+403744.4 & 16h40m19.7s & +40d37m44.4s & 0.151(1) & 10.5~~~~~~ &  $<$0.199 & 10.76$\pm$0.08 \\
5MUSES-284 & 5MUSES$\_$J164043.69+413310.0 & 16h40m43.7s & +41d33m10.0s & 0.155(2) & 5.7~~~~~~ & 0.720$\pm$0.006 & 11.14$\pm$0.04 \\
5MUSES-285 & 5MUSES$\_$J164046.60+412522.6 & 16h40m46.6s & +41d25m22.6s & 0.096(2) & 20.7~~~~~~ & 0.098$\pm$0.003 & 10.78$\pm$0.05 \\
5MUSES-286 & 5MUSES$\_$J164101.35+411850.6 & 16h41m01.4s & +41d18m50.7s & 0.099(2) & 22.1~~~~~~ & 0.072$\pm$0.013 & 10.67$\pm$0.05 \\
5MUSES-287 & 5MUSES$\_$J164115.38+410320.7 & 16h41m15.4s & +41d03m20.7s & 0.138(2) & 5.6~~~~~~ & 0.519$\pm$0.006 & 11.14$\pm$0.02 \\
5MUSES-288 & 5MUSES$\_$J164135.27+413807.3 & 16h41m35.3s & +41d38m07.3s & 0.395(2) & 5.3~~~~~~ & 0.072$\pm$0.003 & 11.58$\pm$0.07 \\
5MUSES-289 & 5MUSES$\_$J164153.76+405842.5 & 16h41m53.8s & +40d58m42.6s & 0.327(2) & 5.9~~~~~~ & 0.119$\pm$0.004 & 11.43$\pm$0.03 \\
5MUSES-290 & 5MUSES$\_$J164211.92+410816.7 & 16h42m11.9s & +41d08m16.8s & 0.144(2) & 11.7~~~~~~ & 0.546$\pm$0.013 & 11.36$\pm$0.04 \\
5MUSES-291 & 5MUSES$\_$J164214.47+405129.0 & 16h42m14.5s & +40d51m29.0s & 0.104(2) & 14.1~~~~~~ &  $<$0.058 & 10.62$\pm$0.01 \\
5MUSES-292 & 5MUSES$\_$J171033.21+584456.8 & 17h10m33.2s & +58d44m56.7s & 0.281(2) & 6.1~~~~~~ & 0.325$\pm$0.001 & 11.39$\pm$0.07 \\
5MUSES-293 & 5MUSES$\_$J171124.22+593121.4 & 17h11m24.2s & +59d31m21.5s & 1.489(2) & 5.6~~~~~~ &  $<$0.080 & 12.92$\pm$0.02 \\
5MUSES-294 & 5MUSES$\_$J171232.34+592125.9 & 17h12m32.4s & +59d21m26.2s & 0.210(2) & 8.7~~~~~~ & 0.507$\pm$0.006 & 11.59$\pm$0.04 \\
5MUSES-295 & 5MUSES$\_$J171233.38+583610.5 & 17h12m33.4s & +58d36m10.3s & 1.663(1) & 5.1~~~~~~ &  $<$0.113 & 13.18$\pm$0.04 \\
5MUSES-296 & 5MUSES$\_$J171233.77+594026.4 & 17h12m33.7s & +59d40m26.8s & 0.217(2) & 5.~~~~~~1 & 0.983$\pm$0.067 & 11.29$\pm$0.03 \\
5MUSES-297 & 5MUSES$\_$J171316.50+583234.9 & 17h13m16.6s & +58d32m34.9s & 0.079(2) & 6.7~~~~~~ & 0.780$\pm$0.020 & 10.34$\pm$0.04 \\
5MUSES-298 & 5MUSES$\_$J171325.18+590531.1 & 17h13m25.2s & +59d05m31.2s & 0.126(1) & 9.4~~~~~~ &  $<$0.189 & 10.33$\pm$0.06 \\
5MUSES-299 & 5MUSES$\_$J171414.81+585221.5 & 17h14m14.8s & +58d52m21.6s & 0.167(1) & 9.0~~~~~~ & 0.780$\pm$0.006 & 11.19$\pm$0.03 \\
5MUSES-300 & 5MUSES$\_$J171419.98+602724.6 & 17h14m20.0s & +60d27m24.8s & 2.990(1) & 5.6~~~~~~ &  \nodata  & 13.79$\pm$0.09 \\
5MUSES-301 & 5MUSES$\_$J171430.76+584225.4 & 17h14m30.8s & +58d42m25.4s & 0.562(2) & 8.3~~~~~~ &  $<$0.075 & 11.70$\pm$0.06 \\
5MUSES-302 & 5MUSES$\_$J171446.47+593400.1 & 17h14m46.4s & +59d33m59.8s & 0.129(1) & 7.5~~~~~~ & 0.637$\pm$0.002 & 11.11$\pm$0.02 \\
5MUSES-303 & 5MUSES$\_$J171447.31+583805.9 & 17h14m47.3s & +58d38m05.8s & 0.257(2) & 5.4~~~~~~ & 0.836$\pm$0.012 & 11.60$\pm$0.04 \\
5MUSES-304 & 5MUSES$\_$J171513.88+594638.1 & 17h15m13.8s & +59d46m38.3s & 0.248(1) & 5.1~~~~~~ & 0.338$\pm$0.091 & 11.21$\pm$0.04 \\
5MUSES-305 & 5MUSES$\_$J171544.03+600835.3 & 17h15m44.0s & +60d08m35.2s & 0.157(2) & 6.9~~~~~~ &  $<$0.190 & 10.72$\pm$0.04 \\
5MUSES-306 & 5MUSES$\_$J171550.50+593548.8 & 17h15m50.5s & +59d35m48.7s & 0.066(2) & 9.1~~~~~~ & 0.073$\pm$0.005 & 10.16$\pm$0.04 \\
5MUSES-307 & 5MUSES$\_$J171614.48+595423.8 & 17h16m14.5s & +59d54m23.6s & 0.153(2) & 8.6~~~~~~ & 0.827$\pm$0.009 & 11.29$\pm$0.03 \\
5MUSES-308 & 5MUSES$\_$J171630.23+601422.7 & 17h16m30.2s & +60d14m22.7s & 0.107(1) & 8.6~~~~~~ & 0.833$\pm$0.133 & 10.75$\pm$0.05 \\
5MUSES-309 & 5MUSES$\_$J171650.58+595751.4 & 17h16m50.6s & +59d57m52.0s & 0.182(1) & 6.8~~~~~~ &  $<$0.313 & 10.74$\pm$0.10 \\
5MUSES-310 & 5MUSES$\_$J171711.11+602710.0 & 17h17m11.1s & +60d27m10.0s & 0.110(1) & 9.5~~~~~~ & 0.488$\pm$0.053 & 10.78$\pm$0.06 \\
5MUSES-311 & 5MUSES$\_$J171747.51+593258.1 & 17h17m47.5s & +59d32m58.1s & 0.248(2) & 5.3~~~~~~ &  $<$0.093 & 10.76$\pm$0.12 \\
5MUSES-312 & 5MUSES$\_$J171754.62+600913.8 & 17h17m54.6s & +60d09m13.4s & 4.270(1) & 9.1~~~~~~ &  \nodata  & 14.59$\pm$0.13 \\
5MUSES-313 & 5MUSES$\_$J171852.71+591432.0 & 17h18m52.7s & +59d14m32.1s & 0.322(2) & 14.0~~~~~~ & 0.112$\pm$0.010 & 11.85$\pm$0.05 \\
5MUSES-314 & 5MUSES$\_$J171913.57+584509.1 & 17h19m13.5s & +58d45m08.9s & 0.318(2) & 8.8~~~~~~ &  $<$0.243 & 11.42$\pm$0.12 \\
5MUSES-315 & 5MUSES$\_$J171933.37+592742.8 & 17h19m33.3s & +59d27m42.7s & 0.139(2) & 7.6~~~~~~ & 0.495$\pm$0.005 & 11.28$\pm$0.07 \\
5MUSES-316 & 5MUSES$\_$J171944.91+595707.7 & 17h19m44.9s & +59d57m07.1s & 0.069(2) & 14.4~~~~~~ & 0.753$\pm$0.005 & 10.73$\pm$0.05 \\
5MUSES-317 & 5MUSES$\_$J172043.28+584026.6 & 17h20m43.3s & +58d40m26.9s & 0.125(2) & 9.7~~~~~~ & 0.498$\pm$0.006 & 11.14$\pm$0.03 \\
5MUSES-318 & 5MUSES$\_$J172044.86+582924.0 & 17h20m44.9s & +58d29m23.9s & 1.697(1) & 5.3~~~~~~ &  $<$0.094 & 13.07$\pm$0.05 \\
5MUSES-319 & 5MUSES$\_$J172159.43+595034.3 & 17h21m59.3s & +59d50m34.2s & 0.028(2) & 9.7~~~~~~ & 0.387$\pm$0.031 & 9.78$\pm$0.03 \\
5MUSES-320 & 5MUSES$\_$J172219.58+594506.9 & 17h22m19.6s & +59d45m07.0s & 0.272(2) & 7.8~~~~~~ &  $<$0.133 & 11.24$\pm$0.02 \\
5MUSES-321 & 5MUSES$\_$J172228.04+601526.0 & 17h22m28.2s & +60d15m26.2s & 0.742(2) & 7.2~~~~~~ &  $<$0.111 & 12.40$\pm$0.03 \\
5MUSES-322 & 5MUSES$\_$J172238.73+585107.0 & 17h22m38.8s & +58d51m07.0s & 1.624(1) & 6.7~~~~~~ &  $<$0.062 & 13.12$\pm$0.04 \\
5MUSES-323 & 5MUSES$\_$J172313.06+590533.1 & 17h23m13.1s & +59d05m33.1s & 0.108(2) & 6.2~~~~~~ & 0.750$\pm$0.037 & 10.85$\pm$0.03 \\
5MUSES-324 & 5MUSES$\_$J172355.58+601301.7 & 17h23m55.5s & +60d13m01.1s & 0.175(2) & 5.4~~~~~~ & 0.905$\pm$0.034 & 11.13$\pm$0.02 \\
5MUSES-325 & 5MUSES$\_$J172355.97+594047.6 & 17h23m56.0s & +59d40m47.4s & 0.030(2) & 5.2~~~~~~ & 0.518$\pm$0.098 & 9.35$\pm$0.04 \\
5MUSES-326 & 5MUSES$\_$J172402.11+600601.4 & 17h24m02.1s & +60d06m01.2s & 0.156(2) & 8.0~~~~~~ & 0.461$\pm$0.024 & 11.13$\pm$0.03 \\
5MUSES-328 & 5MUSES$\_$J172546.80+593655.3 & 17h25m46.8s & +59d36m55.3s & 0.035(2) & 26.0~~~~~~ & 0.554$\pm$0.041 & 10.49$\pm$0.04 \\
5MUSES-329 & 5MUSES$\_$J172551.35+601138.9 & 17h25m51.3s & +60d11m38.9s & 0.029(1) & 27.3~~~~~~ & 0.454$\pm$0.005 & 10.25$\pm$0.03 \\
5MUSES-330 & 5MUSES$\_$J172619.80+601600.1 & 17h26m19.8s & +60d16m00.0s & 0.924(1) & 6.5~~~~~~ &  $<$0.039 & 12.35$\pm$0.08 \\
  \enddata
\tablenotetext{a}{The redshifts obtained from NASA/IPAC Extragalactic Database are indicated with ``1'' while the redshifts derived from IRS spectra are indicated with ``2''.}
\end{deluxetable}

\begin{deluxetable}{lcccc}
  \setlength{\tabcolsep}{0.1in}
  \tablecaption{Median Luminosity Ratios of the Sample\label{l_ir_tab}}
  \tablewidth{0pc}
  \tablehead{
  \colhead{} & \colhead{SB} & \colhead {composite} & \colhead{AGN} & \colhead{Whole Sample}
  }
  \startdata
  log(L$_{\rm PAH 6.2\mu m}$/L$_{\rm IR}$)            &   -2.03$\pm$0.13\tablenotemark{a}    &    -2.11$\pm$0.13    &   -2.22$\pm$0.14    &   -2.06$\pm$0.14  \\
  log(L$_{\rm PAH 7.7\mu m}$/L$_{\rm IR}$)            &   -1.51$\pm$0.15    &    -1.54$\pm$0.18    &   -1.76$\pm$0.28    &   -1.53$\pm$0.20   \\
  log(L$_{\rm PAH 11.3\mu m}$/L$_{\rm IR}$)           &   -2.04$\pm$0.13    &    -2.14$\pm$0.20    &   -2.20$\pm$0.18    &   -2.08$\pm$0.17   \\
  log(L$_{\rm PAH 6.2+7.7+11.3\mu m}$/L$_{\rm IR}$)    &   -1.29$\pm$0.14    &    -1.33$\pm$0.12    &   -1.39$\pm$0.19    &   -1.31$\pm$0.14   \\
  log(L$_{\rm 5.8\mu m}$/L$_{\rm IR}$)               &   -1.44$\pm$0.15    &    -1.36$\pm$0.29    &   -0.77$\pm$0.32    &   -1.33$\pm$0.42   \\
  log(L$_{\rm IRAC 8\mu m}$/L$_{\rm IR}$)             &   -0.90$\pm$0.13    &    -0.96$\pm$0.15    &   -0.66$\pm$0.23    &  -0.86$\pm$0.22   \\
  log(L$_{\rm 14\mu m}$/L$_{\rm IR}$)                &   -1.19$\pm$0.10    &    -1.11$\pm$0.21    &   -0.57$\pm$0.19    &  -1.12$\pm$0.31   \\
  log(L$_{\rm 24\mu m}$/L$_{\rm IR}$)                &   -0.87$\pm$0.14    &    -0.82$\pm$0.23    &   -0.49$\pm$0.15    &  -0.81$\pm$0.22   \\  
 \enddata
\tablenotetext{a}{The dispersion is the 1$\sigma$ deviation for each group of objects.}
\end{deluxetable}

\begin{deluxetable}{cccccc}
  \setlength{\tabcolsep}{0.1in}
  \tablecaption{Median PAH strengths and Continuum Ratios of the Sample\label{pahew_slope}}
  \tablewidth{0pc}
  \tablehead{
  \colhead{log(f$_{\rm 30}$/f$_{\rm 15}$)\tablenotemark{a}} & \colhead{log(f$_{\rm 70}$/f$_{\rm 24}$)} & \colhead{6.2\,$\mu$m PAH EW\tablenotemark{b}} & \colhead{log(L$_{\rm PAH}$/L$_{\rm IR}$)\tablenotemark{c}} & \colhead{a} & \colhead{b}
  }
  \startdata
  {\bf$>$0.898}           &   0.99$^{+0.08}_{-0.17}$   &   0.57$^{+0.18}_{-0.16}$ (56)\tablenotemark{d}  &   -1.40 & 0.65$\pm$0.10  &  0.94$\pm$0.14 \\
  {\bf0.793$-$0.898}      &   0.97$^{+0.11}_{-0.18}$   &   0.62$^{+0.21}_{-0.13}$ (56)  &   -1.31 & 0.79$\pm$0.33  &  0.74$\pm$0.61 \\
  {\bf0.650$-$0.793}      &   1.04$^{+0.05}_{-0.35}$   &   0.58$^{+0.17}_{-0.28}$ (55)  &   -1.29 & 0.70$\pm$0.28  &  0.94$\pm$0.65 \\
  {\bf0.376$-$0.650}      &   0.57$^{+0.42}_{-0.21}$   &   0.19$^{+0.36}_{-0.09}$ (55)  &   -1.54 & 0.41$\pm$0.07  &  1.72$\pm$0.31 \\
  {\bf$<$0.376}           &   0.25$^{+0.21}_{-0.29}$   &   0.08$^{+0.05}_{-0.04}$ (53)  &   -2.09 & 0.50$\pm$0.03  &  1.04$\pm$0.18 \\
  \hline
  0.76$^{+0.13}_{-0.23}$  &   {\bf$>$1.053}            &  0.61$^{+0.15}_{-0.14}$ (56)   &   -1.29 & 0.62$\pm$0.07  &  0.57$\pm$0.10 \\
  0.69$^{+0.18}_{-0.35}$  &   {\bf0.954$-$1.053}       &  0.61$^{+0.15}_{-0.14}$ (56)   &   -1.31 & 0.33$\pm$0.10  &  0.99$\pm$0.18 \\
  0.85$^{+0.18}_{-0.42}$  &   {\bf0.733$-$0.954}       &  0.60$^{+0.18}_{-0.21}$ (55)   &   -1.36 & 0.41$\pm$0.02  &  0.52$\pm$0.06 \\
  0.61$^{+0.32}_{-0.31}$  &   {\bf0.381$-$0.733}       &  0.16$^{+0.25}_{-0.08}$ (56)   &   -1.60 & 0.47$\pm$0.01  &  0.52$\pm$0.06 \\
  0.57$^{+0.31}_{-0.30}$  &   {\bf$<$0.381}            &  0.08$^{+0.08}_{-0.03}$ (52)   &   -2.07 & 0.44$\pm$0.02  &  0.23$\pm$0.04 \\
 \enddata
 \tablenotetext{a}{We sort the spectra by the continuum slope and divide the objects into five groups, in which each group have the same number of sources (56).}
 \tablenotetext{b}{The upper limits are also included.}
 \tablenotetext{c}{The PAH luminosity is the sum of the 6.2, 7.7 and 11.3\,$\mu$m PAH luminosities measured from the composite spectra.}
 \tablenotetext{d}{The number in the parenthesis indicates the number of objects that we have measured the 6.2\,$\mu$m PAH EW or PAH luminosity.}
\end{deluxetable}

\begin{figure}
  \epsscale{1.0}
  \plotone{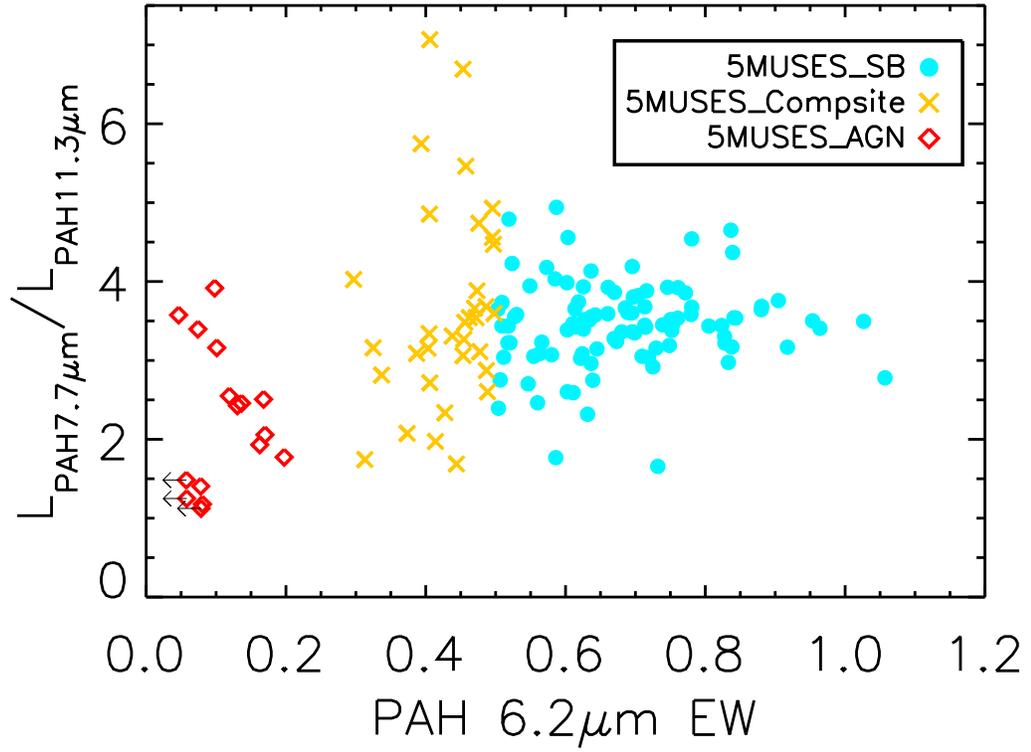}
  \caption{The PAH luminosity ratio of L$_{\rm PAH 7.7\mu
      m}$/L$_{\rm PAH 11.3\mu m}$ versus the 6.2\,$\mu$m PAH EW for
    5MUSES. The AGN-dominated sources appear to have lower L$_{\rm PAH
      7.7\mu m}$/L$_{\rm PAH 11.3\mu m}$ ratios than the composite or
    SB dominated sources. The mean ratios are 3.45$\pm$0.55,
    3.65$\pm$1.28 and 2.26$\pm$0.89 respectively for SB, composite and
    AGN. }
  \label{fig:5muses_77_11}
\end{figure}

\begin{figure}
  \epsscale{1.}
  \plotone{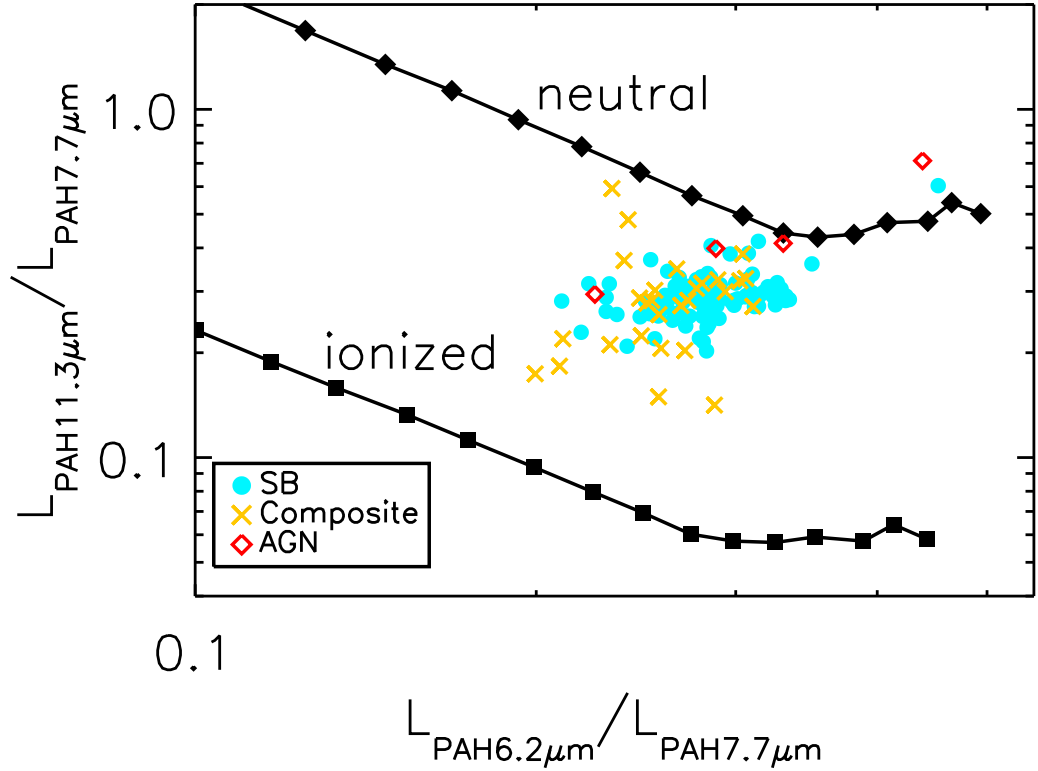}
  \caption{The PAH band-to-band ratios of L$_{\rm PAH 11.3\mu
      m}$/L$_{\rm PAH 7.7\mu m}$ versus L$_{\rm PAH 6.2\mu m}$/L$_{\rm
      PAH 7.7\mu m}$. The lines represent the expected ratios for
    neutral (upper line) and ionized (lower line) PAHs from model
    predictions. Note we have fewer objects in this Figure than in
    Figure \ref{fig:5muses_77_11}, because we require the source to
    have S/N$>$3 for all three PAH features (6.2, 7.7 and
    11.3\,$\mu$m) to be included.}
  \label{fig:pah627711}
\end{figure} 

\end{document}